\documentclass{ws-ijmpa}

\newcommand{\cN}{{\cal N}}
\newcommand{\Tr}{{\rm Tr\;}}

\newcommand{\etab}{\overline{\eta}}
\newcommand{\half}{\frac{1}{2}}

\def\beq{\begin{equation}}
\def\eeq{\end{equation}}
\def\bea{\begin{eqnarray}}
\def\eea{\end{eqnarray}}
\def\nn{\nonumber}
\def\bec{\begin{center}}
\def\eec{\end{center}}

\newcommand{\cO}{{\cal O}}
\newcommand{\cA}{{\cal A}}

\newcommand{\cAb}{{\overline{\cal A}}}
\newcommand{\cF}{{\cal F}}
\newcommand{\cFb}{{\overline{\cal F}}}
\newcommand{\cD}{{\cal D}}
\newcommand{\cDb}{{\overline{\cal D}}}
\newcommand{\cQ}{{\cal Q}}
\newcommand{\cU}{{\cal U}}

\newcommand{\cUb}{{\overline{\cal U}}} 
\newcommand{\lambdabar}{\overline{\lambda}}

\newcommand{\psib}{{\overline{\psi}}}
\newcommand{\phib}{{\overline{\phi}}}

\newcommand{\ba}{{\boldsymbol a}}

\newcommand{\bn}{{\boldsymbol n}}
\newcommand{\br}{{\boldsymbol r}}
\newcommand{\bR}{{\boldsymbol R}}
\newcommand{\hatbe}{\widehat{\boldsymbol e}}

\newcommand{\hatbg}{\widehat{\boldsymbol g}}
\newcommand{\hatbmu}{\widehat{\boldsymbol {\mu}}}

\newcommand{\vx}{ {\bf x} }
\newcommand{\vk}{ {\bf k} }
\newcommand{\vq}{ {\bf q} }
\newcommand{\vp}{ {\bf p} }

\newcommand{\cbar}{{\overline{c}}}

\begin{document}

\markboth{Anosh Joseph}
{Supersymmetric Yang--Mills theories with exact supersymmetry on the lattice}

%
\catchline{}{}{}{}{}
%

\title{Supersymmetric Yang--Mills theories with exact supersymmetry on the lattice\footnote{Preprint no. LA-UR-11-11648}}

\author{ANOSH JOSEPH}

\address{Theoretical Division, Los Alamos National Laboratory, Los Alamos, NM 87545, USA\\
anosh@lanl.gov}

\maketitle

\begin{abstract}
Inspired by the ideas from topological field theory it is possible to rewrite the supersymmetric charges of certain classes of extended supersymmetric Yang--Mills (SYM) theories in such a way that they are compatible with the discretization on a Euclidean spacetime lattice. Such theories are known as {\it maximally twisted} SYM theories. In this review we discuss the construction and some applications of such classes of theories. The one-loop perturbative renormalization of the four-dimensional lattice $\cN=4$ SYM is discussed in particular. The lattice theories constructed using twisted approach play an important role in investigating the thermal phases of strongly coupled SYM theories and also the thermodynamic properties of their dual gravitational theories.  

\end{abstract}

\ccode{PACS numbers: 12.60.Jv, 11.25.Tq, 11.10.Gh, 11.15.Ha}

\section{Introduction}	
\label{sec:intro}

Formulation of discretized supersymmetric Yang--Mills (SYM) theories in a way ensuring compatibility with the lattice is an old problem in lattice field theory. Recent developments based on the ideas taken from topological field theories and also orbifold/deconstruction techniques have created promising pathways into the solution of this long standing problem. In this review we focus on a class of lattice gauge theories constructed from {\it topologically twisting} the continuum SYM theories. The twisted SYM theories on the lattice turn out to be local, doublers free, and exact gauge-invariant theories, which are invariant under one or more supersymmetries. These theories, in principle, are qualified enough to form the basis for a truly non-perturbative definition of the continuum SYM theories. 

To make this review as self-contained as possible, we have included a set of introductory topics, such as the construction of SYM theories in various dimensions, general properties of topological field theories, and their connections to SYM theories and lattice formulations, and the geometric structure of the resultant lattices.

We introduce the four-dimensional $\cN=4$ SYM theory in Section 2. To maintain the symmetry between the number of fermion and boson degrees of freedom in a given Yang--Mills theory coupled to spin-1/2 fermions in arbitrary dimensions, various conditions can be imposed on Dirac fermions in various dimensions. These conditions lead to Weyl, Majorana, and Weyl--Majorana fermions. The method of dimensional reduction is then applied to the ten-dimensional $\cN=1$ SYM theory to obtain the $\cN=4$ SYM theory in four dimensions. 

In Section 3, we introduce topological field theories, BRST invariance in gauge theories, and then discuss topological field theories of Witten type, which are the focus of our interest. 

We show how to twist the supersymmetries of SYM theories with extended supersymmetries in Section 4. The method of maximal twisting is discussed, and then its relevance to the lattice constructions is explained. We give the twisted versions of the two-dimensional $\cN=2$ and four-dimensional $\cN=4$ SYM theories, exposing the nilpotent scalar supersymmetries appearing as a consequence of the twist. We also write down the action and scalar supersymmetries of these theories. 

In Section 5, we introduce supersymmetric lattices, their geometric structure, orientation of the field operators, covariant derivatives on the lattice, connection to Dirac--K\"ahler fermions, and discretized supersymmetries and actions of the twisted SYM theories. 

After these introductory sections, in Section 6, we review the $\cN=4$ SYM at one-loop on a four-dimensional lattice. The lattice formulation of the $\cN=4$ SYM retains one exact supersymmetry at non-zero lattice spacing. This feature, combined with gauge-invariance and the large point group symmetry of the lattice theory, can be used to show that the only counterterms that appear at any order in perturbation theory correspond to the renormalizations of existing terms in the bare lattice action. Also it can be shown that mass terms are not generated at any finite order of perturbation theory. The one-loop renormalization coefficients of the theory exhibit a common logarithmic divergence that can be absorbed by a single wavefunction renormalization. This implies that for the lattice $\cN=4$ theory, at one-loop, only a fine tuning of the finite parts is required to regain full supersymmetry in the continuum limit. 

In Section 7, we review some of the applications of the lattice constructions of twisted SYM theories in the context of the AdS/CFT correspondence. We discuss recent results obtained from lattice simulations of the SYM quantum mechanics and the two-dimensional SYM theory at finite temperature. Both the gauge theories appear in the generalizations of the holographic duality conjecture. The dual gravitational theory of SYM quantum mechanics contains black holes. As the temperature changes, there is a transition between black holes (described by a confined phase in the gauge theory) and a gas of hot strings (described by a deconfined phase in the gauge theory) in the dual gravitational theory. The transition is continuous and it passes through a point known as the Horowitz--Polchinski correspondence point. In the case of the two-dimensional supersymmettic Yang--Mills theory, the dual gravitational theory has an even more interesting phase structure. At large $N$, the (1+1)-dimensional sixteen supercharge $SU(N)$ Yang--Mills is believed to describe a dual description of the decoupling limit of $N$ coincident D1-branes on a circle. The D1-branes have charge and mass, and they describe certain black hole geometry. In the dual gravitational theory, at large $N$ and strong coupling, it has been proposed that, there is a phase transition related to the Gregory--Laflamme phase transition connecting a localized black hole solution and a uniform black hole solution wrapping around the compactified direction. The study of the time and space Polyakov loops in the two-dimensional gauge theory may show evidence supporting this transition. This gauge theory is analyzed on the lattice using twisted construction, and indeed the behavior the time and space Polyakov loops strongly suggests such a transition. In particular, at strong coupling, the transition has the parametric dependence on coupling predicted by the gravitational theory. The temperature corresponding to the Gregory--Laflamme phase transition can be estimated from the lattice study of the gauge theory which, interestingly, is not yet known directly in the gravitational dual. 

\section{$\cN=4$ super Yang--Mills theory} 
\label{sec:SYMtheory}

Supersymmtric Yang--Mills (SYM) theories form an interesting class of quantum field theories. Among them, the four-dimensional SYM theory with sixteen supersymmetries is a very special quantum field theory in its own right. This theory exhibits many interesting properties. For zero theta angle, the four-dimensional SYM theory with a simple gauge group has just a single dimensionless coupling parameter, the gauge coupling parameter $g$. The classical version of this theory exhibits superconformal invariance, owing to the dimensionless nature of its coupling parameter. Its beta function vanishes identically to all orders in perturbation theory and the same is believed to be true at the nonperturbative level. This theory, therefore, is finite, with no renormalization at all. Its coupling parameter does not run, unlike most gauge theories, different values of $g$ really give different theories, rather than being transmuted to a change of scale. Another interesting property exhibited by this theory is exact electric-magnetic duality, that is, the invariance under the interchange of electric and magnetic quantum numbers. The theory is also invariant under the replacement of $g$ with $4 \pi/g$, that is, the theory with a weak gauge coupling $g$ is fully equivalent to the one with a strong gauge coupling $4 \pi/g$. 
 
In 1997, Maldacena proposed\cite{Maldacena:1997re} a new duality relating Type II supergravity (a certain low energy limit of string theory) in $(d+1)$-dimensional anti-de Sitter (AdS) space and $d$-dimensional super conformal theories. This is known as the holographic principle. The $\cN=4$ SYM theory takes part in the most successful realization of holographic principle. This theory can be realized as the gauge theory living on a D3-brane of Type IIB superstring theory in $AdS_5 \times S^5$ space. 

The action of $\cN=4$ SYM theory was given for the first time in 1977 in Refs. \refcite{Brink:1976bc} and \refcite{Gliozzi:1976qd} within the framework of string theory toroidal compactifications. This theory has the maximal amount of supersymmetry, sixteen real supercharges, for a four-dimensional field theory with global supersymmetry. 

There exist different types of construction schemes for four-dimensional $\cN=4$ SYM theory. We follow the original work of Brink, Schwarz and Scherk given in Ref. \refcite{Brink:1976bc}, where it is constructed by dimensional reduction from ten dimensions.

\subsection{Yang--Mills theory with fermions}

We are interested in constructing a Yang--Mills theory coupled to spin-$\frac{1}{2}$ fermions in $d$ spacetime dimensions with an additional symmetry: the number of bosonic and fermionic degrees of freedom are equal. We will call this symmetry supersymmetry. 

A massless gauge potential in $d$ dimensions has $d-2$ on-shell real degrees of freedom. A Dirac spinor in $d$ dimensions has $2^{[d/2]}$ on-shell real degrees of freedom, where $[d/2]$ represents the integral part. These two numbers do not match in any dimension. In order to demand the additional symmetry, we will have to reduce the number of fermionic degrees of freedom by requiring the spinor to satisfy some supplementary conditions.

Let us consider a Yang--Mills theory coupled to massless spin-$\frac{1}{2}$ particles on a $d$-dimensional flat Minkowski space ${\mathbb R}^{1, (d-1)}$ with signature $g_{mn} = \textrm{diag}(-,+,+,\cdots,+)$, where $m, n=0, 1, 2, \cdots, (d-1)$. The metric is 
\beq
ds^2 = \sum_{m, n} g_{mn} dx^m dx^n = -(dx^0)^2 + (dx^1)^2 + \cdots + (dx^{(d-1)})^2~.
\eeq

The theory has a gauge field $A_m$ taking values in the real Lie algebra of a compact gauge group $G$. The gauge field takes values in anti-hermitian matrices, in the adjoint representation of $G$. The covariant derivative $D_m$ is
\beq
D_m = \partial_m + A_m~,
\eeq
The corresponding curvature $F_{mn}$ is
\beq
F_{mn} = [D_m, D_n] = \partial_m A_n - \partial_n A_m +[A_m, A_n]~.
\eeq
We add a fermionic term to the $d$-dimensional Yang--Mills action. The fermions are contained in a Dirac spinor $\lambda$ taking values in the Lie algebra of $G$. 

The action is
\beq
\label{eq:YM-Dirac-action-d}
S = \Tr \int d^d x \Big(-\frac{1}{4} F_{mn} F^{mn} + i \overline{\lambda} \Gamma^mD_m \lambda\Big)~,
\eeq
where $\Gamma^m$ are the $d$-dimensional gamma matrices. Since anti-hermitian matrices generate the Lie algebra in our case, the trace $\Tr$is negative definite.

We examine the dimensions in which the action (\ref{eq:YM-Dirac-action-d}) permits the extra symmetry - supersymmetry - between the gauge bosons and the fermions without the addition of other fields. The requirement of same number of bosonic and fermionic degrees of freedom is essential for field theories that transform as linear representations of supersymmetry. Since the spinor degrees of freedom grow faster than that of gauge bosons, we will reduce the number of spinor degrees of freedom by imposing some additional conditions on the fermions. Before we choose those conditions, a familiarization with spinor representations in arbitrary dimensions would be useful.  

\subsection{Spinors in higher dimensions}

The Lorentz group, the symmetry group of Minkowski space, admits finite-dimensional representations. Spinors appear as fields that transform under finite-dimensional representations of the Lorentz group. 

We use the language of Clifford algebras to discuss the spinor representations in $d$ dimensions. A Clifford algebra is a set of matrices (we call them gamma matrices) satisfying the anticommutation relations:
\beq
\{ \Gamma_m, \Gamma_n \} = 2 g_{mn}~,
\eeq
where $m,n = 0,1,\cdots (d-1)$.

Given such a set of matrices, we see that the following antisymmetric matrices,
\beq
\Sigma^{mn} = -\frac{i}{4}[\Gamma^m, \Gamma^n] = -\Sigma^{nm}~,
\eeq
satisfy the commutation relations of the Lorentz group generators:
\beq
i[\Sigma^{mn}, \Sigma^{sr}] = \eta^{ns}\Sigma^{mr} + \eta^{mr}\Sigma^{ns} - \eta^{nr}\Sigma^{ms} - \eta^{ms}\Sigma^{nr}~.
\eeq
The matrices $\Sigma^{mn}$ give a $d$-dimensional representation of the Lorentz algebra. They are a set of antisymmetric tensors transforming according to the $d$-dimensional Lorentz vector representation of $SO(1, d-1)$. They act on the space of fields called Dirac spinors. The algebra generated by $\Sigma^{mn}$ yield the spinor representation of $SO(1, d-1)$.

The $d$-dimensional representation of the Lorentz algebra generated by $\Sigma^{mn}$ is not always an irreducible representation. To see if a given representation is reducible or not, we need to consider separately the case where $d$ is an odd or even dimension.

We begin with the construction of gamma matrices in even dimensions, $d=2k+2$, where $k=1, 2, \cdots$. We group the gamma matrices into a set of raising and lowering operators\cite{Polchinski:1998rr,Weinberg:2000cr}
\bea
u^{\pm}_0 &=& \frac{1}{2}(\pm \Gamma_0 + \Gamma_1)~, \nn \\
u^{\pm}_a &=& \frac{1}{2}(\Gamma_{2a} \pm i \Gamma_{2a+1})~,
\eea
where $a=1,\cdots,k$. These operators satisfy the following anticommutation relations:
\bea
\{ u_i^+, u_j^- \} &=& \delta_{ij}~, i,j=0, 1, \cdots, k \nn \\
\{ u_i^+, u_j^+ \} &=& \{ u_i^-, u_j^- \} =0~,
\eea
along with the conditions:
\bea
(u_i^+)^2 &=& (u_i^-)^2 = 0~.
\eea
We can let $u_i^-$ operators act repeatedly on any spinor state to reach a spinor $|\xi\rangle$ annihilated by all $u_i^-$'s
\beq
u_i^- |\xi\rangle = 0,~~~\textrm{for all}~i~.
\eeq
Now we can let the creation operator $u_i^+$ act on $|\xi\rangle$, at most once each, in all possible ways to obtain a spinor representation. The spinor states obtained in that way are given in Table \ref{tb:spinor states}.

\begin{table}[ph]
\tbl{Spinor states and their corresponding numbers.}
{\begin{tabular}{|r || c | c | c | c | c |}
\hline
states & $|\xi \rangle$ & $u_i^+|\xi \rangle$ & $u_i^+u_j^+|\xi \rangle$ & $\cdots$ & $(u_k^+u_{(k-1)}^+ \cdots u_0^+)|\xi \rangle$ \\ \hline 
 & & & & & \\
number & 1 & $k+1$ & $_{k+1} {\Large C}_2$ & $\cdots$ & 1 \\ \hline
\end{tabular} \label{tb:spinor states}}
\end{table}

The total number of states is 
\beq
1 + (k+1) + {\ }_{k+1} {\Large C}_2 + \cdots + 1 = \sum_{n=0}^{k+1} {\ }_{k+1} {\Large C}_n = 2^{k+1} = 2^{d/2}~.
\eeq
This representation has dimension $2^{k+1}$. The spinor representation is given by 
\beq
|s_0 s_1 \cdots s_k \rangle = (u_k^+)^{s_k + \half} \cdots (u_0^+)^{s_0+\half}|\xi \rangle~,
\eeq
where each of $s_i$ is $\pm \half$. The $|\xi \rangle$ we started with contains all $s_i = -\half$.

The matrix elements of $\Gamma^m$ can be derived from the definitions and the anticommutation relations by taking the $|s_0 s_1 \cdots s_k \rangle$ as a basis. 

The generators $\Sigma_{2i, 2i-1}$ form a commuting set. We consider the operator
\beq
S_i \equiv \Sigma_{2i, 2i-1} =  u_i^+ u_i^- - \half~.
\eeq
The basis vectors  $|s_0 s_1 \cdots s_k \rangle$ defined above form simultaneous eigenstates of all the $S_i$'s with eigenvalues $s_i$,
\beq
S_i |s_0 s_1 \cdots s_k \rangle = s_i |s_0 s_1 \cdots s_k \rangle~.
\eeq
The half-integer eigenvalues show that this is a spinor representation. The spinors form the $2^{k+1}$-dimensional Dirac representation of the Lorentz algebra $SO(1, 2k+1)$. For example, in $d=4$, the states $|\pm \half, \pm \half \rangle$ form a four component Dirac spinor.

Noting that increasing $d$ by two doubles the size of Dirac matrices, we can give an iterative expression for gamma matrices in even dimensions starting in $d=2$.

The gamma matrices in $d=2$ are:
\beq
\Gamma^0 =  \left( \begin{array}{cc}
0 & 1 \\
-1 & 0 \end{array} \right),~~~~\Gamma^1 =  \left( \begin{array}{cc}
0 & 1 \\
1 & 0 \end{array} \right)~.
\eeq

Then in $d=2k+2$ with $k=1, 2, \cdots$ we have,
\bea
&&\Gamma^m =  \gamma^m \otimes \left( \begin{array}{cc}
-1 & 0 \\
0 & 1 \end{array} \right),~~~m=0,\cdots,d-3~, \nn \\
&&\Gamma^{(d-2)} =  {\mathbb I} \otimes \left( \begin{array}{cc}
0 & 1 \\
1 & 0 \end{array} \right), \nn \\
&&\Gamma^{(d-1)} =  {\mathbb I} \otimes \left( \begin{array}{cc}
0 & -i \\
i & 0 \end{array} \right)~,
\eea
with $\gamma^m$ the $2^k \times 2^k$ Dirac matrices in $d-2$ dimensions and ${\mathbb I}$ the $2^k \times 2^k$ identity. The $2 \times 2$ matrices act on the index $s_k$, which is added in going from $2k$ to $2k+2$ dimensions.

For representations in odd dimensions, we need to add a new gamma matrix $\Gamma_{d+1}$ to the $\Gamma_m$ matrices. Let us define $\Gamma_{d+1}$ in the following section. 

\subsubsection{Weyl spinors}

Since the generators $\Sigma^{mn}$ are quadratic in the gamma matrices, the spinor states $|s_0 s_1 \cdots s_k \rangle$ with even and odd numbers of $+\half$s do not mix. This indicates that the Dirac representations in even dimensions are reducible representations of the Lorentz algebra. 

We define a new gamma matrix:
\beq
\Gamma_{d+1} = i^{-k}\Gamma_0 \Gamma_1 \cdots \Gamma_{d-1}~,
\eeq
which has the properties: 
\beq
(\Gamma_{d+1})^2 = 1,~~~\{\Gamma_{d+1}, \Gamma^m\} = 0,~~~[\Gamma_{d+1}, \Sigma^{mn}] = 0~.
\eeq
All the Dirac spinor states are eigenstates to $\Gamma_{d+1}$
\beq
\label{eq:chiraliyt-d+1}
\Gamma_{d+1} |s_0 s_1 \cdots s_k \rangle = \pm |s_0 s_1 \cdots s_k \rangle~,
\eeq
with eigenvalue $+1$ for even numbers of $s_i = +\half$ and $-1$ for odd ones. 

Since $\Gamma_{d+1}$ commutes with the generators of the Lorentz algebra $\Sigma^{mn}$ cannot furnish an irreducible representation of $SO(1, d-1)$. The Dirac representation, let us denote it by ${\mathbb S}$, breaks down into two $2^k$-dimensional irreducible representations ${\mathbb S}^+$ and ${\mathbb S}^-$. These representations are called Weyl (or chiral) representations, and they can be obtained by projecting out the two subspaces using $\Gamma_{d+1}$. We define a projection operator:
\beq
{\mathbb P}^{\pm} = \frac{1}{2}({\mathbb I} \pm \Gamma_{d+1})~.
\eeq
The Lorentz generators and representation now split into two parts:
\beq
\Sigma_{mn}^{\pm} = {\mathbb P}^{\pm} \Sigma_{mn},~~{\mathbb S}^{\pm} = {\mathbb P}^{\pm} {\mathbb S}~.
\eeq
The spinors obtained in this way are called Weyl spinors. 

In $d=4$, the Dirac representation is the familiar four-dimensional one, which separates into two two-dimensional Weyl representations distinguished by their eigenvalue under the chirality operator $\Gamma_5$.
\beq
{\bf 4}_{\textrm{Dirac}} = {\bf 2} + {\bf 2}'~.
\eeq
Here we have labeled a representation ${\mathbb S}$ by its dimension (in boldface). In $d=10$, the representations are: 
\beq
{\bf 32}_{\textrm{Dirac}} = {\bf 16} + {\bf 16}'~.
\eeq

To get representations in odd dimensions, $d=2k+3$, we simply add $\Gamma_{d+1}$ to the gamma matrices for $d=2k+2$. The set of creation and annihilation operators is the same as that of $d=2k+2$. This is now an irreducible representation of the Lorentz algebra because $\Sigma^{md}$ anticommutes with $\Gamma_{d+1}$. Thus, there is a single spinor representation of $SO(1, 2k+2)$, which has dimension $2^{k+1}$. There is no chirality in odd dimensions. 

For $k$ even, the Weyl irreducible representations are equivalent to complex conjugates of each other. While for $k$ odd each Weyl representation is equivalent to its own complex conjugate. The Weyl representations can only be real for $k=1$ (mod) 4 and must be pseudo-real for $k=3$ (mod) 4.

The Lorentz generators $\Sigma_{mn}$ in odd-dimensional case furnish an irreducible representation of the Lorentz group by themselves. In each odd dimension, the fundamental spinor representation is either real or pseudo-real. 

\subsubsection{Majorana spinors}

The above construction of the irreducible representations of gamma matrices shows that, in even dimensions, $d=2k+2$, the irreducible representations are unique up to a change of basis. That is, for any set of gamma matrices $\{\Gamma^m\}$ and $\{\Gamma^{m'}\}$ both satisfying the Clifford algebra, there exists a nonsingular matrix $M$, such that
\beq
\Gamma_m = M \Gamma_m'M^{-1},~~~\textrm{for all}~m=0,1, \cdots d-1~.
\eeq
Thus, the matrices $(\Gamma^{m})^*$ and $-(\Gamma^{m})^*$ satisfy the same Clifford algebra as $\Gamma^m$. This implies that the Dirac representation is its own conjugate in even dimensions.

We can impose a condition that relates the spinor state $|\xi \rangle^*$ to $|\xi\rangle$. This condition must be consistent with Lorentz transformations and so must have the form:
\beq
|\xi\rangle^* = B |\xi\rangle~,
\eeq
with $B$, a nonsigular matrix satisfying
\beq
B \Sigma^{mn} B^{-1} = - (\Sigma^{mn})^*~.
\eeq
Such a condition, called the Majorana (or reality) condition, is consistent only if $BB^* = 1$.

Using the reality and anticommutation properties of the gamma matrices, one finds
\beq
B^*B = (-1)^{k(k+1)/2}~~\textrm{or}~~(-1)^{k(k-1)/2}~.
\eeq
Thus, a Majorana condition is possible only if $k=0$ or 3 (mod) 4 for the first case, and for $k=0$ or 1 (mod) 4 for the second case. If $k=0$, both conditions are possible, but they are physically equivalent. They are related to each other through a similarity transformation.

The Majorana condition on a Dirac spinor $\lambda$ is: 
\beq
\lambda = C \overline{\lambda}^T~,
\eeq
where $C$ is the charge conjugation matrix. It transforms the Lorentz representation matrices in the following way:
\beq
C \Sigma^{mn}C^{-1} = -\Sigma^{mnT}~.
\eeq
{\ }\\
\begin{table}[ph]
\tbl{We can impose various conditions on $SO(1, d-1)$ Dirac spinors in various dimensions. For the Weyl representation, it is indicated whether these are conjugate to themselves (self) or to each other (complex). The smallest representation in each dimension, counting the number of real components, is given in the final column. A dash indicates that the condition cannot be imposed.}
{\begin{tabular}{c c c c c c}
\hline
 $d$ & Majorana & Weyl & Weyl--Majorana & min. rep. \\ \hline
 2 & yes & self & yes & 1  \\
 3 & yes & - & - & 2  \\
 4 & yes & complex & - & 4 \\
 5 & - & - & - & 8 \\
 6 & - & self & - & 8 \\
 7 & - & - & - & 16  \\
 8 & yes & complex & - & 16 \\
 9 & yes & - & - & 16  \\
 10 & yes & self & yes & 16 \\
 \hline
\end{tabular} \label{tb:conditions}}
\end{table}

\subsubsection{Weyl--Majorana spinors}
Imposing a Majorana condition on a Weyl spinor requires the Weyl spinor representation to be conjugate to itself. For $k$ odd, which is $d=0$ or 4 (mod) 8, it is therefore not possible to impose both the Majorana and Weyl conditions on a spinor: one can impose one or the other. Precisely for $k=0$ (mod) 4, which is $d=2$ (mod) 8, a spinor can simultaneously satisfy the Majorana and Weyl conditions.

\begin{table}[ph]
\tbl{Conditions on spinor degrees of freedom in various dimensions. Note the bold numbers indicating the matching of fermion and gauge field degrees of freedom.}
{\begin{tabular}{ c c c c c c }
\hline
 $d$ & $A_m$ & $\lambda_D$ & $\lambda_M$ & $\lambda_W$ & $\lambda_{WM}$ \\ \hline
 3 & {\bf 1} & 2 & {\bf 1} & - & - \\
 4 & {\bf 2} & 4 & {\bf 2} & {\bf 2} & - \\
 6 & {\bf 4} & 8 & - & {\bf 4} & - \\
 10 & {\bf 8} & 32 & 16 & 16 & {\bf 8} \\
 \hline
\end{tabular} \label{tb:conditions1}}
\end{table}

We can have Majorana spinors in $d=2, 3, 4, 8, 9, 10$, and Weyl spinors in $d=2, 4, 6, 8, 10$. For the cases $d=4, 8$, while one can, in principle, impose a Majorana condition, this condition is incompatible with the Weyl condition and, thus, there are no Weyl--Majorana spinors for $d=4, 8$. For $d=2, 10$, we can impose both Majorana and Weyl conditions, that is, we have Weyl--Majorana spinors (See Table \ref{tb:conditions}). The Weyl--Majorana spinors in $d=2$ and $d=10$ are particularly important because of their relevance in string theory.

Imposing a Majorana or Weyl condition on the spinor, though, reduces its degrees of freedom, each by a factor of one half. Starting with $d=3$, the various possibilities for matching the degrees of freedom of a gauge field $A_m$ to those of Dirac ($\lambda_D$), Majorana ($\lambda_M$), Weyl ($\lambda_W$) and Weyl--Majorana ($\lambda_{WM}$) spinors are shown in Table \ref{tb:conditions1}.

Note that in $d>10$ there are no solutions to our matching problem on fermion-gauge boson degrees of freedom. That is, $d=10$ is the highest dimension in which we can have a SYM action (on a flat spacetime without adding extra fields).

\subsection{Supersymmetric Yang--Mills theory in ten dimensions}
\label{sec:SYM-10d}

The SYM action in ten dimensions has the form
\beq
\label{eq:YM-action-d}
S = \Tr \int d^{10} x \Big(-\frac{1}{4} F_{mn} F^{mn} + i \overline{\lambda} \Gamma^mD_m \lambda\Big)~,
\eeq
where $F_{mn}$ is the ten-dimensional curvature, $m,n = 0, 1, \cdots, 9$; $\lambda$ is a Weyl--Majorana spinor (it is known as a gaugino) with its 8 degrees of freedom matching with those of the ten-dimensional gauge field $A_m$. The Dirac spinor in ten dimensions has 32 degrees of freedom. This can be reduced to 16 by imposing the Weyl condition (decomposing $\lambda$ in to chiral and antichiral parts $\lambda_{\pm}$ by applying the projection operator ${\mathbb P}_{\pm}$). Imposing the Majorana condition $\lambda = C \overline{\lambda}^T$ on this Weyl spinor further reduces the number of degrees of freedom down to 8. Thus we obtain a Weyl--Majorana spinor with 8 degrees of freedom matching with those of the gauge field. 

The action (\ref{eq:YM-action-d}) is invariant under a set of transformations of the fields, called the supersymmetry transformations
\bea
\label{eq:susy-transform-10d}
\delta_S A_m &=& i \overline{\alpha} \Gamma_m \lambda~, \nn \\
\delta_S \lambda &=& \Sigma_{mn} F^{mn} \alpha~,
\eea
where the constant spinor field parameter $\alpha$ is a single Weyl--Majorana spinor parameterizing the supersymmetry transformations. This is referred to as $\cN=1$ supersymmetry. In $d=10$, there are 16 real supercharges corresponding to these transformations.

The symbol $\delta_S$ stands for the {\it supersymmetric variation}. For a generic field $\Phi$, it means:
\beq
\delta_S \Phi = \sum_{a=1}^{16}[\epsilon^a Q_a, \Phi \}~,
\eeq
where $Q_a$ are the sixteen supersymmetries. The symbol $[X, Y\}$ denotes, the graded commutator, $XY - (-1)^{|X||Y|}YX$. We have $|X| = 1$ when the field $X$ is fermionic and $|X| = 0$ when it is bosonic. 

\subsection{Dimensional reduction to four dimensions}
\label{sec:dim-red-four}

We are interested in constructing $\cN=4$ SYM theory in four dimensions. To obtain this theory, we dimensionally reduce the ten-dimensional $\cN=1$ SYM theory down to four dimensions. 

\subsubsection{The method of dimensional reduction}

Let us consider compactifying one spatial dimension of ${\mathbb R}^{1, (d-1)}$ on a circle of radius $R$, that is,
\beq
{\mathbb R}^{1, (d-1)} \longrightarrow {\mathbb R}^{1, (d-2)} \times S^1.
\eeq
The coordinates $x^m$, $m=0,1,\cdots, (d-1)$ of ${\mathbb R}^{1, (d-1)}$ decompose into ($x^\mu$, $z$), where $x^{\mu}$ are the coordinates of ${\mathbb R}^{1, (d-2)}$ and $z$ the coordinate on the compactified spatial dimension $S^1$. The limit $R \rightarrow 0$, in which the compactified dimension shrinks to zero size, is called `dimensional reduction.' To understand what happens to spacetime fields under this action, we consider the simplest case of a complex scalar field $\varphi$ with periodic boundary conditions on the compactified direction $S^1$. This field has the Fourier expansion:
\beq
\varphi(x^{\mu}, z) = \frac{1}{\sqrt{2 \pi R}}\sum_{n \in {\mathbb Z}} \varphi_n(x^{\mu}) e^{inz/R}~.
\eeq
The kinetic part of the action for this field becomes: 
\bea
S_{KE}^{(d)} &=&\int d^d x^m \varphi^{\dagger}\Big(\Box_{(d)} - m^2\Big)\varphi\nn \\
&=& \int d^{(d-1)} x^{\mu} \int dz \varphi^{\dagger}\Big(\Box_{(d-1)} + \frac{\partial^2}{\partial z^2} - m^2\Big)\varphi \nn \\
&=& \int d^{(d-1)} x^{\mu} \sum_{n \in {\mathbb Z}} \varphi^{\dagger}_n(x^{\mu}) \Big(\Box_{(d-1)}- m^2 - \frac{n^2}{R^2}\Big) \varphi_n(x^{\mu})~.~~~~
\eea
The Fourier modes $\varphi_n(x^{\mu})$ acquire curvature dependent masses $m^2 + n^2/R^2$. In the limit $R \rightarrow 0$, the modes $\varphi_n$ for $n \neq 0$ become infinitely massive. It would cost an infinite amount of energy to excite such modes, and they therefore decouple from the theory. The only mode that survives in this limit is the zero mode $\varphi_0$, with the kinetic action:
\beq
S_{KE}^{(d-1)} = \int d^{(d-1)} x^{\mu} \varphi_0^{\dagger} \Big(\Box_{(d-1)} - m^2\Big) \varphi_0~.
\eeq
We can extend the method of dimensional reduction to more than one space dimensions. Consider compactification on a torus ${\mathbb T}^k = S^1 \times S^1 \cdots S^1$, $k$ times, with each circle of radius $R$. The spacetime becomes:
\beq
{\mathbb R}^{1, (d-1)} \longrightarrow {\mathbb R}^{1, d-1-k} \times {\mathbb T}^k~.
\eeq
The Lorentz group splits in the following way:
\beq
SO(1, d-1) \rightarrow SO(1, d-1-k) \times \textrm{isometries on}~{\mathbb T}^k~.
\eeq
The representations of the fields also take new forms. The covariant derivative, $D_m = \partial_m + A_m$, acting on the zero mode of a field simply reduces to $D_{\mu}$ for $m = \mu$. The components of the covariant derivative in the reduced directions, $D_i$ for $i=1, \cdots, k$, acting on a zero mode is just $A_i$ on the mode. The gauge field components in the reduced dimensions, $A_i$ $i = 1, \cdots k$, become a collection of scalar fields.

In the limit $R \rightarrow 0$, the isometries on ${\mathbb T}^k$ become the rotations on ${\mathbb R}^k$, and we have the splitting:
\beq
\label{spinor-gp-splitting}
SO(1, d-1) \rightarrow SO(1, d-1-k) \times SO(k)~.
\eeq
A spinor field decomposes into direct sums of representations of $SO(1, d-1-k)$ because of the tensor product structure of the Clifford algebra. The Lorentz group splitting is the same as in (\ref{spinor-gp-splitting}).

\subsubsection{From $\cN=1$, $d=10$ SYM to $\cN=4$, $d=4$ SYM}

Dimensional reduction of ten-dimensional $\cN=1$ SYM theory down to four dimensions leads to an $\cN=4$ SYM with the same number of supersymmetries. 

The Lorentz group $SO(1, 9)$ splits according to 
\beq
SO(1, 9) \rightarrow SO(1, 3) \times SO(6)~.
\eeq
We will also be using the notation of $Spin$ group, the double cover of the Lorentz group, in the later sections. The double cover splits according to
\beq
Spin(1, 9) \rightarrow Spin(1, 3) \times Spin(6) \approx Spin(1, 3) \times SU(4)~.
\eeq

Dimensional reduction of the theory on a six-dimensional torus ${\mathbb T}^6$ gives rise to a multiplet of four-dimensional fields possessing an additional $SO(6) \sim SU(4)$ global symmetry. This internal rotational symmetry is known as the $R$-symmetry, $SO_R(6)$, of the dimensionally reduced theory. 

After dimensional reduction, the ten-dimensional gauge field reduces to a four dimensional real vector $A_{\mu}$, $\mu = 0, 1, 2, 3$, transforming under the $SO(1, 3)$ symmetry. The reduced components of the gauge field $A_i$, $i=1, 2, \cdots, 6$  become six real scalars.  The $SO_R(6)$ becomes an internal symmetry mixing between these scalars. They transform as the second rank complex self-dual ${\bf 6}$ of $SU(4)$.

The Clifford algebra splits up as follows:
\beq
\Gamma_{\mu} = \gamma_{\mu} \otimes {\mathbb I}_8,~~~\Gamma_i \approx \gamma_{pq} = \gamma_5 \otimes \left( \begin{array}{cc}
0 & \rho_{pq} \\
\rho^{pq} & 0 \end{array} \right)~,
\eeq
where $\gamma_{\mu}$, $\mu = 0, 1, 2, 3$, are the ordinary $4 \times 4$ gamma matrices, and the $4 \times 4$ $\rho$ matrices, with $p, q = 1, 2, 3, 4$, are given by
\beq
(\rho_{pq})_{rs} = \epsilon_{pqrs},~~~(\rho^{pq})_{rs} = \frac{1}{2}\epsilon^{pqkl}\epsilon_{klrs}~,
\eeq
and the chirality matrix $\Gamma_{11}$, in terms of our usual $\gamma_5$ is: 
\beq
\Gamma_{11} = \Gamma_0 \cdots \Gamma_9 = \gamma_5 \otimes {\mathbb I}_8~.
\eeq

Finally, the ten-dimensional charge conjugation matrix is related to the four-dimensional matrix $C$ by
\beq
C_{10} = C \otimes \left( \begin{array}{cc}
0 & {\mathbb I}_4 \\
{\mathbb I}_4 & 0 \end{array} \right)~.
\eeq

Imposing both Majorana and Weyl conditions on the Dirac spinor results in the structure
\beq
\lambda = \left( \begin{array}{c}
L \chi^s\\
 \\
R \widetilde{\chi}_s \end{array} \right)~,
\eeq
where $L = \frac{1}{2}({\mathbb I} + \gamma_5)$ and $R=\frac{1}{2}({\mathbb I} - \gamma_5)$; $s = 1, 2, 3, 4$; and $\widetilde{\chi}_s = C \overline{\chi}^{sT}$. We have four left-handed and four right-handed (Weyl) spinors.

The spinor index ${\bf 16}$ separates into $({\bf 2}, {\bf 4}) + (\overline{{\bf 2}}, \overline{{\bf 4}})$ under $SO(1, 3) \times SO(6)$. The ten-dimensional spinor becomes four Weyl spinors. 

Thus the dimensionally reduced action is
\bea
\label{eq:n4d4-action}
S &=& \int d^4 x \Tr \Big(-\frac{1}{4}F_{\mu \nu}F^{\mu \nu} - \frac{1}{2}D_{\mu}A_i D^{\mu}A_i + \frac{1}{4}[A_i, A_j]^2\Big)\nn \\
&&- \frac{i}{2}\Tr (\overline{\lambda}\Gamma^{\mu}D_{\mu}\lambda + i\overline{\lambda}\Gamma_i[A_i, \lambda])~.
\eea

The supersymmetry transformation laws take the following form after dimensional reduction
\bea
\delta A_{\mu} &=& -i\overline{\alpha}\Gamma_{\mu} \lambda~, \nn \\
\delta A_i &=& -i\overline{\alpha}\Gamma_i \lambda~, \nn \\
\delta \lambda &=& \Big(\frac{1}{2}F_{\mu \nu}\Gamma^{\mu \nu} + D_{\mu}A_j \Gamma^{\mu j} + \frac{i}{2}[A_i, A_j] \Gamma^{ij}\Big)\alpha~.
\eea

\section{Topological field theory}

Supersymmetric field theories naively break supersymmetry when they are discretized on a lattice. Topological field theories provide a crucial insight into establishing the compatibility between SYM theories and lattice discretization. Certain supersymmetric field theories with extended supersymmetries can be discretized on a lattice while preserving at least one supersymmetry. The continuum limit of these discretized theories turn out to have a structure similar to that of topological field theories. In this section we briefly introduce a class of topological field theories and show how their structure is compatible with discretization on the lattice.

As the name suggests, topological field theories are characterized by observables (correlation functions) which depend only on the topology (global features) of the space on which these theories are constructed. The non-dependence on local features implies that the observables of topological field theories are independent of the metric of the space on which they are defined. 

The origin of topological field theories goes back to the work of Schwarz and Witten. In 1978, Schwarz showed\cite{Schwarz:1978cn} that Ray-Singer torsion\footnote{Ray-Singer torsion is a particular topological invariant of Riemannian manifolds.} could be represented as the partition function of a certain quantum field theory. In 1982, the work of Witten\cite{Witten:1982im} provided a framework for understanding Morse theory\footnote{Morse theory is a method to determine the topology of a manifold from the critical points of only one suitable function on the manifold.} in terms of supersymmetric quantum mechanics. These two field theory constructions represent the prototype of all known topological field theories.  

There are two general classes of topological field theories: they are known as Witten and Schwarz type. In Witten type topological field theories, the classical action is trivial (zero or a topological invariant). In Schwarz type theories, classical actions are non-trivial. The prototype example of a Witten type theory is the Donaldson theory\footnote{Donaldson theory is the study of smooth 4-manifolds using gauge theory techniques.}; the best known example of Schwarz type theory is the Chern--Simons theory. 

\subsection{Yang--Mills theory and BRST invariance}
\label{sec:brst-sym}

Let us begin our brief description of topological field theory focusing only on Witten type theory, as only this type eventually leads to a discretization on the lattice. 

We look at a conventional nonabelian field theory with gauge symmetry. The best example is Yang--Mills theory in four dimensions. The classical action is a combination of gauge field Lagrangian and Dirac Lagrangian. It is 
\beq
S_c = \int d^4x \Tr \Big(-\frac{1}{4}F_{\mu \nu}F^{\mu \nu} + \overline{\psi} (\Gamma^{\mu}D_{\mu} - m) \psi\Big)~,
\eeq
where the trace is over the generators of the gauge group $G$ and the fermion multiplet $\psi$ belongs to an irreducible representation of $G$. The field strength is: 
\beq
F_{\mu \nu}^A = \partial_{\mu} A_{\nu}^A -  \partial_{\mu} A_{\nu}^A + f^{ABC} A_{\mu}^B A_{\nu}^C~,
\eeq
where $f^{ABC}$ are the structure constants of $G$. The covariant derivative is defined in terms of the representation matrices $T^A$ by 
\beq
D_{\mu} = \partial_{\mu} + A_{\mu}^A T^A~.
\eeq
The gauge-fixed (quantum) action after Faddeev--Popov gauge-fixing is
\beq
\label{eq:fp-gauge-fixed}
S_q = S_c + \int d^4 x \Tr \Big(\frac{1}{2\xi} (\partial^{\mu}A_{\mu}^A)^2  + \overline{c}^A (-\partial^{\mu}D_{\mu}^{AC})c^C\Big)~,
\eeq
where $\xi$ is a gauge parameter, and $c$ and $\overline{c}$ are the Faddeev--Popov ghost and anti-ghost fields.

The Faddeev--Popov ghost fields serve as negative degrees of freedom to cancel the effects of unphysical time-like and longitudinal polarization states of gauge bosons $A_{\mu}$, and thus make the gauge theory a complete interacting theory. 

There is a beautiful formal tool to implement this cancellation, known as the BRST formulation\cite{Becchi:1975nq}$^,$\cite{Iofa:1976je}. 

Let us rewrite the gauge-fixed action by introducing a new commuting scalar field $B^A$ to expose the the symmetry associated with the BRST technique: 
\beq
S_q = S_c + \int d^4 x \Tr \Big(- \frac{\xi}{2}(B^A)^2 + B^A \partial^{\mu}A_{\mu}^A + \overline{c}^A (-\partial^{\mu}D_{\mu}^{AC})c^C\Big)~.
\eeq
The new field $B^A$ is not a normal propagating field, as it has a quadratic term without derivatives. These type of fields, which appear in the functional integral part but have no independent dynamics, are called auxiliary fields. We can eliminate them by using the equations of motion. We could also get rid of the dependence on $B$ by integrating it in a functional integral with a standard Euclidean measure $[dB]$. This would bring us back to (\ref{eq:fp-gauge-fixed}), the Faddeev--Popov gauge-fixed action.

The BRST symmetry has a continuous parameter that is an anticommuting number. Let us denote it by $\epsilon$ (we call this the BRST parameter), and consider the following infinitesimal transformation of the fields in the action:
\bea
\delta A_{\mu}^A &=& \epsilon D_{\mu}^{AC}c^C \nn \\
\delta \psi &=& i\epsilon c^A T^A \psi \nn \\
\delta c^A &=& -\half \epsilon f^{ABC} c^B c^C \nn \\
\delta\overline{c}^A &=& \epsilon B^A \nn \\
\delta B^A &=& 0
\eea
The BRST transformation above is a global symmetry of the gauge-fixed action for any value of the gauge parameter $\xi$. 

The BRST transformation has one more remarkable feature, which is a natural consequence of its anticommuting nature. Let $Q \Phi$ be the BRST transformation of the generic field $\Phi$ of the theory:
\beq
\delta \Phi = \epsilon Q \Phi~.
\eeq
Then the BRST variation of $Q\Phi$ vanishes:
\beq
Q^2 \Phi = 0~.
\eeq
That is, the BRST operator $Q$ is nilpotent. 

The BRST operator gives a precise relation between  the unphysical gauge boson polarization states and anti-ghosts as positive and negative degrees of freedom. We can use the principle of BRST symmetry to remove the unphysical gauge boson polarizations in nonabelian gauge theories. The complete quantum action, $S_q$, which comprises the classical action $S_c$ together with the necessary gauge-fixing and ghost terms, is, by construction, $Q$-invariant.

The change in gauge field $A_{\mu}^A$ involves the ghost field $c^A$. In the infinitesimal gauge symmetry (Yang--Mills symmetry), transformation for $A_{\mu}^A$ given by $A_{\mu}^A \rightarrow A_{\mu}^A - (D_{\mu}\theta)^A$ we can replace the gauge parameter $-\theta^A$ by the ghost field $c^A$. That is, gauge-invariant quantities are also BRST-invariant. We also see that all observables are given by BRST-invariant expressions, since all observables in a gauge theory must be gauge-invariant. 

With the BRST transformations, we can write: 
\beq
S_q = S_c + \cQ \int d^4x \Big[\overline{c}^A \Big(\partial^{\mu}A_{\mu}^A - \frac{\xi}{2}B^A \Big)\Big]~.
\eeq

We can show that the vacuum expectation value of $\cQ \cO$ for any (not necessarily $\cQ$ invariant) functional $\cO$ is zero. We write:
\bea
\langle \cO \rangle &=& \int [dA][dc][d\cbar]~ \cO(A, c, \cbar) e^{-S_q(A, c, \cbar)}~.
\eea
Let us rename the variables of integration in the following way:
\bea
A_{\mu}^{A'} &=& A_{\mu}^A + \delta A_{\mu}^A~, \nn \\
c^{A'} &=& c^A + \delta c^A~, \nn \\
\cbar^{A'} &=& \cbar^A + \delta \cbar^A~.
\eea
where $\delta = \epsilon \cQ$ is the BRST variation with $\epsilon$ an arbitrary Grassmann number. The vacuum expectation value of $\cO$ becomes:
\bea
\langle \cO \rangle&=& \int [dA'][dc'][d \cbar']~ \cO(A', c', \cbar') e^{-S_q(A', c', \cbar')}~.
\eea
Assuming that the measure of integration is invariant, which it should be for consistency of the theory, we get the vacuum expectation value of $\cO$:
\bea
\langle \cO \rangle &=& \int [dA][dc][d \cbar]~(\cO + \delta \cO) e^{-S_q - \delta S_q}~,\nn \\
&=& \langle \cO \rangle + \langle \delta \cO \rangle~,
\eea
since $\delta S_q = 0$. The change $\delta\cO$ is the BRST variation of the operator $\cO$. We can thus write the above equation as 
\beq
\langle \cQ \cO \rangle = 0~.
\eeq

\subsection{Introducing topological field theory}
\label{sec:tft-intro}

Now that we are familiar with the basics of BRST quantization of gauge theories, we can move on to introducing topological field theories. Following Ref. \refcite{Birmingham:1991ty} we define a topological field theory as a field theory that consists of
\begin{romanlist}[(ii)]
\item A collection of fields $\Phi$ (which are Grassmann graded) defined on a Riemannian manifold $(M,g)$,
\item A nilpotent operator $Q$, which is odd with respect to the Grassmann grading,
\item Physical states defined to be $Q$-cohomology classes,
\item An energy-momentum tensor which is $Q$-exact, i.e.,
\beq
T_{\alpha \beta} = Q V_{\alpha \beta}(\Phi, g)~,
\eeq
for some functional $V_{\alpha \beta}$ of the fields and the metric.
\end{romanlist}

The collective field content of the theory $\Phi$ includes the gauge field, ghosts, and multipliers. The theory has local gauge symmetry, and, as we briefly discussed before in the case of Yang--Mills, we can construct a BRST type operator $Q$ that is nilpotent. We denote the variation of any functional $\cO (\Phi)$ as:
\beq
\delta \cO = Q \cO~.
\eeq

The physical Hilbert space is defined by the condition: 
\beq
Q~|\textrm{phys}\rangle = 0~.
\eeq
Furthermore, a physical state of the form: 
\beq
|\textrm{phys}\rangle’ = |\textrm{phys}\rangle + Q |\chi \rangle
\eeq
is equivalent to $|\textrm{phys}\rangle$, for any state $|\chi\rangle$. A state is called  $Q$-closed if it is annihilated by $Q$, while a state is called $Q$-exact if it is of form $Q|\chi\rangle$. Thus the physical Hilbert space splits into different equivalence classes called {\it Q-cohomology classes}.

We take $Q$ to be metric independent, which is the simplest situation to deal with, also the best choice of connecting SYM theories with global supersymmetries. For a theory defined on some manifold $M$, with a metric $g_{\alpha \beta}$, the energy-momentum tensor $T_{\alpha \beta}$ is defined by the change in the action under an infinitesimal deformation of the metric:
\beq
\delta_g S_q = \frac{1}{2}\int_M d^n x \sqrt{g} \delta g^{\alpha \beta} T_{\alpha \beta}~.
\eeq
We assume that the functional measure in the path integral is both $Q$-invariant and metric independent.

We now consider the change in the partition function:
\beq
Z = \int [d\Phi] e^{-S_q}~,
\eeq
under the infinitesimal change in the metric:
\bea
\delta_g Z &=& \int [d\Phi] e^{-S_q} (\delta_g S_q)~,\nn \\
&=& \int [d\Phi] e^{-S_q} \Big(-\frac{1}{2}\int_M d^n x \sqrt{g} \delta g^{\alpha \beta} T_{\alpha \beta}\Big)~, \nn \\
&=& \int [d\Phi] e^{-S_q} \Big(-\frac{1}{2}\int_M d^n x \sqrt{g} \delta g^{\alpha \beta} Q V_{\alpha \beta} \Big)~.
\eea
Let us denote:
\beq
\chi = -\frac{1}{2}\int_M d^n x \sqrt{g} \delta g^{\alpha \beta} V_{\alpha \beta}~.
\eeq
Thus, we have 
\bea
\delta_g Z &=& \int [d\Phi] e^{-S_q} Q \chi  = \langle Q \chi \rangle = 0~.
\eea
That is, the partition function $Z$ is independent of metric deformations. It depends not on the local structure of the manifold, but only on global properties. Thus $Z$ can be considered as a topological invariant of the theory.

We can now move on to finding other metric independent correlation functions in the theory. Let us consider the vacuum expectation value of an observable $\cO(\Phi)$:
\beq
\langle \cO \rangle = \int [d\Phi] e^{-S_q} \cO~,
\eeq
and look for the conditions that are sufficient for this expectation value to be a topological invariant, that is, for $\delta_g \langle \cO \rangle$ to be zero. 

Following the steps as before, we find:
\beq
\delta_g \langle \cO \rangle =  \int [d\Phi] e^{-S_q} (\delta_g \cO - \delta_g S_q \cdot \cO)~.
\eeq
Assuming that $\cO$ enjoys the properties:
\beq
\label{eq:vacuum-exp-O}
\delta_g \cO = Q R~\textrm{and}~Q \cO = 0~,
\eeq
for some $R$, we have that
\beq
\label{eq:vacuum-exp-O-2}
\delta_g \langle \cO \rangle = \langle Q R \rangle + \langle  \cQ (\chi \cO) \rangle = 0~.
\eeq
Now, it is clear that if $\cO = Q \cO'$, for some $\cO'$, we automatically have $\langle \cO \rangle = 0$. Thus, BRST invariant operators that are not $Q$-exact are topological invariants if they satisfy the condition $\delta_g \cO = Q R $. 

In the case of Witten type theories, the complete quantum action $S_q$, which comprises the classical action plus all the necessary gauge-fixing and ghost terms, can be written as a BRST commutator, that is,
\beq
S_q = Q V ~,
\eeq
for some functional $V(\Phi, g)$ of the fields, and $Q$ is the nilpotent BRST charge. 

By using the $Q$-exact nature of the action, we can prove that the partition function $Z$ and the above class of topological invariant correlators are also exact at the semi-classical level. Let us introduce a dimensionless parameter $\beta$ to rescale the action $S_q \rightarrow \beta S_q$ and then consider the variation of the partition function under a change in $\beta$:
\bea
\delta_{\beta} Z &=& - \int [d\Phi] e^{-\beta S_q} S_q \delta \beta \nn \\
&=& - \int [d\Phi] e^{-\beta S_q} (Q V)~\delta \beta = 0~.
\eea
This shows that $Z$ is independent of $\beta$, as long as $\beta$ is non-zero\footnote{We cannot set $\beta$ to zero, as the path integral requires a damping factor.}. We can evaluate the partition function in the large-$\beta$ limit. Such a limit corresponds to the semi-classical approximation, in which the path integral is dominated by fluctuations around the classical minima. In Witten type theories, such an approximation is exact. We can also establish the semi-classical exactness of the topologically invariant correlation functions in a similar way.

It should be noted that topological field theories do not admit dynamical excitations. That is, these theories have no propagating degrees of freedom. In Witten type theories, the BRST operator $Q$ plays the role of a supersymmetry charge as well. The classical action for Witten type theories is: 
\beq
S_c = \left\{
  \begin{array}{l}
    0 \\
    \textrm{or a topological invariant}\\
  \end{array} \right.
\eeq
This action admits a large amount of topological shift symmetry:
\beq
A_{\mu}^{A'} = A_{\mu}^A + \epsilon_{\mu}^A~.
\eeq
From the structure of the topological shift symmetry, we can see that each bosonic field has a $Q$-superpartner. We have defined our theory by the requirement that physical states are annihilated by $Q$. Hence, the superpartners are interpreted as ghosts, leading to a total of zero degrees of freedom. The energy of any physical state in these theories is zero, and, hence, there are no physical excitations.

Thus, the number of degrees of freedom in a Witten type topological field theory and a conventional supersymmetric field theory are quite different. There are no physical degrees of freedom at all in Witten type theories. This may seem a little strange since from what we have described above topological field theories are also supersymmetric theories in their own right with supersymmetry charge $\cQ$. If we think of them as topological field theories, they have to satisfy the requirement that they have no degrees of freedom, while, on the other hand, if we think of them as supersymmetric field theories, we require them to have both bosonic and fermionic states. These two requirements do not contradict with each other if we look at these theories from the point of view of the so-called {\it twisting} of the supersymmetry. (We will describe the details of twisting in Section 3.) In the context of the lattice supersymmetry constructions, we are strongly dependent on this view point. 

We can construct topological field theories from SYM theories through the twisting process. The zero degrees of freedom restriction would then be equivalent to a projection to the vacuum states of the supersymmetric gauge theory. Once $\cQ$ is chosen, we can change the physical interpretation of the supersymmetric gauge theory in the following way to make it a topological field theory: We restrict our interest to $\cQ$-invariant path integrals, observables, and states, and we consider anything of the form $\cQ \cO$, for any operator $\cO$, to be trivial. Thus, the interesting observables or states lie in the cohomology groups of $\cQ$. Theories obtained after these restrictions are topological field theories.

Since we will be interested in dynamical excitations of the (twisted) supersymmetric gauge theories, we will not impose these restrictions on path integrals, observables, and states, but treat the theory as merely a twisted version of the original supersymmetric theory that exposes a nilpotent supersymmetry explicitly.

\subsection{Constructing a topological field theory}
\label{sec:constructing-tft}

The original construction of topological quantum field theory by Witten\cite{Witten:1988ze} showed that Donaldson theory can be realized as a four-dimensional ``twisted" $\cN=2$ SYM theory. There are different approaches to deriving the action of Witten's theory and a topological field theory in general. We briefly describe them below. We will be interested in the third method, the method of twisting.

\subsubsection{Gauge-fixing topological shift symmetry}

Beaulieu and Singer\cite{Baulieu:1988xs}, and Brooks, Montano and Sonnenschein\cite{Brooks:1988jm} noted that Witten's theory can be derived by gauge-fixing the local transformation
\beq
\delta A^A_{\mu} = \theta^A_{\mu}~,
\eeq
where $A_{\mu}^A$ is a gauge field in the adjoint representation. The gauge-fixing has two steps: BRST gauge-fixing to expose fermionic symmetries and Yang--Mills gauge-fixing of the gauge field. They started with a classical action that is BRST and Yang--Mills gauge-invariant. The set of classical actions that satisfy these conditions are the trivial classical action $S_c=0$ and actions that are topological invariants (such as the theta term). 

The gauge-fixing condition that leads to Witten's theory corresponds to gauge field configurations with vanishing instanton curvature,
\beq
F^+_{\mu \nu} = F_{\mu \nu} + \widetilde{F}_{\mu \nu} = 0~.
\eeq
A series of topological gauge-fixing steps generate a set of ghosts and ghost for ghost fields, leading to Witten's $\cN=2$ SYM action in four dimensions\cite{Witten:1988ze}.

\subsubsection{Quantization through Batalin--Vilkovisky procedure} 

The basic idea here\cite{Labastida:1988qb} is to regard the instanton equation $F^+_{\mu \nu} = 0$ as arising from a suitable classical action involving a linear combination of the $F^+_{\mu \nu}$ and an auxiliary self-dual field $G_{\mu \nu}$. The equation of motion for $G_{\mu \nu}$ becomes the Langevin equation for the system. This theory has an on-shell reducibility. Quantizing the theory with this on-shell reducibility requires us to make use of the Batalin--Vilkovisky quantization procedure\cite{Batalin:1984jr}. The result is the quantum action of $\cN=2$, $d=4$ SYM theory given by Witten\cite{Witten:1988ze}. 

\subsubsection{Twisting the supercharges of Yang--Mills theory}

There is yet another way to understand the origin of the action given in\cite{Witten:1988ze}. This is the most useful way for us in the context of lattice supersymmetry. The motivation here is to obtain the (scalar) BRST supercharge by ``twisting" a set of conventional (spinorial) supercharges. After twisting, we obtain an action that bears a formal similarity to that of Witten's four-dimensional $\cN = 2$ SYM theory. The twisting procedure can obviously be applied to various classes of SYM theories with extended supersymmetries.

It is most natural to use Euclidean signature in constructing topological quantum field theories. Twisting of the supersymmetries is problematic in Lorentz signature.

\section{Twisted super Yang--Mills theories}

In Section 4 we briefly mentioned that we can twist the supersymmetries of SYM theories to derive topological field theories. Since topological field theories are most naturally related to Euclidean signature, we will be focusing on SYM theories on Euclidean spacetime. Our interest in constructing SYM theories on the lattice also require these theories to have a flat Euclidean signature. Although twisting is not consistent with Lorentz signature, we can usually return to Lorentz signature, if the theory is constructed on a manifold of type $M = {\mathbb R} \times W$, by simply taking Lorentz signature on ${\mathbb R}$.

We are specifically interested in the method of twisting, as it provides a way of studying a class of SYM theories on a flat Euclidean spacetime lattice. All SYM theories do not admit twisting; only SYM theories with extended supersymmetries ($\cN > 1$) undergo twisting. Among the set of extended SYM theories, we focus on a special class of SYM theories that can be {\it maximally twisted}. 

In Section 2, we showed that SYM theories can be constructed only in certain spacetime dimensions. The theories we construct in that way are $\cN=1$ SYM theories. We can construct SYM theories with extended supersymmetries through the method of dimensional reduction. In Table \ref{tb:dim-red-N1-SYM}, we show how a set of SYM theories with extended supersymmetries can be obtained through the dimensional reduction of a set of $\cN=1$ theories in higher dimensions.

\begin{table}[ph]
\tbl{Dimensional reduction of a set of $\cN=1$ SYM theories and their daughter theories in lower dimensions. Here $(a, a)$ represents left- and right-handed supersymmetries.}
{\begin{tabular}{| c | c | c |}
\hline 
  &  &  \\
 $\cN=1$, $d=10$ & $\cN=1$, $d=6$ & $\cN=1$, $d=4$ \\
 $\downarrow$ & $\downarrow$ & $\downarrow$ \\
 $\cN=2$, $d=6$ & $\cN=2$, $d=4$ & $\cN=2$, $d=3$ \\
 $\downarrow$ & $\downarrow$ & $\downarrow$ \\
 $\cN=4$, $d=4$ & $\cN=4$, $d=3$ & $\cN=(2, 2)$, $d=2$ \\
 $\downarrow$ & $\downarrow$ & \\
 $\cN=8$, $d=3$ & $\cN=(4, 4)$, $d=2$ &  \\
 $\downarrow$ &  &   \\
 $\cN=(8, 8)$, $d=2$ &  & \\
   &  &  \\
\hline
\end{tabular} \label{tb:dim-red-N1-SYM}}
\end{table}

Theories with extended supersymmetries in $d$ dimensions contain a (Euclidean) spacetime rotation group $SO(d)$ and an $R$-symmetry group, which we denote by $G_R$. Supersymmetric theories typically have global chiral symmetries that do not commute with the supercharges. They are called ``$R$-symmetries." These symmetries turn out to play a crucial role in twisting. We are interested in the full twist of the Lorentz group - called {\it maximal twisting}. Construction of a manifestly supersymmetric $d$-dimensional Yang--Mills theory through twisting requires the $R$-symmetry group to contain $SO(d)$ as a subgroup. That is, there should exist a nontrivial homomorphism from the Euclidean Lorentz group $SO(d)_E$ to the $R$-symmetry group $G_R$. In Table \ref{tb:SYM-symm-twist} we list a set of Euclidean SYM theories with their Lorentz and $R$-symmetries, and the existence of maximal twist in each case.

\begin{table}[ph]
\tbl{A set of Euclidean SYM theories with symmetry groups and the possibilities of maximal twist.}
{\begin{tabular}{|c| c| c| c|}
\hline
 Theory & Lorentz symmetry & $R$-symmetry & Maximal twist\\ \hline
 $d=2$, $\cN=2$ & $SO(2)$ & $SO(2) \times U(1)$ & Yes\\
 $d=2$, $\cN=4$ & $SO(2)$ & $SO(4) \times SU(2)$ & Yes\\
 $d=2$, $\cN=8$ & $SO(2)$ & $SO(8)$ & Yes\\ \hline
 $d=3$, $\cN=1$ & $SO(3)$ & $U(1)$ & No\\
 $d=3$, $\cN=2$ & $SO(3)$ & $SO(3) \times SU(2)$ & Yes\\
 $d=3$, $\cN=4$ & $SO(3)$ & $SO(7)$ & Yes\\ \hline
 $d=4$, $\cN=1$ & $SO(4)$ & $U(1)$ & No\\
 $d=4$, $\cN=2$ & $SO(4)$ & $SO(2) \times SU(2)$ & No\\
 $d=4$, $\cN=4$ & $SO(4)$ & $SO(6)$ & Yes \\
 \hline
\end{tabular} \label{tb:SYM-symm-twist}}
\end{table}

The constraint on the minimal size of the $R$-symmetry group forbids maximally twisted lattice formulation of some interesting class of theories, such as the $\cN = 2$ SYM (the Seiberg--Witten theory) in four dimensions and the generic $\cN = 1$ supersymmetric QCD theories. 

The well known $\cN=4$ SYM in four dimensions can be twisted in three different ways\cite{Yamron:1988qc,Vafa:1994tf,Marcus:1995mq} but only one of them, introduced by Marcus\cite{Marcus:1995mq}, undergo maximal twisting and, thus, leads to a lattice construction of this theory. The other two twists cannot be implemented on a lattice in a gauge covariant way.

The twists of three-dimensional $\cN = 4$ and $\cN = 8$ and two-dimensional $\cN = (8, 8)$, $\cN = (4, 4)$ theories are presented by Blau and Thompson\cite{Blau:1996bx}.

\subsection{Twisting in $d$ dimensions}
\label{sec:twisting}

In this section, we briefly review the maximal twists of extended SYM theories in the continuum formulation on ${\mathbb R}^d$. From the list we created above we see that the $R$-symmetry group possess an $SO(d)_R$ subgroup for six of the theories. The theories that allow maximal twisting have the property:
\beq
SO(d)_E \times SO(d)_R \subset SO(d)_E \times G_R~.
\eeq
To construct the twisted theory, we embed a new rotation group $SO(d)'$ into the diagonal sum of $SO(d)_E \times SO(d)_R$, and declare this $SO(d)'$ as the new Lorentz symmetry of the theory. This is called the {\it twisted rotation group}.  

The details of twist construction are slightly different in each case. We focus on the general idea of twisting first and then go on to the special cases of interest in later sections. Let us assume that a fermionic field, which is a spacetime spinor, is in the spinor representation of the $R$-symmetry group $SO(d)_R$ as well\footnote{It is the spin group $Spin(d)$ to be more precise, but using $SO(d)$ will also lead to same results.}. After twisting, the fermions become integer spin representations of the twisted rotation group $SO(d)'$, since the product of two half-integer spins is always an integer spin. The fermions still preserve their Grassmann odd nature, but they are now irreducible antisymmetric tensor fields of the twisted rotation group. They can be expressed as a direct sum of scalars, vectors, anti-symmetric tensors, and other higher $p$-forms. 

The bosons of the theory, Grassmann even fields, transform as vectors {\bf d} under the $SO(d)'$ - the gauge bosons $V_{\mu}$ transform as $({\bf d}, {\bf 1})$, and the scalars $B_{\mu}$ transform as $({\bf 1}, {\bf d})$ under the $SO(d)_E \times SO(d)_R$. If there are more than $d$ scalars in the untwisted theory (for example, $\cN=4$, $d=4$ theory has six scalars), they become either $0$-forms or $d$-forms under $SO(d)'$. 

It is clear now why we have used the the name maximal twist for this type of twisting. The twisting procedure involves the twisting of the full Lorentz symmetry group instead of twisting a subgroup of it. The four-dimensional $\cN = 2$ theory can only admit a {\it half twisting} as its $R$-symmetry group is not as large as the Lorentz rotation group $SO(4)_E$. The other two theories, $\cN = 1$ in $d = 4$ and $\cN = 1$ in $d = 3$ do not admit a nontrivial twisting as there is no nontrivial homomorphism from their Euclidean rotation group to their $R$-symmetry group.

The supersymmetries also take new forms under the twisted rotation group. They also transform like twisted fermions, in integer spin representations of the twisted rotation group. The scalar component $\cQ$ of the twisted supersymmetries is nilpotent
\beq
\cQ^2 = 0~.
\eeq

The twisted superalgebra implies that the momentum $P_a$ is now the $Q$-variation of something. That is, it is $Q$-exact. This fact renders it plausible that the entire energy momentum tensor may be $Q$-exact in twisted theories. This, in turn, implies that the entire action of the theory could be written in a $Q$-exact form $S = Q \Lambda$. (In some cases, for example, $\cN=4$ in $d=4$ case, the twisted action is a sum of $Q$-exact and $Q$-closed terms.) The subalgebra $Q^2 = 0$ of the twisted supersymmetry algebra does not produce any spacetime translations. We can use this fact to carry the twisted theory easily onto the lattice.

On a flat Euclidean spacetime, the twisted theory is merely a rewriting of the physical theory, and, indeed, possesses all supersymmetries of the physical theory. The twisted SYM theory can be made topological by interpreting the scalar supercharge $\cQ$ as a BRST operator. Then the observables of the physical theory are restricted only to a set of topological observables, appropriately defined correlators of the twisted operators.

Although the twisted formulation of supersymmetry goes back to Witten\cite{Witten:1988ze} in the topological field theory construction of four-dimensional $\cN=2$ SYM theory in the context of Donaldson invariants, this formulation had been anticipated in earlier lattice work using Dirac--K\"ahler fields\cite{Elitzur:1982vh,Sakai:1983dg,Kostelecky:1983qu,Scott:1983ha,Aratyn:1984bc}. The precise connection between Dirac--K\"ahler fermions and topological twisting was found by Kawamoto and collaborators\cite{Kawamoto:1999zn,Kato:2003ss,D'Adda:2004jb}. They observed that the 0-form supercharge that arises after twisting is a scalar that squares to zero and constitutes a closed subalgebra of the full twisted superalgebra. It is this scalar supersymmetry that can be made manifest in the lattice action, even at finite lattice spacing\cite{Catterall:2004zy,Catterall:2004np,Catterall:2005fd,Catterall:2005eh,Catterall:2009it}.

\subsection{Twisted $\cN=2$, $d=2$ SYM theory}
\label{sec:twistedN2d2}

We begin with a simple example of the twist construction: the two-dimensional $\cN=2$ SYM theory. This theory can be obtained by the dimensional reduction of four-dimensional $\cN=1$ SYM theory. The global symmetry of the four-dimensional theory:
\beq
SO(4)_E \times U(1)~,
\eeq
where $SO(4)_E$ is the Euclidean Lorentz symmetry and $U(1)$ is the chiral symmetry, splits in the following way, after dimensional reduction, to become the global symmetry of the two-dimensional theory
\beq
G = SO(2)_E \times SO(2)_{R_1} \times U(1)_{R_2}~.
\eeq
Here, $SO(2)_E$ is the Euclidean Lorentz symmetry; $SO(2)_{R_1}$ is rotational symmetry along reduced dimensions and $U(1)_{R_2}$ is the chiral $U(1)$ symmetry of the theory. We rewrite the symmetry group of the theory as:
\beq
SO(2)_E \times SO(2)_{R_1} \times U(1)_{R_2} \sim SO(2)_E \times SO(2)_{R_1} \times SO(2)_{R_2}~.
\eeq

Since the internal symmetry group contains two $SO(2)$'s, we can maximally twist this theory in two ways. They are called the $A$-model and the $B$-model twists\cite{Witten:1993yc}. In the $A$-model twist, the twisted rotation is defined as the diagonal $SO(2)$ subgroup of the product of the Lorentz rotation $SO(2)_E$ and the (chiral) $SO(2)_{R_2}$ symmetry. In the $B$-model twist, the twisted rotation group is the diagonal $SO(2)$ subgroup of the product of the Lorentz rotation $SO(2)_E$ and the (internal) $SO(2)_{R_1}$ symmetry.

We will be focusing on the $B$-model twist picture (it is also known as {\it self-dual twist}), since the form of the twisted action resembles that of the orbifold constructions \cite{Kaplan:2002wv,Cohen:2003xe,Cohen:2003qw,Kaplan:2005ta}, a complementary and equivalent approach to lattice supersymmetry. 

The fermions and supersymmetries are now decomposed into integer spin representations of the twisted rotation group - there is a $0$-form $\eta$, a 1-form $\psi_a$ and a 2-form $\chi_{ab}$: 
\bec
  \begin{tabular}{ l  c  c  c } 
    $\textrm{supercharges: }$ & $\cQ$ & $\cQ_a$ & $\cQ_{ab}$ \\  
    $\textrm{fermions: }$ & $\eta$ & $\psi_a$ & $\chi_{ab}$ \\  
    $\textrm{number of fields: }$ & 1 & 2 & 1 \\  
  \end{tabular}
\eec 
The twisted fermions transform under the twisted symmetry, $SO(2)' \times U(1)_{R_2}$, in the following integer spin representation
\beq
\eta \oplus \psi_a \oplus \chi_{ab} \longrightarrow {\bf 1}_{\frac{1}{2}} \oplus {\bf 2}_{-\frac{1}{2}} \oplus {\bf 1}_{\frac{1}{2}}
\eeq
The gauge field $A_a$ transform as $({\bf 2}, {\bf 1})_0$, and the scalars $B_a$ transform as $({\bf 1}, {\bf 2})_0$ under the rotation group $SO(2)_E \times SO(2)_{R_1} \times U(1)_{R_2}$. In the new rotation group $SO(2)' \times U(1)_{R_2}$, they transform as $({\bf 2})_0$. Naturally we can combine the gauge field and scalars to obtain a complexified gauge field in this type of twist, that is,
\beq
\cA_a = A_a + iB_a~\textrm{and}~\cAb_a = A_a - iB_a.
\eeq
Thus, the complexified gauge bosons transform as
\beq
\cA_a \oplus \cAb_a \longrightarrow {\bf 2}_{0} \oplus {\bf 2}_{0}~.
\eeq
\subsubsection{Supersymmetry transformations and twisted action}
The twisting process produces a nilpotent supercharge $\cQ$; it acts on the twisted fields in the following way:
\bea
&& \cQ \cA_a = \psi_a \nn \\
&& \cQ \psi_a = 0 \nn \\
&& \cQ \cAb_a = 0 \nn \\
&& \cQ \chi_{ab} = - \cFb_{ab} \nn \\
&& \cQ \eta = d \nn \\
&& \cQ d = 0
\eea
where $d$ is an auxiliary field introduced for the off-shell completion of the supersymmetry algebra. It has equations of motion:
\beq
d = [\cDb_a, \cD_a]~.
\eeq
The twisted theory has complexified covariant derivatives and field strengths. For a generic field $\Phi$, we have:
\beq
\cD_a \Phi \equiv \partial_a \Phi + [\cA_a, \Phi],~~\cDb_a \Phi \equiv \partial_a \Phi + [\cAb_a, \Phi]~.
\eeq
The field strength takes the form:
\beq
\cF_{ab} = [\cD_a, \cD_b],~~~\cFb_{ab} = [\cDb_a, \cDb_b]~.
\eeq

The action of the twisted theory can be expressed in a $\cQ$-exact form:
\bea
S &=& \cQ \int d^2 x \Tr \Lambda \nn \\
&=& \cQ \int d^2x \Tr \Big(\chi_{ab} \cF_{ab} + \eta[\cDb_a, \cD_a] - \frac{1}{2}\eta d \Big)~.
\eea
After $\cQ$-variation and integrating out the field $d$ yields
\beq
S = \int d^2x \Tr \Big(-\cFb_{ab} \cF_{ab} + \frac{1}{2}[\cDb_a, \cD_a]^2 - \chi_{ab}\cD_{[a}\psi_{b]} - \eta \cDb_a \psi_a \Big)~.
\eeq
The action is $\cQ$-invariant by construction
\beq
\cQ S = \cQ^2 \Lambda = 0~.
\eeq
This theory can be made topological by regarding $\cQ$ as a BRST charge.

\subsubsection{The twisted supersymmetry algebra}

The two-dimensional supersymmetry algebra of the untwisted $\cN = 2$ theory has the form
\beq
\{ Q_{\alpha i}, Q_{\beta j} \} = 2 \delta_{ij} \gamma^a_{\alpha \beta} P_a~,
\eeq
where $Q_{\alpha i}$ is supercharge, the left-indices $\alpha(= 1, 2)$ and the right-indices $i(= 1, 2)$ are Lorentz spinor and internal spinor suffixes labeling two different $\cN = 2$ supercharges, respectively. We can take these operators to be Majorana in two dimensions. $P_a$ is the generator of translation. 

The process of twisting leads to the decomposition of the above supercharges with double spinor indices into scalar, vector and pseudo-scalar components:
\beq
\label{eq:matrix-form}
Q_{\alpha i} = ({\mathbb I}\cQ + \gamma^a \cQ_a + \gamma^5 \widetilde{\cQ})_{\alpha i},~~~\widetilde{\cQ} = \epsilon_{ab}\cQ_{ab}~.
\eeq
These are the twisted supercharges of the two-dimensional $\cN = 2$ SYM theory. The supersymmetry relations can be rewritten by the twisted generators in the following form:
\bea
&&\{\cQ, \cQ_a \} = P_a,~~~\{\widetilde{\cQ}, \cQ_a \} = -\epsilon_{ab}P^b~, \\
&&\cQ^2 = \widetilde{\cQ}^2 = \{\cQ, \widetilde{\cQ} \} = \{\cQ_a, \cQ_b \} = 0~.
\eea
This is the twisted $\cN = d = 2$ supersymmetry algebra.

\subsubsection{Connection with Dirac--K\"ahler fermions}

The supercharges and fermions become tensorial in their representations as a result of twisting. The twisted fermions appearing in the matrix form (\ref{eq:matrix-form}) can be considered as components of a geometrical object called a Dirac--K\"ahler field\cite{D'Adda:2004jb}
\beq
\Psi = (\eta, \psi_a, \chi_{ab})~.
\eeq
If we take a standard free fermion action for a theory with two degenerate Majorana species and replace the fermions by matrices, we find that the action can be easily written as\cite{D'Adda:2004jb}
\beq
S_F = \Tr \Psi^{\dagger} \gamma_a \partial_a \Psi~.
\eeq
Expanding the matrices into (real) components $(\eta, \psi_a, \chi_{ab})$ and doing the trace yields
\beq
S_F = \frac{1}{2} \eta \partial_a \psi_a + \chi_{ab} \partial_{[a} \psi_{b]}~.
\eeq
This geometrical rewriting of the fermionic action yields the so-called Dirac--K\"ahler action, which is most naturally rewritten using the language of differential forms as\cite{Banks:1982iq} 
\beq
S_F = \langle \Psi \cdot (d - d^{\dagger}) \Psi \rangle~.
\eeq
Here $d$ and $d^{\dagger}$ are the usual exterior derivative and its adjoint. Their action of $d$ on general rank $p$-antisymmetric tensors (forms) $\omega_{[\mu_1\cdots \mu_p]}$ yields a rank $p+1$ tensor with components $\omega_{[\mu_1\cdots \mu_p \mu_{p+1}]}$ and the square bracket notation indicates complete antisymmetrization between all indices. The dot notation just indicates that corresponding tensor components are multiplied and integrated over space. The operator $d^{\dagger}$ maps rank $p$ tensors to rank $p-1$. This recasting of the action in geometrical terms not only yields a nilpotent supersymmetry but allows us to discretize the action without inducing fermion doubles\cite{Rabin:1981qj}. 

The choice of maximal twisting gives rise to twisted fermions that are just sufficient to saturate a single Dirac--K\"ahler field\cite{D'Adda:2005zk} and, thus, leads to a lattice construction that does not suffer from the fermion doubling problem.

\subsection{Twisted $\cN=4$, $d=4$ SYM theory}
\label{sec:twistedN4d4}

We begin with looking at the symmetries of the ten-dimensional $\cN=1$ SYM theory, as the theory we are interested in is obtained by the dimensional reduction of it down to four dimensions. Taking spinors into consideration, the rotational symmetry group of the ten-dimensional theory is $Spin(10)$. The ten-dimensional Dirac spinors are in the spin representations ${\mathbb S}^+$ and ${\mathbb S}^-$ of rank 16. These representations are complex conjugates of each other in Euclidean spacetime. We can define a Euclidean chirality operator $\Gamma^E_{11}$ in ten dimensions. It acts on the spin representations by a multiplication by $\mp i$. (In (\ref{eq:chiraliyt-d+1}), the chirality operator acts on Lorentz representations of Dirac spinor.), that is,
\beq
\label{Egamma-acts-on-S}
\Gamma^E_{11} {\mathbb S}^{\pm} = \mp i ~{\mathbb S}^{\pm}~.
\eeq
If $\epsilon$ is the infinitesimal Grassmann valued parameter generating supersymmetry transformations then
\beq
\label{Egamma-acts-on-e}
\Gamma_{11}^E \epsilon = -i \epsilon~.
\eeq

After dimensional reduction, the ten-dimensional Euclidean rotation symmetry group reduces to 
\bec
$Spin(10)_E \rightarrow Spin(4)_E \times Spin(6)_R$~,
\eec
where $Spin(4)_E \sim SU(2) \times SU(2)$ is the four-dimensional rotational symmetry group on ${\mathbb R}^4$ and $Spin(6)_R \sim SU(4)_R$ is the global $R$-symmetry group of the dimensionally reduced theory. 

The ten-dimensional chirality operator also splits into two $\Gamma^E_{11} \rightarrow \widehat{\Gamma}^E\widetilde{\Gamma}^E$, where $\widehat{\Gamma}^E$ measures the $Spin(4)$ chirality and $\widetilde{\Gamma}^E$ measures the $Spin(6)$ chirality. Thus, in four dimensions, the chirality condition becomes
\beq
\Gamma^E_{11} \epsilon = \widehat{\Gamma}^E\widetilde{\Gamma}^E \epsilon~.
\eeq

The complexification of $Spin(4)$ is $SL(2, {\mathbb C}) \times SL(2, {\mathbb C})$ and the two spin representations corresponding to the two eigenvalues of $\widehat{\Gamma}^E$ are $({\bf 2}, {\bf 1})$ and $({\bf 1}, {\bf 2})$ of $SL(2, {\mathbb C}) \times SL(2, {\mathbb C})$\footnote{The two-dimensional representation of the first $SL(2, {\mathbb C})$ tensored with the trivial one-dimensional representation of the second $SL(2, {\mathbb C})$ gives $({\bf 2}, {\bf 1})$, and vice versa gives $({\bf 2}, {\bf 1})$.}. They are pseudo-real in Euclidean dimensions. The spin representations of $Spin(6)$ are the defining four-dimensional representation ${\bf 4}$ of $SU(4)_R$ and its dual $\overline{\bf 4}$. Thus, the four-dimensional fermion fields transform under
\beq
Spin(4) \times Spin(6) \sim SL(2, {\mathbb C}) \times SL(2, {\mathbb C}) \times Spin(6)
\eeq
as
\beq
\label{10d-decomposition}
({\bf 2}, {\bf 1}, \overline{{\bf 4}}) \oplus ({\bf 2}, {\bf 1}, {\bf 4})~.
\eeq
The supersymmetries also transform the same way under $Spin(4) \times Spin(6)$.

Now we introduce the maximal twisting of this theory. This twist was originally introduced by Marcus\cite{Marcus:1995mq}. This twist plays a crucial role in the Geometric Langlands program as well. See Ref. \refcite{Kapustin:2006pk} for a natural description of the geometric Langlands program using the (Marcus) twisted $\cN = 4$ SYM theory in four dimensions compactified on a Riemann surface.

There is a nontrivial homomorphism from the four-dimensional rotation group $Spin(4)$ to the $R$-symmetry group $Spin(6)$ of the theory. That means there exists maximal twisting of the theory. We replace the $Spin(4)$ rotation group with a different subgroup $Spin'(4)$ of $Spin(4) \times Spin(6)$. Though the new $Spin'(4)$ group is isomorphic to the original rotational symmetry $Spin(4)$, and acts on ${\mathbb R}^4$ the same way that $Spin(4)$ does, it acts differently on the $\cN=4$ gauge theory. 

We choose the homomorphism from $Spin(4)$ to $Spin(6)$, such that the action of $Spin'(4)$ on ${\mathbb S}^+$ has a non-zero invariant vector. Since the supersymmetry generator $\epsilon$ takes values in ${\mathbb S}^+$ (See (\ref{Egamma-acts-on-S}) and (\ref{Egamma-acts-on-e}) above), a choice of an invariant vector in ${\mathbb S}^+$ will give us a $Spin'(4)$-invariant supersymmetry. We will call it $\cQ$. This is a scalar symmetry under the $Spin'(4)$ group, and it will automatically obey $\cQ^2 = 0$. 

We describe below how the fields transform under the twisted rotation group. From the twist construction, we want the ${\bf 4}$ of $Spin(6)$ ($=SU(4)_R$) to transform as $({\bf 2}, {\bf 1}) \oplus ({\bf 1}, {\bf 2})$ of $Spin(4)$($=SU(2) \times SU(2)$). The $\overline{\bf 4}$ of $Spin(6)$, which is the complex conjugate of the ${\bf 4}$, transforms the same way under $Spin(4)$, since the $({\bf 2}, {\bf 1})$ and $({\bf 1}, {\bf 2})$ of $Spin(4)$ are pseudo-real.

We can embed the $Spin(4)$(=$SU(2) \times SU(2)$) in $Spin(6)$ (= $SU(4)_R$). This embedding commutes with the additional $U(1)$ group. So our embedding is such that the ${\bf 4}$ of $Spin(6)$ transforms under $SU(2) \times SU(2) \times U(1)$ as $({\bf 2}, {\bf 1})_1 \oplus ({\bf 1}, {\bf 2})_{-1}$. The  $\overline{\bf 4}$ transforms as the complex conjugate of this, or  $({\bf 2}, {\bf 1})_{-1} \oplus ({\bf 1}, {\bf 2})_1$.

We could also use the language of $SO$ groups to describe the twist instead of $Spin$ groups. To do so, we use the fact that the fundamental six-dimensional vector representation ${\bf 6}$ of $SO(6)$ is, in terms of $Spin(6) = SU(4)_R$, the same as antisymmetric part of ${\bf 4} \otimes {\bf 4}$. So ${\bf 6}$ is the antisymmetric part of $({\bf 2}, {\bf 1})_1 \oplus ({\bf 1}, {\bf 2})_{-1}$, which is $({\bf 2}, {\bf 2})_0 \oplus ({\bf 1}, {\bf 1})_2 \oplus ({\bf 1}, {\bf 1})_{-2}$. Here $({\bf 2}, {\bf 2})$ is the same as the vector representation ${\bf 4}$ of $SO(4)$. So the ${\bf 6}$ of $SO(6)$ decomposes into the sum of a vector ${\bf 4}$ and two scalars of $SO(4)$.

We can likewise analyze how the supersymmetries transform under $Spin'(4)$. The $\overline{\bf 4}$ of $Spin(6)$ transforms as $({\bf 2}, {\bf 1})_{-1} \oplus ({\bf 1}, {\bf 2})_1$ of $Spin'(4) \times U(1)$, and the ${\bf 4}$ as $({\bf 2}, {\bf 1})_1 \oplus ({\bf 1}, {\bf 2})_{-1}$. So using (\ref{10d-decomposition})
\beq
({\bf 2}, {\bf 1}, \overline{{\bf 4}}) \oplus ({\bf 2}, {\bf 1}, {\bf 4})~,\nn
\eeq
the supersymmetries that transform as $({\bf 2}, {\bf 1})$ of $Spin(4)$ transform under $Spin'(4) \times U(1)$ as 
\beq
({\bf 2}, {\bf 1})_0 \otimes \Big[({\bf 2}, {\bf 1})_{-1} \oplus ({\bf 1}, {\bf 2})_1\Big] = ({\bf 1}, {\bf 1})_{-1} \oplus ({\bf 3}, {\bf 1})_{-1} \oplus ({\bf 2}, {\bf 2})_1~,
\eeq
and the supersymmetries that transform as $({\bf 1}, {\bf 2})$ of $Spin(4)$ transform under $Spin'(4) \times U(1)$ as
\beq
({\bf 1}, {\bf 2})_0 \otimes \Big[({\bf 2}, {\bf 1})_{-1} \oplus ({\bf 1}, {\bf 2})_1\Big] = ({\bf 1}, {\bf 1})_{-1} \oplus ({\bf 1}, {\bf 3})_{-1} \oplus ({\bf 2}, {\bf 2})_1~.
\eeq
Thus, the supercharges and fermions transform under the new rotation group 
\beq
SU(2)' \times SU(2)' \times U(1)\nn
\eeq
as
\beq
({\bf 1}, {\bf 1})_{-1} \oplus ({\bf 2}, {\bf 2})_1 \oplus [({\bf 3}, {\bf 1}) \oplus ({\bf 1}, {\bf 3})]_{-1} \oplus ({\bf 2}, {\bf 2})_1 \oplus ({\bf 1}, {\bf 1})_{-1}~,
\eeq
or equivalently under the rotation group
\beq
SO(4)' \times U(1) \nn
\eeq
as
\beq
{\bf 1}_{-1} \oplus {\bf 4}_1 \oplus {\bf 6}_{-1} \oplus {\bf 4}_1 \oplus {\bf 1}_{-1}~.
\eeq
As a result of this choice of embedding, the twisted theory contains supersymmetries and fermions in integer spin representations. They transform as scalars, vectors and higher rank $p$-form tensors:
\bec
  \begin{tabular}{ l  c  c  c  c  c }
    $\textrm{supercharges: }$ & $\cQ$ & $\cQ_{\mu}$ & $\cQ_{\mu \nu}$ & $\bar{\cQ}_{\mu}$ & $\bar{\cQ}$\\ 
    $\textrm{fermions: }$ & $\eta$ & $\psi_{\mu}$ & $\chi_{\mu \nu}$ & $\bar{\psi}_{\mu}$ & $\bar{\eta}$\\
    $\textrm{number of fields: }$ & 1 & 4 & 6 & 4 &1 \\ 
  \end{tabular}
\eec
The four gauge bosons transform as $({\bf 2}, {\bf 2})_0$ under the twisted rotation group. We label them as a vector field $A_{\mu}$. Similarly, four of the six scalars of the theory are now elevated to the same footing as the gauge bosons; they also transform as $({\bf 2}, {\bf 2})_0$ under the twisted rotation group. We label them as a vector field $B_{\mu}$. The two other scalars remain as singlets under the twisted rotation group. We label them by $\phi$ and $\bar{\phi}$. Thus the bosons of the twisted theory transform as:
\beq
SU(2)' \times SU(2)' \times U(1) \rightarrow ({\bf 1}, {\bf 1})_1 \oplus ({\bf 2}, {\bf 2})_0 \oplus ({\bf 2}, {\bf 2})_0 \oplus ({\bf 1}, {\bf 1})_{-1}~,
\eeq
or equivalently
\beq
SO(4)' \times U(1) \rightarrow {\bf 1}_1 \oplus {\bf 4}_0 \oplus {\bf 4}_0 \oplus {\bf 1}_{-1}~.
\eeq
We parametrize the bosonic field content of the theory by 
\bec
  \begin{tabular}{ l  c  c  c  c }
    $\textrm{bosons: }$ & $\phi$ & $A_{\mu}$ & $B_{\mu}$ & $\bar{\phi}$\\ 
    $\textrm{number of fields: }$ & 1 & 4 & 4 &1 \\
  \end{tabular}
\eec

\subsection{Supersymmetry transformations and twisted action}
\label{subsec:susy-trans-tw-ac}

We have seen that the two vector fields $A_{\mu}$ and $B_{\mu}$ of the twisted $\cN=4$, $d=4$ theory transform the same way under the twisted rotation group. We can take the complex combination of these two vector fields to describe the twisted theory in a compact way\cite{Marcus:1995mq}. Now there are two complexified connections in the theory:
\bea
\cA_{\mu} \equiv A_{\mu} + iB_{\mu}~, \nn\\
\cAb_{\mu} \equiv A_{\mu} - iB_{\mu}~.
\eea
It is possible to define three covariant derivatives and field strengths\footnote{We employ an anti-hermitian basis for the generators $U(N)$.} using these connections:
\bea
D_{\mu} ~\cdot \equiv \partial_{\mu} + [A_{\mu}, ~\cdot~ ],~~~ F_{\mu \nu} \equiv [D_{\mu}, D_{\nu}]~,\\
\cD_{\mu} ~\cdot \equiv \partial_{\mu} + [\cA_{\mu}, ~\cdot~ ],~~~ \cF_{\mu \nu} \equiv [\cD_{\mu}, \cD_{\nu}]~,\\
\cDb_{\mu} ~\cdot \equiv \partial_{\mu} + [\cAb_{\mu}, ~\cdot~ ],~~~ \cFb_{\mu \nu} \equiv [\cDb_{\mu}, \cDb_{\nu}]~.
\eea
To make contact with the lattice construction, which we will discuss in Section 5, we will go one more step further. We assemble the complexified gauge fields and the two scalar fields, $\phi$ and $\overline{\phi}$, into a single five-component complexified connection:
\beq
\cA_{a} = \Big(\cA_{\mu} \equiv A_{\mu} + i B_{\mu},~~\cA_{5} \equiv A_5 + i B_5 \Big)~, ~~~a = 1, \cdots, 5\; ; \mu = 1, \cdots, 4
\eeq
where the fifth component $\cA_5 = \phi$ and $\cAb_5 = \phib$. Correspondingly, we package the fermions in the $SU(5) \times U(1)$ representation (which is a subgroup of $SO(10)$, the Lorentz symmetry group of the ten-dimensional theory) - they become five-dimensional scalar, vector and antisymmetric tensors $(\eta,\psi_a,\chi_{ab})$. The original twisted theory will then be obtained by simple dimensional reduction of a theory in five dimensions. A similar language arises in the orbifold construction of this theory\cite{Kaplan:2005ta} where the fermions and bosons transform in the representations of $SU(5)\times U(1)$:
\bec
  \begin{tabular}{ l  c  c  l}
    $\textrm{bosons: }$ & ${\bf 10}$ & $\rightarrow$ & ${\bf 5} \oplus \overline{\bf 5}$ \\ 
    $\textrm{fermions: }$ & ${\bf 16}$ & $\rightarrow$ & ${\bf 1} \oplus {\bf 5} \oplus \overline{\bf 10}$ \\
  \end{tabular}
\eec

In addition to these fields, we introduce one auxiliary bosonic scalar field $d$ for off-shell completion of the scalar supersymmetry.

The nilpotent scalar supersymmetry $\cQ$ now acts on these fields in a simple manner
\bea
\cQ \cA_{a} &=& \psi_{a} \nn \\
\cQ \psi_a &=& 0 \nn \\
\cQ \cAb_a &=& 0 \nn \\
\cQ \chi_{ab} &=& -\cFb_{ab} \nn \\
\cQ \eta &=& d \nn \\
\cQ d &=& 0
\eea
The action of the twisted theory can now be expressed in a compact five-dimensional form, as a linear combination of $\cQ$-exact and $\cQ$-closed terms:
\beq
S = \cQ \Lambda + S_{\cQ{\rm -closed}}~,
\eeq
where
\beq
\Lambda = \int\Tr \Big(\chi_{ab} \cF_{ab} + \eta [\cDb_a, \cD_a] - \frac{1}{2} \eta d\Big)~,
\eeq 
and 
\beq
S_{\cQ{\rm -closed}} = - \frac{1}{2} \int ~\Tr \epsilon_{abcde} \chi_{de} \cDb_{c} \chi_{ab}~.
\eeq
The invariance of the $\cQ$-closed term is a result of the Bianchi identity (or Jacobi identity for covariant derivatives) 
\beq
\epsilon_{abcde}\cDb_{c}\cFb_{de} =  \epsilon_{abcde}{[} \cDb_c, {[} \cDb_d, \cDb_e]] = 0~.
\eeq
Carrying out the $\cQ$-variation and subsequently eliminating the auxiliary field $d$ using the equation of motion, we can write down the action in terms of the propagating fields:
\bea
\label{eq:compact-five-d-action-0}
S &=& \int\Tr \Big( -\cFb_{ab} \cF_{ab} + \frac{1}{2} [\cDb_a, \cD_a]^2 - \chi_{ab} \cD_{[a}\psi_{b]} - \eta\cDb_a \psi_a \nn \\
&&~~~~~~~~~~~- \frac{1}{2}\epsilon_{abcde} \chi_{de} \cDb_{c} \chi_{ab}\Big).~~~
\eea

We can obtain the twisted theory in four dimensions by dimensional reduction of this theory along the 5th direction. We write down the decomposition of five-dimensional fields into four-dimensional fields as follows
\bea
\cA_a &\rightarrow& \cA_{\mu} \oplus \phi \nn \\
\cF_{ab} &\rightarrow& \cF_{\mu \nu} \oplus \cD_{\mu} \phi \nn \\
{[}\cDb_a, \cD_a ] &\rightarrow& {[}\cDb_{\mu}, \cD_{\mu}] \oplus [\overline{\phi}, \phi] \nn \\
\psi_a &\rightarrow& \psi_{\mu} \oplus \overline{\eta} \nn \\
\chi_{ab} &\rightarrow& \chi_{\mu \nu} \oplus \overline{\psi}_{\mu}
\eea
The action (\ref{eq:compact-five-d-action-0}), after dimensional reduction, yields:
\bea
S &=& \int \Tr \Big(-\cF_{\mu \nu} \cF_{\mu \nu} + \frac{1}{2} [\cDb_{\mu}, \cD_{\mu}]^2 + \frac{1}{2}{[}\phib, \phi]^2 + (\cDb_{\mu}\phib) (\cD_{\mu}\phi) - \chi_{\mu \nu} \cD_{[\mu} \psi_{\nu]} \nn \\
\label{eq:susy-lattice-twist-action}
&& -\psib_{\mu} \cD_{\mu} \etab - \psib [\phi, \psi_{\mu}] - \eta \cDb_{\mu} \psi_{\mu} - \eta {[}\phib, \etab] - \chi^*_{\mu \nu} \cDb_{\mu} \psib_{\nu} - \frac{1}{2}\chi^*_{\mu \nu} {[}\phib, \chi_{\mu \nu}]\Big),
\eea
where the last two terms arise from the dimensional reduction of the $\cQ$-closed term with $\chi^*$, the Hodge dual of $\chi$,
defined as $\chi^*_{\mu \nu} = \frac{1}{2} \epsilon_{\mu \nu \rho \lambda} \chi_{\rho \lambda}$ and $\psib_\mu=\frac{1}{2}\chi_{5\mu}$.  

The twisted supersymmetry transformations take the following form after dimensional reduction to four dimensions:
\bea
&& \cQ \cA_{\mu} = \psi_{\mu},~~~\cQ \psi_{\mu} = 0 \nn \\
&& \cQ \cAb_{\mu} = 0,~~~\cQ \chi_{\mu \nu} = -\cFb_{\mu \nu} \nn \\
&& \cQ \eta = d,~~~\cQ d = 0,~~~\cQ \phi = \overline{\eta} \nn \\
&& \cQ \overline{\eta} = 0,~~~\cQ \overline{\psi}_{\mu} = \cDb_{\mu} \phi \nn \\
&& \cQ \overline{\phi} = 0 
\eea

\section{Supersymmetric lattices}

Continuum supersymmetric field theories when naively discretized on the lattice break supersymmetry completely. The attempt to arrive at the continuum limit of the renormalized lattice theories, by carefully tuning the coefficients of the counterterms, turns out to be unnatural and practically impossible in most of the cases. For this reason supersymmetric field theories resisted discretization on the lattice for a long time since they were discovered. 

We can easily identify the problem with discretization just by looking at the supersymmetry algebra. It naively breaks on the lattice. The algebra $\{Q,\overline{Q}\} = \gamma_a P_a$ is already broken at the classical level as there are no infinitesimal translation generators on a discrete spacetime. Another way to realize this difficulty is by looking at the supersymmetry variation on the lattice. A naive supersymmetry variation of a naively discretized supersymmetric theory cannot yield zero as a consequence of the failure of the Leibniz rule when applied to lattice difference operators. 

At present there exist a new set of theoretical tools and ideas to construct a family of lattice models that retain exactly some of the continuum supersymmetry at non-zero lattice spacing. The focus is to maintain a particular subalgebra of the full supersymmetry algebra in the lattice theory that contain some supersymmetric charges that are exact on the lattice theory. This exact symmetry will constrain the effective lattice action and protect the theory from dangerous supersymmetry violating counterterms. The resultant supersymmetric lattice theories turns out to be local and free of doublers, and, in the case of Yang--Mills theories, also possess exact gauge-invariance. These lattice models, in principle, form the basis for a truly non-perturbative definition of the continuum supersymmetric field theories.

Having a lattice formulation of supersymmetric gauge theories is very advantageous, as it opens up a large arena of theoretical and numerical investigations. For example, the availability of a supersymmetric lattice construction for the four-dimensional $\cN=4$ SYM theory is clearly very exciting from the point of view of exploring the connection between gauge theories and string/gravitational theories. The lattice formulation of this theory is important in its own right, even without the connection to string theory -- it provides a non-perturbative formulation of a supersymmetric gauge theory.

The geometric structure of twisted SYM theories allows them to be easily transported onto the lattice. The fermions manifest themselves in integer spin representations of the twisted rotation group. They carry the structure of anti-symmetric tensor fields. They also fill out the right number of ingredients to build a single Dirac-K\"ahler field. Such a construction suitably evades the fermion doubling problem on the lattice. The nilpotent supercharge exposed by the process of twisting does not generate translations. This property makes the twisted theory to be discretized keeping the nilpotent scalar supersymmetry unbroken. All these unique features make the twisted continuum theory well qualified to undergo discretization. We follow a geometric discretization scheme to construct lattice versions of the twisted SYM theories \cite{Catterall:2004np,Rabin:1981qj,Becher:1981cb,Becher:1982ud,Aratyn:1984bd,Damgaard:2008pa} and it is detailed in Sub. sec. 5.1.

There also exist other variants of exact lattice supersymmetry formulations in the literature. In Refs \refcite{Hanada:2010kt} and \refcite{Hanada:2010gs} lattice formulation of four-dimensional $\cN=4$ SYM theory, which requires no fine-tuning is constructed. Similarly, the four-dimensional $\cN=2$ SYM theory can be constructed\cite{Hanada:2011qx}. Interestingly, these formulations regularize the four-dimensional space-time by a two-dimensional lattice and fuzzy 2-sphere. Lattice formulation of two-dimensional $\cN=(2,2), (4,4)$ SYM theories preserving nilpotent supercharges is presented in Refs. \refcite{Sugino:2003yb} and \refcite{Sugino:2004qd}. Although it is not based on the geometric construction we will be discussing, it has been nonperturbatively shown that continuum limit of the formulation for two-dimensional $\cN=(2,2)$ SYM\cite{Kanamori:2008bk,Hanada:2009hq} gives the same physics as the orbifold constructions\cite{Hanada:2010qg}.

\subsection{Geometric structure of continuum and lattice action}
\label{cont-latt-action}

We begin the description of the lattice formulation of supersymmetric gauge theories by looking at the general structure of the continuum gauge theory. The bosonic and fermionic fields are in integer spin representations of the twisted rotation group. The fermions are $p$-forms, that is, they are tensor fields in general. We take the gauge group to be $U(N)$ and represent all the fields in the adjoint representation of this gauge group. The continuum action, defined on a $d$-dimensional flat Euclidean spacetime has the following properties.

The action is Lorentz invariant, and it consists of complex covariant derivatives $\cD_a$ and $\cDb_a$ associated with a complex (not hermitian) connection $\cA_a$ and its complex conjugate $\cAb_a$, respectively, and a set of (bosonic and/or fermionic) tensor fields, $\{f^{(\pm)}_{a_1 \cdots a_p} \}$, that is,
\beq
S_{\textrm{cont}} = S_{\textrm{cont}} \Big( \cD_a, \cDb_a, \{f^{(\pm)}_{a_1 \cdots a_p} \} \Big)~.
\eeq

The covariant derivatives can act on the tensor fields in a curl-like or a divergence-like operation. The curl-like operation gives
\bea
\cD_a \{f^{(\pm)}_{a_1 \cdots a_p}(x) \} &=& \partial_a \{f^{(\pm)}_{a_1 \cdots a_p}(x) \} + {[}\cA_a(x), \{f^{(\pm)}_{a_1 \cdots a_p}(x) \} {]}~, \nn \\
\label{eq:curl-like-2}
\cDb_a \{f^{(\pm)}_{a_1 \cdots a_p}(x) \} &=& \partial_a \{f^{(\pm)}_{a_1 \cdots a_p}(x) \} + {[}\cAb_a(x), \{f^{(\pm)}_{a_1 \cdots a_p}(x) \} {]}~,
\eea
while the divergence-like operation gives
\bea
\cD_{a_i} \{f^{(-)}_{a_1 \cdots a_p}(x) \} &=& \partial_{a_i} \{f^{(-)}_{a_1 \cdots a_p}(x) \} + [\cA_{a_i}(x), \{f^{(-)}_{a_1 \cdots a_p}(x) \}]~, \nn \\
\label{eq:div-like-2}
\cDb_{a_i} \{f^{(+)}_{a_1 \cdots a_p}(x) \} &=& \partial_{a_i} \{f^{(+)}_{a_1 \cdots a_p}(x) \} + [\cAb_{a_i}(x), \{f^{(+)}_{a_1 \cdots a_p}(x) \}]~,
\eea
where $(1 \leq i \leq p)$.

We choose a hypercubic abstract lattice to write down the lattice versions of the SYM theories\footnote{Later we will see that there are more exotic lattice choices that expose the maximum amount of symmetry and thus impose stronger constraints on the counterterms on the lattice. We can write down a set of transformation rules that connects the basis vectors of such lattices with those of the hypercubic lattice.}. The $p$-form fields are mapped to lattice fields living on $p$-cells of the lattice. The $p$-cell lattice field can have two possible orientations. This orientation is physical and determines how the lattice fields are gauge rotated on the lattice. So we need to choose an orientation that respects gauge symmetry on the lattice. We choose the fields to be positively oriented, that is, the orientation of the field corresponds to the one in which the link vector has positive components with respect to the coordinate basis.

We replace the complexified connections $\cA_a$ and $\cAb_a$ with the following link fields on the lattice:
\bea
\cA_a(x) &\rightarrow& e^{\cA_a(\bn)} \equiv \cU_a(\bn)~, \nn \\
\cAb_a(x) &\rightarrow& e^{\cAb_a(\bn)} \equiv \cUb_a(\bn)~,
\eea
where $\bn$ denotes the integer valued lattice site.

The lattice action contains a set of site, link and $p$-form fields:
\beq
S_{\textrm{latt}} = S_{\textrm{latt}} \Big( \cU_a(\bn), \cUb_a(\bn), \{f^{(\pm)}_{a_1 \cdots a_p}(\bn) \} \Big)~.
\eeq

The fields on the lattice can be regarded as variables living on orientable links. As a result of this prescription the lattice variables $\cU_a(\bn)$, $\cUb_a(\bn)$, $\{f^{(+)}_{a_1 \cdots a_p}(\bn) \}$,  $\{f^{(-)}_{a_1 \cdots a_p}(\bn) \}$ live on links $(\bn, \bn+\hatbmu_a)$, $(\bn + \hatbmu_a, \bn)$, $(\bn, \bn+\hatbmu_{a_1}+\cdots+\hatbmu_{a_p})$ and $(\bn+\hatbmu_{a_1}+\cdots+\hatbmu_{a_p}, \bn)$ respectively. A site variable $\eta(\bn)$ lives on a link $(\bn, \bn)$.

For $G(\bn) \in U(N)$, the lattice variables translate under the gauge transformations in the following way:
\bea
\cU_a(\bn) &\rightarrow& G(\bn) \cU_a(\bn) G^{\dagger}(\bn + \hatbmu_a) \nn \\ 
\cUb_a(\bn) &\rightarrow& G(\bn + \hatbmu_a) \cUb_a(\bn)G^{\dagger}(\bn) \nn \\ 
\{f^{(+)}_{a_1 \cdots a_p}(\bn) \} &\rightarrow& G(\bn) \{f^{(+)}_{a_1 \cdots a_p}(\bn) \} G^{\dagger}(\bn + \hatbmu_{a_1} + \cdots+\hatbmu_{a_p} ) \nn \\ 
\label{eq:gauge-lattice-4}
\{f^{(-)}_{a_1 \cdots a_p}(\bn) \} &\rightarrow& G(\bn + \hatbmu_{a_1} + \cdots+\hatbmu_{a_p} )\{f^{(-)}_{a_1 \cdots a_p}(\bn) \} G^{\dagger}(\bn)
\eea
Notice that these transformations respect the $p$-cell and orientation assignments of the lattice fields.

The covariant derivatives $\cD_a$ ($\cDb_a$) in the continuum become forward and backward covariant differences $\cD^{(+)}_a~(\cDb^{(+)}_a)$ and $\cD^{(-)}_a~(\cDb^{(-)}_a)$, respectively. They act on the lattice fields $f^{(\pm)}_{a_1 \cdots a_p}(\bn)$ in the following way:
\bea
&&\cD_b^{(+)}f^{(+)}_{a_1 \cdots a_p}(\bn) \equiv \cU_b(\bn)f^{(+)}_{a_1 \cdots a_p}(\bn + \hatbmu_b)-f^{(+)}_{a_1 \cdots a_p}(\bn) \cU_b(\bn+\hatbmu)~~~~~~~~\nn \\
&&\cD_b^{(+)}f^{(-)}_{a_1 \cdots a_p}(\bn) \equiv \cU_b(\bn+\hatbmu)f^{(-)}_{a_1 \cdots a_p}(\bn + \hatbmu_b)-f^{(-)}_{a_1 \cdots a_p}(\bn) \cU_b(\bn)~~~~~~~~\nn \\
&&\cDb_b^{(+)}f^{(+)}_{a_1 \cdots a_p}(\bn) \equiv f^{(+)}_{a_1 \cdots a_p}(\bn + \hatbmu_b)\cUb_b(\bn+\hatbmu)-\cUb_b(\bn)f^{(+)}_{a_1 \cdots a_p}(\bn)~~~~~~~~\nn \\
&&\cDb_b^{(+)}f^{(-)}_{a_1 \cdots a_p}(\bn) \equiv f^{(-)}_{a_1 \cdots a_p}(\bn + \hatbmu_b)\cUb_b(\bn)-\cUb_b(\bn+\hatbmu)f^{(-)}_{a_1 \cdots a_p}(\bn)~~~~~~~~
\eea
where we have defined $\hatbmu = \sum_{i=1}^p \hatbmu_{a_i}$. 
\subsubsection{Prescription for discretization}
Thus, from a given continuum twisted action in $d$ dimensions, we can construct the lattice action using the following prescription for discretization. 
\begin{itemize}
\item[$(i.)$] For complexified gauge bosons in the continuum $\cA_a(x)$ and $\cAb_a(x)$, we introduce lattice link fields $\cU_a(\bn) = e^{\cA_a(\bn)}$ and $\cUb_a(\bn) = e^{\cAb_a(\bn)}$. 
\item[$(ii.)$] A continuum $p$-form field will be mapped to a corresponding lattice $p$-form field associated with a $p$-dimensional hypercubic lattice. The lattice site $(\bn)$ is spanned by the (positively oriented) unit vectors $\{\hatbmu_{a_1} \cdots \hatbmu_{a_p}\}$. The continuum fields become link variables and live on oriented links. The continuum complex covariant derivatives $\cD_a$ and $\cDb_a$ become link variables $\cU_a(\bn)$ and $\cUb_a(\bn)$, and they live on the links $(\bn, \bn+\hatbmu_a)$ and $(\bn + \hatbmu_a, \bn)$, respectively. The tensor fields $f^{(+)}_{a_1 \cdots a_p}(x)$ and $f^{(-)}_{a_1 \cdots a_p}(x)$ become lattice variables  $f^{(\pm)}_{a_1 \cdots a_p}(\bn)$ living on links $(\bn, \bn+\hatbmu_{a_1} + \cdots + \hatbmu_{a_p})$ and $(\bn+\hatbmu_{a_1} + \cdots + \hatbmu_{a_p}, \bn)$, respectively.
\item[$(iii.)$] The curl-like complex covariant derivatives become forward covariant differences given in (\ref{eq:curl-like-2}).
\item[$(iv.)$] The divergence-like complex covariant derivatives become backward covariant differences given in (\ref{eq:div-like-2}).
\item[$(v.)$] The gauge transformations of lattice variables are given in (\ref{eq:gauge-lattice-4}).
\end{itemize}

\subsection{Two-dimensional lattice $\cN=2$ SYM theory}
\label{sec:latt-N2d2}

As a result of the geometrical discretization prescription the two-dimensional $\cN=2$ lattice SYM theory lives on a two-dimensional square lattice spanned by two orthogonal basis vectors. The fermionic and bosonic fields live on sites, links and body diagonal of the lattice unit cell.

The lattice covariant forward difference operator $\cD_a^{(+)}$ acts on the lattice scalar and vector fields in the following way:
\bea
\cD_a^{(+)}f(\bn) &=& \cU_a(\bn)f(\bn + \hatbmu_a) -f(\bn) \cU_a(\bn)~, \nn \\
\cD_a^{(+)}f_b(\bn) &=& \cU_a(\bn)f_b(\bn + \hatbmu_a) - f_b(\bn) \cU_a(\bn + \hatbmu_b)~,
\eea
where $\hatbmu_a$ is the unit vector along the $\ba$ direction; there are two unit vectors: ($\hatbmu_1, \hatbmu_2$). We have replaced the continuum complex gauge fields $\cA_a$ by non-unitary link fields $\cU_a = e^{\cA_a}$. 

The lattice covariant backward difference operator $\cDb_a^{(-)}$ replaces the continuum covariant derivative in divergence-like operations and its action on (positively oriented) lattice vector fields can be obtained by requiring that it to be the adjoint to $\cD_a^{(+)}$. Thus, its action on lattice vectors is
\beq
\cDb_a^{(-)} f_a (\bn) = f_a(\bn)\cUb_a(\bn) - \cUb_a (\bn - \hatbmu_a) f_a(\bn - \hatbmu_a)~.
\eeq
The nilpotent scalar supersymmetry acts on the lattice fields in the following way:
\bea
\cQ \cU_a(\bn) &=& \psi_a(\bn) \nn \\
\cQ \psi_a(\bn) &=& 0 \nn \\
\cQ \cUb_a(\bn) &=& 0 \nn \\
\cQ \chi_{ab}(\bn) &=& \cF_{ab}^{\dagger}(\bn) \nn \\
\cQ \eta(\bn) &=& d(\bn) \nn \\
\cQ d(\bn) &=& 0
\eea
\begin{figure}[pb]
\centerline{\psfig{file=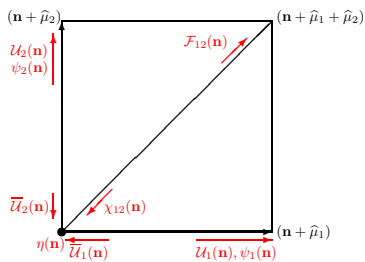,width=4.7cm}}
\vspace*{8pt}
\caption{Orientations of the twisted fields of the two-dimensional $\cN=2$ SYM on a two-dimensional Euclidean lattice. \label{fig:square-fields}}
\end{figure}

The lattice field strength can be written as: 
\beq
\cF_{ab}(\bn) = \cD_a^{(+)}\cU_b(\bn) = \cU_a(\bn)\cU_b(\bn + \hatbmu_a) - \cU_b(\bn)\cU_a(\bn + \hatbmu_b)~.
\eeq
It reduces to the continuum (complex) field strength in the naive continuum limit and is automatically antisymmetric in the indices.

The supersymmetry transformations on the lattice, associated with the nilpotent supersymmetry, imply that the fermion fields $\psi_a(\bn)$ have the same orientation as their superpartners, the gauge links $\cU_a(\bn)$, and run from $\bn$ to $(\bn + \hatbmu_a)$. However, the field $\chi_{ab}(\bn)$ must have the same orientation as $\cF_{ab}^{\dagger}(\bn)$ and hence is to be assigned to the negatively oriented link running from $(\bn + \hatbmu_a + \hatbmu_b)$ to $\bn$. The negative orientation is crucial for allowing us to write down gauge-invariant expressions for the fermion kinetic term. The scalar fields $\eta(\bn)$ and $d(\bn)$ can be taken to transform simply as site fields.

\subsubsection{Gauge transformations on the lattice}

The gauge transformation properties of the lattice fields conveniently summarize these link mappings and orientations:
\bea
\eta(\bn) &\rightarrow& G(\bn)\eta(\bn)G^{\dagger}(\bn) \nn \\ 
\psi_a(\bn) &\rightarrow& G(\bn)\psi_a(\bn)G^{\dagger}(\bn + \hatbmu_a) \nn \\ 
\chi_{ab} (\bn) &\rightarrow& G(\bn + \hatbmu_a + \hatbmu_b)\chi_{ab}(\bn)G^{\dagger}(\bn) \nn \\ 
\cU_a(\bn) &\rightarrow& G(\bn)\cU_a(\bn)G^{\dagger}(\bn + \hatbmu_a) \nn \\ 
\cUb_a (\bn) &\rightarrow& G(\bn + \hatbmu_a)\cUb_a(\bn)G^{\dagger}(\bn)
\eea

The action is again $\cQ$-exact on the lattice: $S= \cQ \Lambda$, where
\beq
\Lambda = \sum_{\bn} \Tr \Big(\chi_{ab}(\bn)\cD_a^{(+)}\cU_b(\bn) + \eta(\bn) \cDb_a^{(-)}\cU_a(\bn) - \frac{1}{2}\eta(\bn) d(\bn) \Big)~.
\eeq
Acting with the $\cQ$ transformation shown above and again integrating out the auxiliary field $d$, we derive the gauge and $\cQ$-invariant lattice action:
\bea
S &=& \sum_{\bn} \Tr \Big(\cF_{ab}^{\dagger}(\bn) \cF_{ab}(\bn) + \frac{1}{2}\Big(\cDb_a^{(-)}\cU_a(\bn)\Big)^2 \nn \\
&&- \chi_{ab}(\bn) \cD^{(+)}_{[a}\psi_{b]}(\bn) - \eta(\bn) \cDb^{(-)}_a\psi_a(\bn) \Big)~.
\eea

It is interesting to see that each term in the action forms a closed loop on the two-dimensional lattice. This is a requirement for preserving the gauge symmetry on the lattice.

\subsection{Four-dimensional lattice $\cN=4$ SYM theory}
\label{sec:latt-N4d4}

The way we discretized the two-dimensional $\cN=2$ theory on a two-dimensional square lattice immediately motivates us to choose the discretization of the four-dimensional $\cN=4$ theory on a four-dimensional hypercubic lattice. The fermions of the four-dimensional theory live on $p$-cells of the hypercubic lattice unit cell, associating themselves with the $p$-form representation of the continuum $SO(4)$ symmetry. The fermionic content of the hypercubic lattice construction manifest themselves as an explicit realization of Dirac--K\"ahler fermions. The bosons are also distributed on this lattice in orientations consistent with those of the fermions. The symmetry of the hypercubic lattice action is $S_4$, much smaller than the symmetry of the hypercube itself, due to the orientation assignment of the fields.  

The gauge link fields $\cU_a(\bn)$, $a=1, 2, 3, 4$, live on elementary coordinate directions in the unit cell of the hypercube pointing in the direction $(\bn, \bn+\hatbmu_a)$. The superpartners of the gauge link fields, $\psi_a(\bn)$, also live on the same links and oriented identically. The field $\cUb_a(\bn)$ is oriented in the opposite direction $(\bn + \hatbmu_a, \bn)$. The complexified field strength $\cF_{ab}(\bn)$ runs along the direction $(\bn, \bn + \hatbmu_a + \hatbmu_b)$. By exact supersymmetry, this implies that the field $\chi_{ab}(\bn)$ (and thus $\cFb_{ab}(\bn)$) runs in the opposite direction.

The assignment of $\cU_5(\bn)$ (and, thus, that of $\psi_5(\bn)$) is not immediately obvious. The Dirac--K\"ahler decomposition demands a 4-form. This motivates assigning the lattice field to the body diagonal of the unit hypercube, which is a 4-cell. It is oriented along the vector $\hatbmu_5=(-1, -1, -1, -1)$. We see that this assignment ensures that $\hatbmu_1 + \hatbmu_2 + \cdots + \hatbmu_5 = 0$, and it is crucial for constructing gauge-invariant quantities on the lattice.

The basis vectors $\hatbmu_a$ of the hypercubic lattice are thus defined as\footnote{These vectors are related to the $\br$-charges defined in the orbifold formulation of the four-dimensional $\cN=4$ lattice SYM theory\cite{Kaplan:2005ta}.}
\bea
\label{eq:mu-vectores}
\hatbmu_1 &=& (1, 0, 0, 0)\nn \\
\hatbmu_2 &=& (0, 1, 0, 0)\nn \\
\hatbmu_3 &=& (0, 0, 1, 0) \nn \\
\hatbmu_4 &=& (0, 0, 0, 1)\nn \\
\hatbmu_5 &=& (-1, -1, -1, -1)
\eea

Though the four-dimensional fields come with five indices they are all taken care of with suitable orientation assignments consistent with the lattice gauge symmetry. 

On the hypercubic lattice the action of the four-dimensional theory takes the following form
\bea
\label{eq:action-on-a4-star}
S &=& \sum_{\bn, a,b,c,d,e} ~\Big\{ \cQ ~\Tr \Big[\chi_{ab} \cD^{(+)}_a \cU_b(\bn) - \eta(\bn) \Big(\cDb^{(-)}_a \cU_a(\bn) - \frac{1}{2}d(\bn) \Big)\Big]\nn \\
&& - \frac{1}{2} \Tr \epsilon_{abcde} \chi_{de}(\bn + \hatbmu_a + \hatbmu_b + \hatbmu_c) \cD^{\dagger(-)}_{c} \chi_{ab}(\bn + \hatbmu_c)\Big\}~.
\eea
where the lattice field strength is given by
\beq
\cF_{ab}(\bn) \equiv \cD^{(+)}_a \cU_b(\bn) = \Big(\cU_a(\bn) \cU_b(\bn + \hatbmu_a) - \cU_b(\bn) \cU_a(\bn + \hatbmu_b)\Big)~.
\eeq
and the covariant difference operators appearing in this expression are given by
\bea
\cD_c^{(+)} f(\bn) &=& \cU_c(\bn) f(\bn + \hatbmu_c) - f(\bn) \cU_c(\bn) \nn \\
\cD_c^{(+)} f_d(\bn) &=& \cU_c(\bn) f_d(\bn + \hatbmu_c) - f_d(\bn) \cU_c(\bn + \hatbmu_d) \nn \\
\cDb_c^{(-)} f_c(\bn) &=& f_c(\bn)\cUb_c(\bn) - \cUb_c(\bn - \hatbmu_c) f_c(\bn - \hatbmu_c) \nn \\
\cDb_c^{(-)} f_{ab} (\bn) &=& f_{ab}(\bn) \cUb_c(\bn - \hatbmu_c) - \cUb(\bn + \hatbmu_a + \hatbmu_b - \hatbmu_c) f_{ab}(\bn - \hatbmu_c)~~~~~~~~~
\eea

The supersymmetry transformations on the lattice fields are almost identical to their continuum counterparts:
\bea
\cQ \cU_a(\bn) &=& \psi_{a}(\bn) \nn \\
\cQ \psi_a(\bn)  &=& 0 \nn \\
\cQ \cUb_a(\bn)  &=& 0 \nn \\
\cQ \chi_{ab}(\bn)  &=& -\cFb_{ab}^L(\bn) \nn \\
\cQ \eta (\bn) &=& d \nn \\
\cQ d (\bn) &=& 0
\eea
After the $\cQ$-variation, as performed in the continuum, and integrating out the auxiliary field $d$, the final lattice action is:
\bea
\label{eq:final-action}
S &=& \sum_{\bn} \Tr \Big[\cF^{L\dagger}_{ab} \cF^L_{ab} + \frac{1}{2} \Big(\cDb^{(-)}_{a} \cU_{a}(\bn)\Big)^2 - \chi_{ab}(\bn) \cD^{(+)}_{[a} \psi_{b]}(\bn)- \eta(\bn) \cD^{\dagger(-)}_{a} \psi_{a}(\bn) \nn \\
&&- \frac{1}{2} \epsilon_{abcde} \chi_{de}(\bn + \hatbmu_a + \hatbmu_b + \hatbmu_c) \cDb^{(-)}_{c} \chi_{ab}(\bn + \hatbmu_c)\Big]~.~~~
\eea
To see that this action targets the continuum twisted theory one needs to expand $\cU_a$ about the unit matrix\cite{Kaplan:2005ta}\footnote{Though this is equivalent to the more conventional expression $\cU_a(\bn)=e^{\cA_a(\bn)}$ at the leading order, the linear representation offers important advantages over the exponential representation.}
\bea
\label{eq:unity-exp}
\cU_a (\bn) &=& {\mathbb I}_N + \cA_a(\bn)~, \nn \\
\label{eq:cu1}
\cUb_a (\bn) &=& {\mathbb I}_N - \cAb_a(\bn)~.
\eea
While the supersymmetric invariance of the $\cQ$-exact term is manifest in the lattice theory it is not immediately clear that the $\cQ$-closed term remains supersymmetric after discretization. Interestingly, this can be shown using a remarkable property of the discrete field strength, which can be shown to satisfy an exact Bianchi identity just as for the continuum\cite{Aratyn:1984bd}.
\beq
\epsilon_{abcde}\cDb^{(-)}_c\cFb_{ab}(\bn + \hatbmu_c)=0~.
\eeq

\subsubsection{The $A_4^*$ lattice construction}
There exists a more symmetric lattice than the hypercubic lattice for the four-dimensional $\cN=4$ theory. This lattice is called the $A_4^*$ lattice. On this lattice, we treat all five basis vectors equally and they are oriented in such a way that the basis vectors connect the center of a 4-simplex to its corners. Having a most symmetric lattice is advantageous because the greater the symmetry is, the fewer relevant or marginal operators will exist on the lattice.  

The lattice possesses an $S^5$ point group symmetry, which is the Weyl group of $SU(5)$. We briefly described the $SU(5) \times U(1)$ decomposition of the fields of the four-dimensional $\cN=4$ SYM theory in Sec. \ref{subsec:susy-trans-tw-ac}. The discretization prescription for such a decomposition of the fields would be the $A_4^*$ lattice. A specific basis for the $A_4^*$ lattice is given in the form of five lattice vectors:
\bea
\hatbe_1 &=&  \Big(\frac{1}{\sqrt{2}}, \frac{1}{\sqrt{6}}, \frac{1}{\sqrt{12}}, \frac{1}{\sqrt{20}}\Big) \nn \\
\hatbe_2 &=& \Big(-\frac{1}{\sqrt{2}}, \frac{1}{\sqrt{6}}, \frac{1}{\sqrt{12}}, \frac{1}{\sqrt{20}}\Big) \nn \\
\hatbe_3 &=& \Big(0, -\frac{2}{\sqrt{6}}, \frac{1}{\sqrt{12}}, \frac{1}{\sqrt{20}}\Big) \nn \\
\hatbe_4 &=& \Big(0, 0, -\frac{3}{\sqrt{12}}, \frac{1}{\sqrt{20}}\Big) \nn \\
\hatbe_5 &=& \Big(0, 0, 0, -\frac{4}{\sqrt{20}}\Big)
\eea
These lattice vectors connect the center of a $4$-simplex to its five corners. They are related to the $SU(5)$ weights of the $5$ representation. The unit cell of the $A_4^*$ lattice is a compound of two $4$-simplices corresponding to the $5$ (formed by the basis vectors $\hatbe_m$) and $\overline{5}$ (formed by the basis vectors $-\hatbe_m$) representations of $SU(5)$. The basis vectors satisfy the relations 
\bea
&&\sum_{m=1}^{5} \hatbe_m = 0;~~\hatbe_m \cdot \hatbe_n = \Big(\delta_{mn} - \frac{1}{5}\Big)\nn \\
&&\sum_{m=1}^{5}(\hatbe_m)_{\mu}(\hatbe_m)_{\nu} = \delta_{\mu \nu}; ~~\mu, \nu = 1, \cdots, 4. \nn
\eea
Notice also that $S^5$ is a subgroup of the twisted rotation symmetry group $SO(4)^\prime$ and that the lattice fields transform in reducible representations of this discrete group - for example, the vector $\cA_a$ decomposes into a four component vector $\cA_\mu$ and a scalar field $\phi$ under $SO(4)^\prime$. Invariance of the lattice theory with respect to these discrete rotations then guarantees that the theory will inherit full invariance under twisted rotations in the continuum limit.

Proceeding in this manner, it is possible to assign all the remaining fields to links on the $A_4^*$ lattice. Since $\psi_a(\bn)$ is a superpartner of $\cU_a(\bn)$ it must also reside on the link connecting $\bn \to \bn+\hatbe_a$. Conversely the field $\cU^{\dagger}_a(\bn)$ resides on the oppositely oriented link from $\bn \to \bn-\hatbe_a$. The ten fermions $\chi_{ab}(\bn)$ are then chosen to reside on new fermionic links $\bn+\hatbe_m+\hatbe_n \to \bn$, while the singlet fermionic field $\eta(\bn)$ is assigned to the degenerate link consisting of a single site $\bn$.

The integer-valued lattice site $\bn$ can be related to the physical location in spacetime using the $A_4^*$ basis vectors $\hatbe_a$. 
\beq
\bR = a \sum_{\nu =1}^4 (\mu_{\nu} \cdot \bn)\hatbe_{\nu} = a \sum_{\nu =1}^{4}n_{\nu}\hatbe_{\nu}~,
\eeq
where $a$ is the lattice spacing. On using the fact that $\sum_{m}\hatbe_m = 0$, we can show that a small lattice displacement of the form $d\bn = \hatbmu_m$ corresponds to a spacetime translation by $(a\hatbe_m)$:
\beq
d\bR = a \sum_{\nu =1}^{4}(\mu_{\nu} \cdot d \bn)\hatbe_{\nu} = a \sum_{\nu =1}^{4} (\hatbmu_{\nu} \cdot \hatbmu_m)\hatbe_{\nu} = a \hatbe_m~.
\eeq

In the next section, we will use the $A_4^*$ lattice construction to study the one-loop renormalization of the four-dimensional $\cN=4$ SYM theory.

\section{Lattice $\cN = 4$ SYM theory at one-loop}

It would be very advantageous to have a lattice formulation of the $\cN = 4$ SYM theory as it would provide a non-perturbative definition of this theory and open up a new window to explore its strong coupling dynamics through lattice simulations. Such a lattice construction would also allow for a systematic study of its dual string theory. In the last section, the lattice version of this theory is derived and showed that the lattice theory on an $A_4^*$ lattice exhibits much more symmetry compared to the one on a four-dimensional hypercubic lattice. 

The $\cN=4$ SYM theory in the continuum has sixteen supersymmetric invariances but the lattice cousine of this theory retains only one supersymmetry exactly on the lattice. This leaves room for the question of how much fine tuning would be required to take the continuum limit of this lattice theory targeting the usual $\cN = 4$ theory. This issue is addressed in this section using both general arguments valid to all orders in perturbation theory and an explicit calculation of the renormalization of the lattice $\cN = 4$ theory to one-loop order. A more detailed discussion can be found in Ref. \refcite{Catterall:2011pd}. 

\subsection{General renormalization structure as revealed by symmetries}
\label{sec:general}

The four-dimensional $\cN=4$ SYM theory on an $A_4^{*}$ lattice exhibits the following set of symmetries\cite{Catterall:2009it}
\begin{itemize}
\item[$i.$] The exact supersymmetry corresponding to the scalar supercharge $\cQ$. 
\item [$ii.$] Lattice gauge symmetry.
\item [$iii.$] The $S_5$ point group symmetry and discrete translations on the lattice.
\end{itemize}
If the gauge group is $U(N)$, the lattice theory possesses an additional fermionic symmetry:
\beq
\eta(\bn) \rightarrow \eta (\bn) + \epsilon {\mathbb I}_N, \qquad  \delta_{\epsilon} (\rm all \; other \; fields) =0~, 
\eeq
where $\epsilon$ is an infinitesimal Grassmann parameter. Thus the set of symmetries contains one more item: 
\begin{itemize}
\item[$iv.$] Fermionic shift symmetry.
\end{itemize}

Following the conventions given in Ref. \refcite{Catterall:2011pd}, we use hermitian basis for the generators of the gauge group satisfying $\Tr(T^AT^B)=\frac{1}{2}\delta^{AB}$ and also explicitly indicate the dependence on the gauge coupling $g$. There are three types of covariant derivatives and field strengths in the continuum theory due to the presence of the complexified connections:
\bea
D_a ~\cdot \equiv \partial_a + ig[A_a, ~\cdot~ ],~~~ F_{ab} \equiv -\frac{i}{g}[D_a, D_b]~, \nn \\
\cD_a ~\cdot \equiv \partial_a + ig[\cA_a, ~\cdot~ ],~~~ \cF_{ab} \equiv -\frac{i}{g}[\cD_a, \cD_b]~, \nn \\
\cDb_a ~\cdot \equiv \partial_a +ig[\cAb_a, ~\cdot~ ],~~~ \cFb_{ab} \equiv -\frac{i}{g}[\cDb_a, \cDb_b]~.
\eea
The continuum action is:
\bea
S &=& \int\Tr \Big( \cFb_{ab} \cF_{ab} + \frac{1}{2g^2} [\cDb_a, \cD_a]^2 - \chi_{ab} \cD_{[a}\psi_{b]} - \eta\cDb_a \psi_a\nn \\
&&- \frac{1}{2}\epsilon_{abcde} \chi_{de} \cDb_{c} \chi_{ab}\Big).~~~~~~~~
\eea
Rescaling of the fields $g\eta \rightarrow \eta$, $g\psi_a \rightarrow \psi_a$, $g\chi_{ab} \rightarrow \chi_{ab}$ and $g\cA_a \rightarrow \cA_a$ provides the extraction of the coupling parameter dependence from the terms in the action:
\bea
S &=& \frac{1}{g^2}\int\Tr \Big( -[\cDb_a, \cDb_b][\cD_a, \cD_b] + \frac{1}{2} [\cDb_a, \cD_a]^2 - \chi_{ab} \cD_{[a}\psi_{b]} - \eta\cDb_a \psi_a \nn \\
&&~~~~~~~~~~~- \frac{1}{2}\epsilon_{abcde} \chi_{de} \cDb_{c} \chi_{ab}\Big).~~~
\label{eq:compact-five-d-action}
\eea

The lattice action take the following $\cQ$-exact form:
\bea
\label{eq:action-on-a4-star1}
S &=& \frac{1}{g^2}\sum_{\bn} ~\Big\{ \cQ ~\Tr \Big[-i\chi_{ab} \cD^{(+)}_a \cU_b(\bn) - \eta(\bn) \Big(i\cD^{\dagger(-)}_a \cU_a(\bn) - \frac{1}{2}d(\bn) \Big)\Big]\nn \\
&& - \frac{1}{2} \Tr \epsilon_{abcde} \chi_{de}(\bn + \hatbmu_a + \hatbmu_b + \hatbmu_c) \cD^{\dagger(-)}_{c} \chi_{ab}(\bn + \hatbmu_c)\Big\}~,
\eea
where the lattice field strength is given by:
\bea
\cF_{ab}(\bn) &\equiv& -\frac{i}{g}\cD^{(+)}_a \cU_b(\bn) \nn \\
&=& -\frac{i}{g}\Big(\cU_a(\bn) \cU_b(\bn + \hatbmu_a) - \cU_b(\bn) \cU_a(\bn + \hatbmu_b)\Big).
\eea

We would be interested to see what the above mentioned set of symmetries of the lattice action tell us about the existence of any relevant or marginal operators - operators whose mass dimension is less than or equal to four - on the lattice. The $\cQ$-invariance of the action restricts the possible counterterms to be either of a $\cQ$-exact form, or of a $\cQ$-closed form. The only one $\cQ$-closed operator permitted by the lattice symmetries is the one already present in the bare lattice action as a fermion kinetic term. Thus a possible renormalization of this term is allowed. There may be a set of $\cQ$-exact counterterms appearing on the lattice.  Any $\cQ$-exact counterterm we add to the action, which respects the lattice symmetries, must be of the form 
\bec
$\mathcal{O}=\cQ\Tr(\Psi f(\cU,\cU^{\dagger}))$~.
\eec
There are, thus, no terms permitted by symmetries with a dimension less than two. Lattice gauge symmetry tells us that each term must correspond to the trace of a closed loop on the lattice. The smallest dimension gauge-invariant operator is then just $\cQ(\Tr \psi_a \cU^{\dagger}_a)$. But this vanishes identically, since both $\cU^{\dagger}_a$ and $\psi_a$ are singlets under $\cQ$. This structure also forbids dimension $\frac{7}{2}$ operators, and so we are left with just dimension four counterterms. Notice, in particular, that lattice symmetries permit no simple fermion bi-linear mass terms. However, gauge-invariant fermion bi-linears with link field insertions are possible, and their effect should be accounted for carefully. Thus we can write down in a schematic way the set of possible dimension four $\cQ$-exact operators
\begin{eqnarray}
L_1&=& g^{-2}\cQ \Tr (\chi_{ab}\cU_a \cU_b)  \nn \\
L_2&=&  g^{-2}\cQ \Tr (\eta \cD^{\dagger}_a \cU_a ) \nn \\
L_3&=&  g^{-2}\cQ \Tr (\eta \cU_a \cU^{\dagger}_a) \nn \\
L_4&=&  g^{-2}\cQ \Tr (\eta)\Tr(\cU_a \cU^{\dagger}_a)
\label{eq:ops}
\end{eqnarray}
The first operator can be simplified to the form $\cQ(\chi_{ab}\cF_{ab})$, by making use of the antisymmetry of $\chi_{ab}$, and it is nothing but one of the continuum $\cQ$-exact terms present in the bare action. The second operator also corresponds to one of the $\cQ$-exact terms in the bare action. However, the third operator $L_3$ and the final double-trace operator $L_4$ are not present in the bare Lagrangian. They both transform non-trivially under the fermionic shift symmetry. However, the linear combination  of the two 
\beq 
D =  L_3 - \frac{1}{N} L_4~,  
\label{eq:ren4}
\eeq
with $N$ the rank of the gauge group $U(N)$, turns out to be invariant under the shift symmetry.

Thus the relevant counterterms correspond to renormalizations of operators that are already present in the bare action together with the new operator $D$. The most general form for the renormalized lattice Lagrangian should be of the form: 
\bea
\label{eq:general-form}
{\cal L} &=& \sum_{\bn,a,b,c,d,e} ~\Big\{ \cQ ~\Tr \Big[-i\alpha_1 \chi_{ab} \cD^{(+)}_a \cU_b(\bn) -i \alpha_2 \eta(\bn) \cD^{\dagger(-)}_a \cU_a(\bn) \nn \\
&&+\frac{\alpha_3}{2}\eta(\bn)d(\bn)\Big] - \frac{\alpha_4}{2} \Tr \epsilon_{abcde} \chi_{de}(\bn + \hatbmu_a + \hatbmu_b + \hatbmu_c) \cD^{\dagger(-)}_{c} \chi_{ab}(\bn + \hatbmu_c)\Big\}\nn \\
&&+ \cQ \beta D~,
\eea
where $(\alpha_i,i=1\ldots 4)$ and $\beta$ are dimensionless numbers taking values $(1,1,1,1)$ and $0$ respectively in the classical lattice theory. Thus, it appears that, at most, four dimensionless ratios of these couplings might need to be tuned to approach $\cN=4$ Yang--Mills in the continuum limit. Furthermore, since these operators are dimension four, we expect this tuning to be, at worst, logarithmic in the cut-off.

We could expand the action around  $\cU_m(\bn)= \frac{1}{a}{\mathbb I}_N$ in order to see the explicit form of the $D$ operator close to the continuum limit.
\beq 
D \sim \frac{1}{a} \Big[ \Tr \eta(\bn)(\sum_{m=1}^5 \psi^m(\bn)) - \frac{1}{N} \Tr \eta(\bn) \Tr (\sum_{m=1}^5 \psi^m(\bn) ) \Big] + \ldots
\eeq
where ellipsis are dictated by supersymmetry. We see that the $S_5$ (and twisted $SO(4)'$) singlet field $(\sum_{a=1}^5 \psi_a)$, which is contained in the reducible representation $\psi_a$, is the only field that could form a fermion mass term by pairing with $\eta$. 

It turns out that the exploration using general lattice symmetry arguments reveals a lot about the possible counter terms in the lattice theory. To determine how the couplings $(\alpha_i, \beta)$ evolve with the cut-off a full perturbative analysis is required. 

\subsection{Lattice propagators and vertices}
\label{sec:prop-vert}

The classical lattice action (\ref{eq:action-on-a4-star1}) is a combination of three parts - bosonic ($S_B$), fermionic ($S_F$) and $\cQ$-closed terms ($S_c$). After rewriting the field strength and covariant derivatives in terms of the bosonic link fields $\cU_a(\bn)$ they take the following form:
\bea
S_B &=& \frac{1}{g^2}\sum_{\bn,a,b} \Tr \Big[\Big(\cD_a^{(+)}\cU_b(\bn)\Big)^{\dagger}\Big(\cD^{(+)}_a \cU_b(\bn)\Big) + \frac{1}{2} \Big(\cD^{\dagger(-)}_{a} \cU_{a}(\bn)\Big)^2\Big]\nn \\
&=& \frac{1}{g^2}\sum_{\bn,a,b} \Tr \Big[\Big(\cU^{\dagger}_b(\bn + \hatbmu_a)\cU^{\dagger}_a(\bn) - \cU^{\dagger}_a(\bn + \hatbmu_b)\cU^{\dagger}_b(\bn)\Big)\nn \\
&& \times \Big(\cU_{a}(\bn)\cU_{b}(\bn + \hatbmu_a) - \cU_b(\bn)\cU_a(\bn + \hatbmu_b)\Big)\nn \\
&&+ \frac{1}{2} \Big(\cU_a(\bn)\cU^{\dagger}_a(\bn) - \cU^{\dagger}_a(\bn - \hatbmu_a)\cU_a(\bn - \hatbmu_a)\Big)^2\Big]~,
\eea
\bea
\label{eq:final-action-U}
S_F &=& -\frac{1}{g^2}\sum_{\bn,a,b,c,d} \Tr \frac{1}{2}(\delta_{ac}\delta_{bd} - \delta_{ad}\delta_{bc}) \Big[\chi_{ab}(\bn)\Big(\cU_c(\bn)\psi_d(\bn + \hatbmu_c) \nn \\
&&- \psi_d(\bn)\cU_c(\bn + \hatbmu_d)\Big)\Big] + \eta(\bn)\Big(\psi_a(\bn)\cU^{\dagger}_a(\bn) \nn \\
&&- \cU^{\dagger}_a(\bn - \hatbmu_a)\psi_a(\bn - \hatbmu_a)\Big)~,
\eea
and
\bea
S_c &=& -\frac{1}{2g^2}\sum_{\bn,a,b,c,d,e} \Tr \epsilon_{abcde}\Big(\chi_{de}(\bn + \hatbmu_a + \hatbmu_b + \hatbmu_c)\nn \\
&&~~~~~~\times \Big[\chi_{ab}(\bn + \hatbmu_c)\cU^{\dagger}_c(\bn) - \cU^{\dagger}_c(\bn + \hatbmu_a + \hatbmu_b)\chi_{ab}(\bn)\Big]\Big)~.~~~
\eea

To proceed further, we expand the $\cU_a(\bn)$ fields around unity
\bea
\label{eq:u}
\cU_a (\bn) &=& \frac{1}{a}{\mathbb I}_N + i\cA_a(\bn)~, \\
\label{eq:cu2}
\cU^{\dagger}_a (\bn) &=& \frac{1}{a}{\mathbb I}_N - i\cAb_a(\bn)~.
\eea
This expansion point is but one of an infinite number of classical vacuum solutions - the full moduli space of the lattice theory corresponds to the set of all bosonic field variables $\cU_a(\bn)$ such that 
\bea
\label{eq:moduli-space-lattice}
0 &=& \sum_{\bn,a,b} \Tr \Big[\Big(\cU^{\dagger}_b(\bn + \hatbmu_a)\cU^{\dagger}_a(\bn) - \cU^{\dagger}_a(\bn + \hatbmu_b)\cU^{\dagger}_b(\bn)\Big)\nn \\&&\times \Big(\cU_{a}(\bn)\cU_{b}(\bn + \hatbmu_a) - \cU_b(\bn)\cU_a(\bn + \hatbmu_b)\Big)\nn \\
&&+ \frac{1}{2} \Big(\cU_a(\bn)\cU^{\dagger}_a(\bn) - \cU^{\dagger}_a(\bn - \hatbmu_a)\cU_a(\bn - \hatbmu_a)\Big)^2\Big]~.
\eea
These equations possess a large class of solutions corresponding to constant diagonal matrices modulo gauge transformations. This additional freedom can be used to compute the one-loop contribution to the effective action of the theory.

\subsubsection{The bosonic propagators}
\label{sec:b-prop}

The derivation of propagators and vertices of the lattice theory becomes easier if we switch to the momentum space. Consider a generic field $\Phi(\vx)$. On the $A_4^*$ lattice it has the Fourier expansion
\beq
\Phi(\vx) = \frac{1}{(La)^4} \sum_{\vp}e^{i \vp \cdot \vx} \Phi_{\vp}~,
\label{ft}
\eeq
where $\vx=a\sum_{a=1}^4 n_a \hatbe_a$ denotes the position on $A_4^*$ lattice\footnote{For simplicity we will adopt the convention that momentum sums $\sum_k$ automatically include the $1/(La)^4$ normalization factor.}. The momenta lie on the dual lattice and are given by $\vp=\frac{2\pi}{La}\sum_{a=1}^4 m_a \hatbg_a$ (for a lattice with spacing $a$ and length $L$). The basis vectors of the dual lattice $\hatbg_a,a=1\ldots 4$ satisfy the relation
\beq
\hatbe_a \cdot \hatbg_b=\delta_{ab}~.
\eeq
Both sets of lattice coordinates $n_a$, $m_a$ take integer values in the range $-L/2+1,\ldots,L/2$ on an $L^4$ lattice. We take the boundary conditions to be periodic in all directions while deriving propagators. On looking at (\ref{ft}) we see that the fields are invariant under translations by a lattice length in any direction. A field shifted by one of the basis vectors can be expressed as
\beq
\Phi(\vx+\hatbe_a)=\sum_{\vp}e^{ip_a}e^{i \vp \cdot \vx} \Phi_{\vp}~,
\eeq
where $p_a=\frac{2\pi}{L}m_a$. Considering the fact that the $A_4^*$ lattice has five basis vectors, we should know how to deal with shifts in the lattice action associated with the additional $\hatbe_5$ vector. It turns out that we can simply replace any $\hatbe_5$ shift encountered in the action by the equivalent shift $-\sum_{a=1}^4 \hatbe_a$ since $\sum_{a=1}^5\hatbe_a=0$. Also it seems like there is an apparent lack of rotational invariance associated with the naive continuum limit of terms in the action which resemble $\sum_{a=1}^5 \sin^2{p_a}$. If we decompose $p_a = \vp \cdot \hatbe_a$ and take the naive continuum limit they become 
\beq
\sum_{a=1}^5 \sin^2{p_a} \rightarrow \sum_{a=1}^5 p_a^2=
\sum_{\mu,\nu}^4\sum_{a=1}^5 p_\mu p_\nu \hatbe^\mu_a \hatbe^\nu_a=
\sum_{\mu}^4 p_\mu^2~,
\eeq
with the correct rotationally invariant form.
 
\begin{figure}[pb]
\centerline{\psfig{file=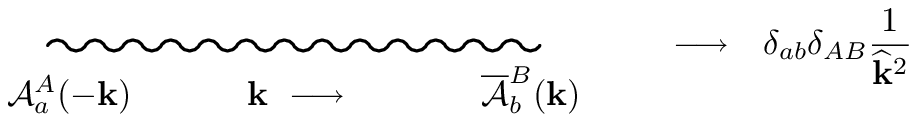,width=9cm}}
\vspace*{8pt}
\caption{\label{bosonic}The bosonic propagators on the lattice.}
\end{figure}

Thus, the bosonic action when expanded around (\ref{eq:u}) and (\ref{eq:cu2}) gives the following second-order term in Fourier space
\bea
S^{(2)}_{B} &\approx& 2\sum_{\vk,a,b} \Tr \Big(\cAb_a(\vk)\Big[\delta_{ab}f_c(\vk)f^*_c(\vk) - f^*_a(\vk)f_b(\vk) \Big]\cA_b(-\vk) \nn \\
&&+ B_a(\vk)~\Big[f^*_a(\vk)f_b(\vk)\Big]~B_b(-\vk)\Big)~,
\eea
where
\beq
f_a(\vk) = (e^{i k_a}-1)~.
\eeq
We need to gauge-fix the bosonic action before deriving the propagators. A natural gauge-fixing choice would be an obvious generalization of Lorentz gauge-fixing\cite{Marcus:1995mq}
\beq
G(\bn) = \sum_{a} \Big(\partial_a^{(-)}\cA_a(\bn) + \partial_a^{(-)}\cAb_a(\bn)\Big)~.
\label{gfixf}
\eeq
This gauge-fixing choice adds the following term to the bosonic action at quadratic order
\beq
S_{GF}= \frac{1}{4 \alpha} \sum_{\bn} G^2(\bn)
= \frac{1}{\alpha} \sum_{\bn, a} \Tr (\partial^{(-)}_a A_a(\bn))^2~,
\eeq
where $\partial^{(-)}_a f(\bn) = f(\bn) - f(\bn - \hatbmu_a)$. On using the relation 
\beq
\sum_{\bn}(\partial^{(+)}_a f(\bn))g(\bn) = -\sum_{\bn}f(\bn)\partial^{(-)}_a g(\bn)~, \nn
\eeq
the gauge-fixing term becomes
\beq
S_{GF} = -\frac{1}{\alpha} \sum_{\bn, a,b} \Tr A_a(\bn)\partial^{(+)}_{a}\partial^{(-)}_bA_b(\bn)~.
\eeq
In momentum space it becomes
\bea
S_{GF} &=&  \frac{1}{\alpha} \sum_{\vk, a, b} \Tr A_a(\vk) f^*_a(\vk)f_b(\vk) A_b(-\vk)~.
\eea
Thus the gauge-fixed bosonic action to quadratic order is
\bea
S^{(2)}_{B} + S_{GF}&\approx& 2 \sum_{\vk, a, b, c} \Tr
\Big(A_a(\vk)~\Big[\delta_{ab} f_c(\vk)f^*_c(\vk) - \Big(1 -
  \frac{1}{2\alpha}\Big) f^*_a(\vk)f_b(\vk)\Big]~A_b(-\vk) \nn \\
&&+ B_a(\vk)~\Big[\delta_{ab} f_c(\vk)f^*_c(\vk)\Big]~B_b(-\vk)\Big)~.
\eea
The choice $\alpha = 1/2$ makes the above expression diagonal
\bea
S^{(2)}_{B} &\approx& 2 \sum_{\vk, a, b, c} \Tr \cAb_a(\vk)~[\delta_{ab} f_c(\vk)f^*_c(\vk)]~\cA_b(-\vk)\nn \\
&=& 2 \sum_{\vk, a, b} \Tr \Big[\cAb_a(\vk)\delta_{ab} \Big( 4 \sum_c \sin^2\Big(\frac{k_c}{2}\Big)\Big)\cA_b(-\vk)\Big]~.
\eea
Putting in the trace (using the convention $\Tr (T^A T^B) = \frac{1}{2}\delta_{AB}$) the quadratic bosonic action can be written as
\beq
S^{(2)}_{B} \approx \sum_{\vk, a, b}\cAb^A_a(\vk)M^{AB}_{ab}(\vk)\cA^B_b(-\vk)~,
\eeq
where $M^{AB}_{ab}(\vk) = \widehat{\vk}^2 \delta_{ab}\delta_{AB}$, with $\widehat{\vk}^2 = 4\sum_c \sin^2\Big(\frac{k_c}{2}\Big)$.
Thus only the $\cA \cAb$ propagator is non-zero and it is given by (See figure~\ref{bosonic}.)
\beq
\langle \cA_a^{A}(-\vk) \cAb_b^B(\vk)\rangle = \delta_{ab}\delta_{AB} \frac{1}{\widehat{\vk}^2}~.
\eeq 

\subsubsection{The fermionic propagators}
The fermionic part of the action, which has the following form on the lattice
\bea
S_F &=& -\frac{1}{g^2}\sum_{\bn, a, b, c, d, e} \Tr \Big( \chi_{ab}(\bn) \cD^{(+)}_{[a}\psi_{b]}(\bn) + \eta(\bn) \cD^{\dagger(-)}_{a} \psi_{a}(\bn) \nn \\
&&+ \frac{1}{2} \epsilon_{abcde}\chi_{de}(\bn+\hatbmu_a+\hatbmu_b+\hatbmu_c)\cD^{\dagger(-)}_{c}\chi_{ab}(\bn+\hatbmu_c) \Big)~,
\eea
when expanded up to second order in the fields using (\ref{eq:u}) and (\ref{eq:cu2}) takes the form
\bea
S^{(2)}_F &\approx& \frac{1}{g^2}\sum_{\vk,a,b,c,d,e} \Tr \chi_{ab}(\vk)\Big[-f^*_a(\vk)\delta_{bc} + f^*_b(\vk)\delta_{ac}\Big]\psi_c(-\vk) + \eta(\vk)f_c(\vk)\psi_c(-\vk) \nn \\
&&+ \frac{1}{2} \epsilon_{abcde}\chi_{de}(\vk)e^{i(k_a+k_b)}f_c(\vk)\chi_{ab}(-\vk)~.
\eea
Upon restricting the sum and rescaling the field $2\chi_{ab} \rightarrow \chi_{ab}$ the fermionic action becomes
\bea
S^{(2)}_F &\approx& \frac{1}{g^2} \sum_{\vk,a<b;c,d<e} \Tr \Big(\chi_{ab}(\vk)\Big[-f^*_a(\vk)\delta_{bc} + f^*_b(\vk)\delta_{ac}\Big]\psi_c(-\vk) + \eta(\vk)f_c(\vk)\psi_c(-\vk) \nn \\
&&+\frac{1}{2}
\epsilon_{abcde}\chi_{de}(\vk)e^{i(k_a+k_b)}f_c(\vk)\chi_{ab}(-\vk)\Big)~.
\eea
It is convenient to write this in the form of a matrix product
\bea
S^{(2)}_F &\approx& \frac{1}{g^2} \sum_{\vk} \left(
\Psi (\vk) \Psi(-\vk)
\right) \left(\frac{1}{4}\right)
\left(
\begin{tabular}{cc}
0 & $M(\vk)$ \\
$-M^T(\vk)$  & 0  \\
\end{tabular}
\right)
\left(
\begin{tabular}{c}
$\Psi(\vk)$ \\
$\Psi(-\vk)$ 
\end{tabular}
\right)
\nn \\ &=&
\frac{1}{4g^2}\sum_{\vk} \Phi(\vk) \mathcal{M} \Phi(\vk)~.
\label{matprod}
\eea
where $\Phi\equiv(\Psi(\vk),\Psi(-\vk))$ and $\Psi_i = (\eta, \psi_1,\dots,\psi_5,\chi_{12},\dots,\chi_{15},\dots,\chi_{45})$. The matrix $M(\vk)$ takes the following block matrix form
\begin{small}
\beq
\left(
\eta \ \psi_{a} \ \chi_{de}
\right)(\vk)
\left(
\begin{tabular}{ccc}
0 & $f_b(\vk)$ & 0 \\
$-f^*_a(\vk)$  & 0 & $ f_g(\vk)\delta_{ha}-f_h(\vk)\delta_{ga}$ \\
0 & $-f^*_d(\vk)\delta_{eb} + f^*_e(\vk)\delta_{db}$ & 
$\epsilon_{ghcde} q_{gh} f_c(\vk)$
\end{tabular}
\right)
\left(
\begin{tabular}{c}
$\eta$ \\
$\psi_{b}$ \\
$\chi_{gh}$
\end{tabular}
\right)(-\vk).\nn
\eeq
\end{small}
where $q_{gh} = e^{i(k_g + k_h)}$. The matrix $M$ also has the properties $M^T(\vk) = -M^*(\vk) = -M(-\vk)$~. 

Upon using the property that $\sum_a \hatbmu_a = 0$, we can square the matrix $M$ to obtain
\beq
M^2(\vk) = -\sum_{a = 1}^5 |e^{ik_a} - 1|^2 {\mathbb I}_{16}  = -4\sum_{a = 1}^5 \sin^2\Big(\frac{k_a}{2}\Big) {\mathbb I}_{16}  = -\widehat{\vk}^2 {\mathbb I}_{16}~.
\eeq
This implies,
\beq
M^{-1} = -\frac{1}{\widehat{\vk}^2}M~,
\eeq
and the inverse of the full fermion matrix is:
\bea
\mathcal{M}^{-1}=-\frac{1}{\widehat{\vk}^2}  
\left(
\begin{tabular}{cc}
0 & $-M^T(\vk)$ \\
$M(\vk)$  & 0  \\
\end{tabular}
\right)~.
\eea
We write the quadratic part of the fermionic action as:
\bea
S_F^{(2)} &=& \frac{1}{4g^2}\sum_{\vk}\Tr\left[ \sum_{ij}\Phi_i(\vk)\mathcal{M}_{ij}(\vk)\Phi_j(\vk) \right] \nn \\ 
&=&\frac{1}{4g^2}\sum_{\vk} \sum_{ij,A,B} \Phi_i^A(\vk)\mathcal{M}_{ij}(\vk)\Phi_j^B(\vk) \Tr(T^A T^B) \nn \\
&=& \frac{1}{8g^2}\sum_{\vk} \sum_{ij,A,B} \Phi_i^A(\vk)\mathcal{M}_{ij}(\vk)\Phi_j^B(\vk) \delta_{AB}~,
\eea
where the fermions are expanded as $\Phi = \Phi^AT^A$ and the property $\Tr(T^AT^B) = \frac{1}{2}\delta_{AB}$ is used. Thus, we write the fermion propagators as: 
\beq
\langle \Phi_i^A(\vk)\Phi_j^B(\vk) \rangle = 2\mathcal{M}^{-1}_{ij}(\vk)\delta_{AB}~,
\eeq
or, alternatively,
\beq
\langle \Psi_i^A(\vk)\Psi_j^B(-\vk) \rangle = \frac{2}{\widehat{\vk}^2} M^{T}_{ij}(\vk)\delta_{AB}~.
\eeq
By switching the fields  and with some relabeling, we have
\bea
\langle \Psi_i^A(-\vk)\Psi_j^B(\vk) \rangle &=& -\langle \Psi_j^B(\vk)\Psi_i^A(-\vk) \rangle \nn \\
&=& -\frac{2}{ \widehat{\vk}^2}M^{T}_{ji}(\vk)\delta_{BA}= -\frac{2}{ \widehat{\vk}^2} M_{ij}(\vk)\delta_{AB}~.
\eea
Notice that if we replace $\vk$ with $-\vk$ we have
\beq
\langle \Psi_i^A(-\vk)\Psi_j^B(\vk) \rangle = \frac{2}{\widehat{\vk}^2} M_{ij}^{T}(-\vk)\delta_{AB}= -\frac{2}{ \widehat{\vk}^2} M_{ij}(\vk)\delta_{AB}~.
\eeq
We must also undo the earlier rescaling of the $\chi$ field. This gives a factor of $\frac{1}{2}$ in the $\psi \chi$ propagators and a factor of $\frac{1}{4}$ in the $\chi \chi$ propagators. It should also be noted that switching the direction of fermion flow in the propagators would lead to an additional minus sign.

\begin{figure}[pb]
\centerline{\psfig{file=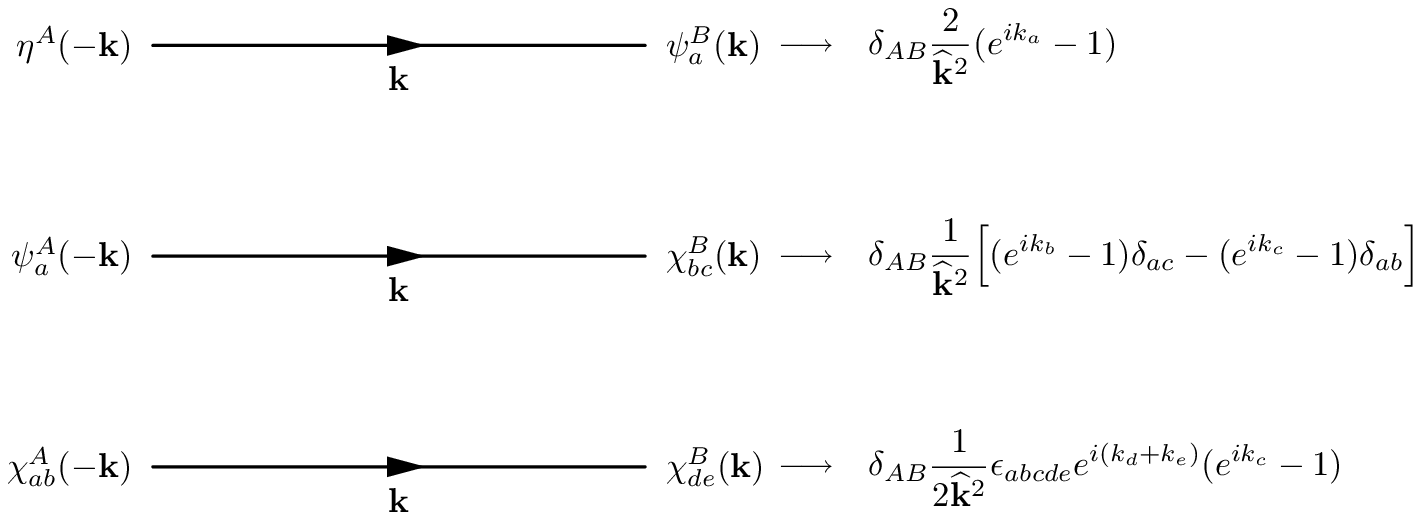,width=10cm}}
\vspace*{8pt}
\caption{\label{fig:propagators}The fermionic propagators on the lattice.}
\end{figure}

\subsubsection{The vertices on the lattice}
Since the expressions for vertices require additional trace contractions of the gauge group generators let us further fix the conventions on the trace algebra. We have for the generators $T^A$ of $U(N)$,
\beq
T^AT^B =  \frac{1}{2}(d_{ABC} + if_{ABC}) T^C~,
\eeq
where $d_{ABC}$ and $f_{ABC}$ are the symmetric and antisymmetric structure constants, respectively. This product formula is consistent with our previous trace convention, $\Tr(T^AT^B) = \frac{1}{2}\delta_{AB}$, and, in addition, it yields the results:
\bea
\Tr(T^AT^BT^C) &=& \Tr\left( \frac{1}{2}(d_{ABD} + if_{ABD}) T^DT^C\right) \\ \nn 
&=& \frac{1}{2}(d_{ABD} + if_{ABD})\Tr[T^DT^C] \\ \nn
&=& \frac{1}{2}(d_{ABD} + if_{ABD})\frac{1}{2}\delta_{DC} \\ \nn
&=& \frac{1}{4}(d_{ABC} + if_{ABC}) = \frac{1}{4}\lambda_{ABC}~.
\eea
We have
\beq
\lambda_{ACB} = \lambdabar_{ABC}~,
\eeq
since $f_{ABC}$ is antisymmetric and $d_{ABC}$ is symmetric.

To derive the expressions for the vertices, we come beck to the original gauge-fixed action for the theory given by
\bea
S &=& \frac{1}{g^2}\sum_{\bn} \Tr \Big[\Big(\cD_a^{(+)}\cU_b(\bn)\Big)^{\dagger}\Big(\cD^{(+)}_a \cU_b(\bn)\Big) + \frac{1}{2} \Big(\cD^{\dagger(-)}_{a} \cU_{a}(\bn)\Big)^2 \nn \\
&&~~~+ 2A_a(\bn)\partial^{(+)}_{a}\partial^{(-)}_bA_b(\bn)-
\Big(\chi_{ab}(\bn) \cD^{(+)}_{[a} \psi_{b]}(\bn) + \eta(\bn) \cD^{\dagger(-)}_{a} \psi_{a}(\bn)\nn \\
&&~~+\frac{1}{2}\epsilon_{abcde} \chi_{de}(\bn + \hatbmu_a + \hatbmu_b + \hatbmu_c) \cD^{\dagger(-)}_{c} \chi_{ab}(\bn + \hatbmu_c)\Big)\Big]~.
\eea
The last three terms of the action give rise to vertices between varying number of $\cA$'s and the fermions $\eta$, $\psi_a$, and $\chi_{ab}$. 

At linear order in $\cA$ there are three vertices. They are:
\bea
V_{\eta\cAb\psi} &=& -\frac{i}{4}\delta_{ab}[\lambda_{ABC} - \lambdabar_{ABC}e^{-i(p_a + q_a)}]~,\nn \\
V_{\chi\cA\psi} &=& -\frac{i}{4}(-\delta_{ac}\delta_{bd} + \delta_{ad}\delta_{bc})[\lambdabar_{ABC}e^{ip_c} - \lambda_{ABC}e^{iq_d}]~,\nn \\
V_{\chi\cAb\chi} &=& -\frac{i}{8}\epsilon_{abcde}\bigg(e^{i(k_a+k_b+k_c)} [\lambda_{ABC}e^{ip_c} - \lambdabar_{ABC}e^{i(q_a + q_b)}] \nn \\
&&- e^{i(p_d+p_e+p_c)}[\lambdabar_{ABC}e^{ik_c} - \lambda_{ABC}e^{i(q_d + q_e)}] \bigg)~.
\eea
\begin{figure}[pb]
\centerline{\psfig{file=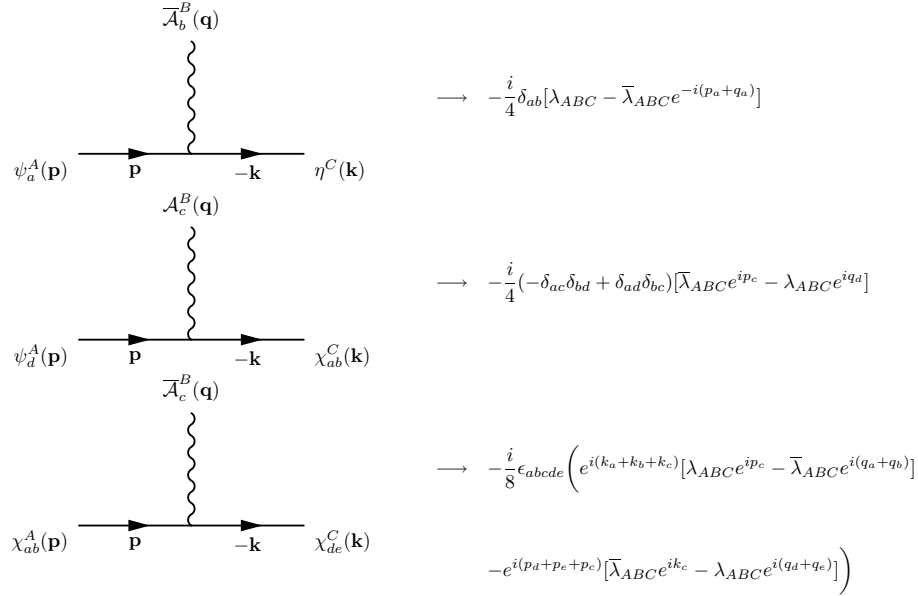,width=12cm}}
\vspace*{8pt}
\caption{\label{fig:vertices}The boson-fermion vertices on the lattice.}
\end{figure}

\subsection{Renormalized fermion propagators at one-loop}
\label{sec:1-loop}

The propagators and vertices described above can be used to construct four different {\em amputated} diagrams at one-loop. The renormalized fermion propagators receive contributions from these amputated diagrams.
\begin{itemize}
\item The amputated $\eta\psi$ diagram. 
\bea
\label{eq:eta-psi}
I_{\eta \psi}(\vp) &=& \sum_{\vk, \vq} \sum_{BC} \sum_{abc} \delta_{-\vp,\vk+\vq}\Big[\frac{1}{\widehat{\vk}^2} [(e^{ik_b}-1)\delta_{ac}-(e^{ik_c} - 1) \delta_{ab}]\Big]\cdot\Big[ \frac{1}{\widehat{\vq}^2}\Big]\nn \\
&& \cdot \Big[\frac{i}{4}[\lambda_{ABC} - \lambdabar_{ABC}e^{i(k_a+q_a)}]\Big]\nn \\
&&\cdot \Big[ \frac{i}{4}(-\delta_{ba}\delta_{cd} +  \delta_{bd}\delta_{ca})[\lambdabar_{BCD}e^{-ip_a} - \lambda_{BCD} e^{iq_d}]\Big]~.
\eea
\item The first amputated $\psi\chi$ diagram.
\bea
I^{1}_{\psi\chi}(\vp) &=& \sum_{\vk,\vq} \sum_{bcdefm} \sum_{BC}\Big[ \frac{1}{2\widehat{\vk}^2}\epsilon_{bcmef}e^{i(k_e + k_f)}(e^{ik_m} - 1)\Big] \nn \\
&&\cdot \Big[ \frac{1}{\widehat{\vq}^2}\Big]\cdot \Big[ -\frac{i}{4}(-\delta_{bd}\delta_{ca} + \delta_{ba} \delta_{cd})[\lambdabar_{ACB}e^{ip_d}-\lambda_{ACB}e^{-iq_a}]\Big]\nn \\
&&\cdot\bigg[\frac{i}{8}\epsilon_{efdgh} \Big(e^{ik_{(d+g+h)}}[\lambdabar_{BCD}e^{-ip_d} - \lambda_{BCD}e^{i(q_g + q_h)}]\Big]\nn \\
&& -e^{-ip_{(d+e+f)}}[\lambda_{BCD}e^{ik_d} - \lambdabar_{BCD}e^{i(q_e + q_f)}]\Big)\bigg]~.
\eea
\item The second amputated $\psi\chi$ diagram.
\bea
I^{2}_{\psi\chi}(\vp) &=& \sum_{\vk,\vq} \sum_{bc} \sum_{BC} \Big[\frac{2}{\widehat{\vk}^2}(e^{ik_c} - 1)\Big]\cdot \Big[ \frac{1}{\widehat{\vq}^2}\Big] \cdot \delta_{ab} \Big[-\frac{i}{4}[\lambda_{ACB} - \lambdabar_{ACB}e^{-i(p_a-q_a)}]\Big]\nn \\
&&\cdot \Big[-\frac{i}{4}(-\delta_{db}\delta_{ec} + \delta_{dc}\delta_{eb})[\lambda_{DCB}e^{ik_b} -\lambdabar_{DCB}e^{iq_c}]\Big]~.
\eea
\item The amputated $\chi\chi$ diagram. 
\bea
\label{eq:chi-chi}
I_{\chi\chi}(\vp) &=&\sum_{\vk,\vq} \sum_{cdef} \sum_{BC} \delta_{\vk+\vq-\vp, 0} \Big[\frac{1}{\widehat{\vk}^2}[(e^{-ik_e} - 1)\delta_{fd} - (e^{-ik_d} - 1)\delta_{fe}]\Big]\cdot\Big[\frac{1}{\widehat{\vq}^2}\Big]\nn \\
&&\cdot\bigg[-\frac{i}{8}\epsilon_{abcde} \Big(e^{-ik_{(a+b+c})}[\lambda_{ACB}e^{ip_c} - \lambdabar_{ACB}e^{-i(q_a + q_b)}]\nn \\
&&-e^{ip_{(c+d+e})}[\lambdabar_{ACB}e^{-ik_c}    - \lambda_{ACB}e^{-i(q_d + q_e)}]\Big)\bigg]\nn \\
&&\cdot\Big[-\frac{i}{4}(-\delta_{gc}\delta_{hf} + \delta_{gf}\delta_{hc})[\lambdabar_{BCD}e^{ik_c} - \lambda_{BCD}e^{iq_f}]\Big]~.
\eea
\end{itemize}

\begin{figure}[pb]
\centerline{\psfig{file=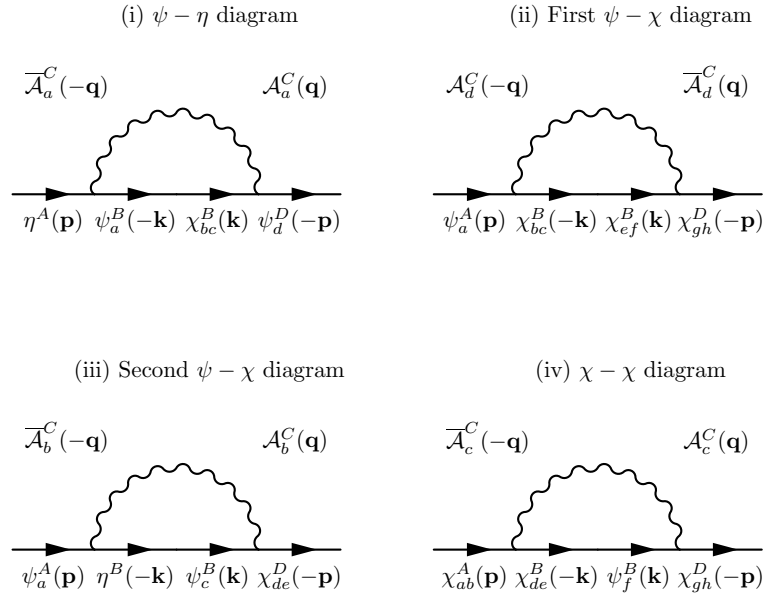,width=10cm}}
\vspace*{8pt}
\caption{\label{fig:one-loop}One-loop diagrams of fermions and complexified gauge fields.}
\end{figure}

Notice that the contributions of these diagrams all vanish in the limit $p \rightarrow 0$. This implies that mass counterterms are not present in the lattice theory at one-loop. In Section~\ref{sec:general}, we argued that the only dangerous mass term involved a coupling of $\eta$ and $\psi_a$, revealed through the new operator $D$. Now we see that this term does not arise at one-loop. In the next section, we will see that this feature persists to all orders and, thus, leads to the conclusion that {\it no mass counterterms are needed at any finite order of perturbation theory}.

\subsection{The effective action}
\label{sec:eff-action}

The above result, the absence of boson and fermion mass terms, can be shown by deriving the effective action of the theory by invoking topological field theory ideas. We are interested in computing the partition function of the lattice theory in one-loop order around an arbitrary classical vacuum state in which the fermions vanish and the bosonic fields correspond to constant commuting matrices. On expanding the fields around such a constant commuting background,
\beq
\cU_a(\bn) = \cU_a + i \cA_a(\bn),~~~~\cU^{\dagger}_a(\bn) = \cU^{\dagger}_a - i \cAb_a(\bn)~,
\eeq
and choosing the gauge $\alpha=1/2$, the quadratic part of the bosonic action takes the form
\beq
S_B = -2 \sum_{\bn, a, b} \Tr \cA_b(\bn) \cD_a^{\dagger(-)} \cD_a^{(+)} \cA_b(\bn)~.
\eeq
The covariant derivatives here depend on the constant commuting classical background $[\cU_a(\bn), \cU^{\dagger}_a(\bn)] = 0$. After integration over the fluctuations in the bosonic fields, the bosonic contribution to the one-loop partition function is
\beq 
{\rm det}^{-5}( \cD_a^{\dagger(-)} \cD_a^{(+)} )~.
\eeq

The gauge-fixing functional (\ref{gfixf}) leads to the quadratic ghost action:
\beq
S_G = \sum_{\bn,a} \Tr \cbar\, \cD_a^{\dagger (-)} \cD_a^{(+)} c~.
\eeq
The quadratic fermionic part of the action is given by the corresponding terms in (\ref{eq:action-on-a4-star1}), except that now the covariant derivatives depend only on the background fields.

Since the background is constant, we can pass to momentum space in which the action separates into terms for each mode $\vk$.  The $16 \times 16$ fermion matrix $M(\vk)$ for the mode $\vk$ then can be shown to satisfy
\beq
\det M(\vk) = \det ( \cD_a^{\dagger(-)}(\vk) \cD_a^{(+)}(\vk) )^8~.
\eeq
Going back to position space, and taking into account the fact that there is a double counting of modes in the matrix form (\ref{matprod}), we obtain 
\beq
{\rm Pf} ({\cal M}) = {\rm det}^4( \cD_a^{\dagger(-)} \cD_a^{(+)} )~.
\eeq
The ghosts add another factor of $\det (\cDb_a^{\dagger(-)} \cD_a^{(+)})$, which is just what is needed to cancel the bosonic contribution given earlier.  

The conclusion is that the effective action of the lattice theory at one-loop obtained by expanding about an arbitrary point in the classical moduli space becomes identically zero. Similar to the continuum case, since the moduli space of the lattice theory is not lifted we can conclude that there can be no boson or fermion masses at one-loop. It is expected that this analysis can be extended to all loops. The topological invariant nature of the partition function allows us compute it exactly in the semi-classical approximation. In Ref. \refcite{Matsuura:2007ec} similar arguments are used to show that the vacuum energy of supersymmetric lattice theories with four and eight supercharges remains zero to all orders in the coupling. The calculation presented above extends that analysis to the case of sixteen supercharges. Thus, the conclusion is that the boson and scalar masses remain zero to all orders in the coupling constant, and in turn, the fermions also remain massless, which is consistent with the explicit one-loop calculation above.

The expressions for the amputated one-loop diagrams derived above, Eq. (\ref{eq:eta-psi})-(\ref{eq:chi-chi}), contribute to the renormalization of the three twisted fermion propagators. This is sufficient to calculate $\alpha_1, \alpha_2$ and $\alpha_4$ that appear in the general action: 
\bea
{\cal L} &=& \frac{1}{g^2}\sum_{\bn,a,b,c,d,e} ~\Big\{ \cQ ~\Tr \Big[-i\alpha_1 \chi_{ab} \cD^{(+)}_a \cU_b(\bn) -i \alpha_2 \eta(\bn) \cD^{\dagger(-)}_a \cU_a(\bn) +\frac{\alpha_3}{2}\eta(\bn)d(\bn)\Big]\nn \\
&& - \frac{\alpha_4}{2} \Tr \epsilon_{abcde} \chi_{de}(\bn + \hatbmu_a + \hatbmu_b + \hatbmu_c) \cD^{\dagger(-)}_{c} \chi_{ab}(\bn + \hatbmu_c)\Big\}~.
\eea
The extraction of the coefficient $\alpha_3$ requires further work. One simple way to extract it is through a computation of the renormalized auxiliary boson propagator.

\subsection{One-loop diagrams for the auxiliary field propagator}
\label{sec:aux-one-loop}

Instead of integrating out the auxiliary field $d$ from the bosonic action, we look at the off-shell form
\beq
S_B = \sum_{\bn,a,b} \Tr \Big( \cF_{ab}^{\dagger}(\bn) \cF_{ab}(\bn)  -\frac{i}{g}d(\bn) \cD^{\dagger(-)}_a \cU_a(\bn) + \frac{1}{2} d^2(\bn)\Big)~,
\eeq 
where $\cF_{ab}(\bn) = -\frac{i}{g}\cD_a^{(+)}\cU_b(\bn)$ to compute the renormalized propagator for the $d$ field.

The Feynman rules for the fermions are identical to those of the previous (on-shell) analysis, but the boson propagators change and so they need to be recomputed in this off-shell scenario. On expanding the link fields around unity and using the same lattice gauge-fixing term as before we find the momentum space form for the gauge fixing term:
\beq
S_{GF}[A] =\frac{1}{\alpha} \sum_{\vk,a,b} \Tr A_a(\vk) f^*_a(\vk)f_b(\vk) A_b(-\vk)~.
\eeq

\begin{figure}[pb]
\centerline{\psfig{file=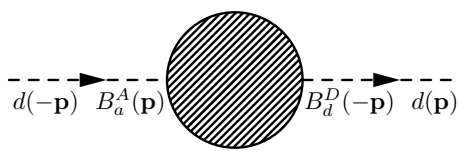,width=6cm}}
\vspace*{8pt}
\caption{\label{Bbubble}The generic diagram contributing to renormalized $d$ propagator.}
\end{figure}

It is convenient in this calculation to work with the real and imaginary parts of the complex gauge field explicitly, thus, 
\beq 
\cA_a = A_a + iB_a~.
\eeq
The gauge-fixed bosonic action on the lattice to quadratic order in fields, with the choice $\alpha = \frac{1}{2}$, is then
\bea
\label{offshellA}
S^{(2)}_B &=&\sum_{\vk,a,b} \Tr 2A_a(\vk)\Big[\delta_{ab}f_c(\vk)f^*_c(\vk)\Big]A_b(-\vk)\nn \\
&&+2B_a(\vk)\Big[\delta_{ab}f_c(\vk)f^*_c(\vk)-f^*_a(\vk)f_b(\vk)\Big]B_b(-\vk) \nn \\
&&-2id(\vk) f_a(\vk) B_a(-\vk) +\frac{1}{2} d(\vk) d(-\vk)~.
\eea 
The $d-B_a$ system decouples from $A_a$ to the order. Its action is given by
\bea
S^{(2)}_B [d, B_a]&\sim& \sum_{\vk,a,b} \Tr 2B_a(\vk)\Big[\delta_{ab}f^*_c(\vk)f_c(\vk)-f^*_a(\vk)f_b(\vk)\Big]B_b(-\vk)\nn \\
&&-2id(\vk) f_a(\vk) B_a(-\vk) +\frac{1}{2} d(\vk) d(-\vk)~.
\eea 
or in matrix form
\bea
\left(\begin{tabular}{cc}$d$ & $B_a$ \end{tabular}\right)(\vk)
\left(\begin{tabular}{cc}
$\frac{1}{2}$ & $-if_b(\vk)$ \\
$-if^*_a(\vk)$  & $M_{ab}(\vk)$
\end{tabular}
\right)
\left(
\begin{tabular}{c}
$d$ \\
$B_b$ 
\end{tabular}
\right)(-\vk)~,
\eea
where $M_{ab}(\vk) = 2[\delta_{ab}\sum_c f_c(\vk)f^*_c(\vk) - f^*_a(\vk)f_b(\vk)]$. Using standard identities for the inverse of a partitioned matrix, we find
\beq
M^{-1} = \left(
\begin{tabular}{cc}
$\frac{1}{2}$ & $-if_b(\vk)$ \\
$-if^*_a(\vk)$  & $M_{ab}(\vk)$
\end{tabular}
\right)^{-1} = \frac{1}{\sum_c f_c(\vk)f^*_c(\vk)}\left(
\begin{tabular}{cc}
0 & $if_b(\vk)$ \\
$if^*_a(\vk)$  & $\frac{1}{2}\mathbf{1}_5$
\end{tabular}
\right)~.
\eeq
We have $\sum_c f_c(\vk)f^*_c(\vk) = 4 \sum_c \sin^2\Big(\frac{\vk_c}{2}\Big)$ and, as before, we define $\widehat{\vk}^2 \equiv 4 \sum_c \sin^2\Big(\frac{\vk_c}{2}\Big)$. Thus the lattice propagators are
\bea
\langle d^A(\vk)d^B(-\vk) \rangle &=& 0~, \\
\langle d^A(\vk)B^B_a(-\vk) \rangle &=& i\delta_{AB}\frac{(e^{-ik_a}-1)}{\widehat{\vk}^2}~, \\
\langle B^A_a(\vk)B^B_b(-\vk) \rangle &=& \delta_{ab}\delta_{AB}\frac{1}{2\widehat{\vk}^2}~.
\eea
From (\ref{offshellA}) the propagator for the $A$ field is also
\beq
\langle A^A_a(\vk)A^B_b(-\vk) \rangle = \delta_{ab}\delta_{AB}\frac{1}{2\widehat{\vk}^2}~.
\eeq
The $d$ field is non-propagating at tree level as it should be since it is an auxiliary field. On using these propagators and those derived earlier for the bosons and fermions, we can now write down the generic Feynman diagram contributing to a renormalization of the auxiliary boson propagator. It is shown in figure~\ref{Bbubble} and represents the set of amputated diagrams possessing two external $B$ field legs. These combine with the external $\langle dB \rangle$ propagators derived above to yield the renormalized propagator for the auxiliary field. The vanishing of the tree level $\langle dd \rangle$ propagators ensures that no amputated diagrams with 2 $d$ field external legs contribute.

The set of all such lattice Feynman diagrams is shown below and corresponds to a subset of the $B$ field vacuum polarization diagrams.

\begin{figure}[pb]
\centerline{\psfig{file=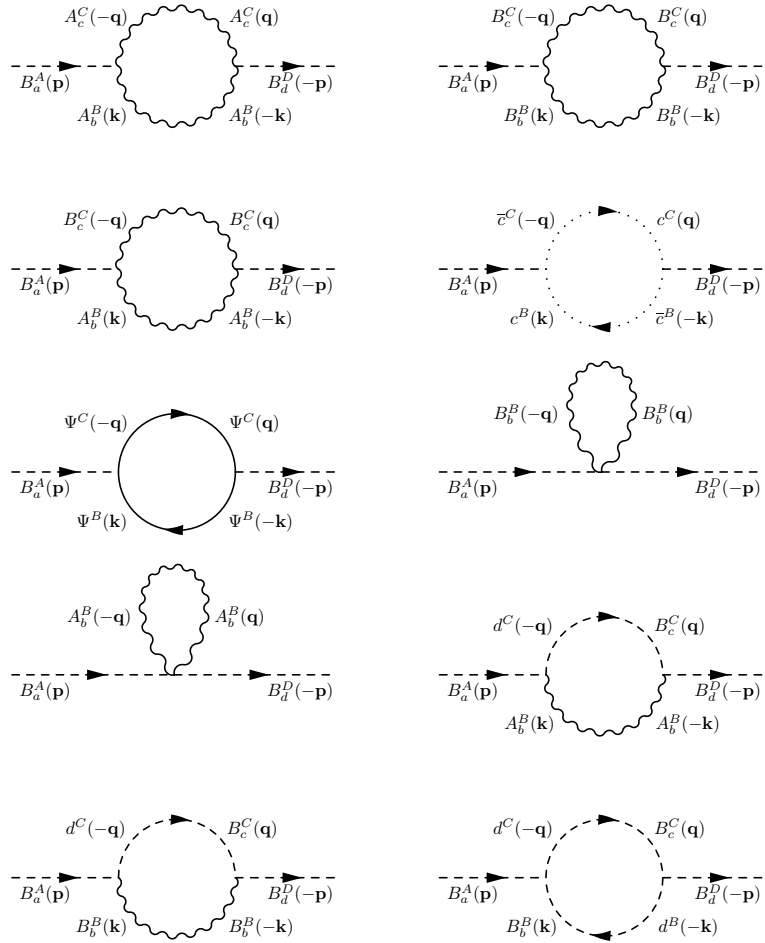,width=10cm}}
\vspace*{8pt}
\caption{\label{AllBbubbles} Set of all lattice amputated Feynman diagrams contributing the renormalized $d$ propagator.}
\end{figure}

It is important to notice that almost all these diagrams appear in the continuum off-shell twisted theory. The exceptions are just the diagrams containing a $BBd$ vertex that corresponds to the lattice vertex
\beq
V_{dBB} = \langle d^A(-\vk-\vq)B^B_a(\vk)B^C_b(\vq) \rangle = \frac{i}{2}\delta_{ab}(\lambda_{ABC}+\lambdabar_{ABC})(1-e^{-i(k_a+q_a)})~.
\eeq
These diagrams do not contribute to the divergent piece in the $\langle dd \rangle$ propagator at this order of perturbation theory since the above vertex vanishes as the lattice spacing is sent to zero.

Hence, we are left with a set of diagrams that correspond to those of the equivalent continuum theory at one-loop order. This fact can be exploited to argue that the leading logarithmic divergences of the lattice theory are shared with the continuum theory.

\subsection{The divergence in the one-loop diagrams}
\label{sec:div-structure}

At this stage, the expressions we have in hand, the amputated one-loop diagrams that determine the renormalization of three fermion propagators and the set of Feynman graphs needed to renormalize the auxiliary bosonic field propagator would in principle, allow us to determine all four coefficients $\alpha_i$ appearing in the renormalized action (\ref{eq:general-form}). The question of how much fine tuning of the lattice theory is required to regain full supersymmetry would be determined by the parts of those expressions that diverge as the lattice spacing is sent to zero. So it is necessary to evaluate those expressions for the one-loop integrals as the lattice spacing goes to zero. 

We first focus on the diagrams contributing to the fermion propagators. The one-loop fermion propagators all vanish as the external momentum goes to zero, which is consistent with the earlier effective action computation showing that no fermionic mass terms can be generated perturbatively. But coming to the amputated one-loop diagrams, there is a difficulty, due to the power counting theorem of Reisz\cite{Reisz:1988a,Reisz:1988b,Reisz:1988c,Reisz:1988d}: we cannot simply take the naive continuum limit of the expressions for the the amputated one-loop diagrams as they have a naive degree of divergence of 1. It turns out that a trick due to Ref. \refcite{Kawai:1981} and detailed in Ref. \refcite{Capitani:2002} can be used to extract the leading divergences from those diagrams without disrespecting Reisz's theorem.

The trick is to split the integral $I(\mathbf{p})$ into two pieces:
\bea
\lim_{a \rightarrow0} I(\mathbf{p})&=& \lim_{a \rightarrow0}\left[ I(\mathbf{p}) -I(\mathbf{0})-\sum_b p_b \frac{\partial I}{\partial p_b} \bigg|_{\mathbf{p}=\mathbf{0}} \right] \nn \\
&&~~~~~~~~~~~~~~~~+\lim_{a \rightarrow0}\left[I(\mathbf{0})+\sum_b p_b \frac{\partial I}{\partial p_b}\bigg|_{\mathbf{p}=\mathbf{0}}\right]~.~~~~~~~~~~
\eea
Now the first term in square brackets can be evaluated in the naive continuum limit and contains no divergence. The divergence is in the second term but since there is no external momenta in the integrand, its evaluation on the lattice becomes simple. Not that, $I(\mathbf{0})$ vanishes for each of the diagrams so the calculation becomes even more simple.

Next step obviously would be the numerical evaluation of the integral for a variety of regulator masses $\mu$\footnote{The regulator mass controls the behavior of the integrand close to the origin of momentum space.} and extract the logarithmic divergence and any constant contributions using a fitting procedure. It turns out that, we could use a simpler approach if we are only interested in the leading log divergences, in which a naive continuum limit can be taken and the expressions evaluated using the dimensional regularization.

As a result of such procedure, we obtain the following expressions for the fermion self-energy diagrams:
\bea
I_{\eta \psi_d}(\mathbf{p}) &\sim& -\frac{i}{8 \pi^2} p_d f_{ABC}f_{BCD} \log \mu a~,\nn \\
I_{\psi_a \chi_{de}}(\mathbf{p}) &\sim& \frac{i}{8 \pi^2}f_{ABC} f_{BCD}(\delta_{da}p_e-\delta_{ea}p_d)\nn \\
I_{\chi_{ab} \chi_{gh}} (\mathbf{p}) &\sim& -\frac{i}{16 \pi^2} f_{ABC} f_{BCD} \sum_d \epsilon_{abdgh} p_d \log \mu a~.
\eea
The cutoff $\frac{1}{a}$ has been inserted inside the logarithm to ensure that it is dimensionless\footnote{Here the case considered is of infinite lattice size which reduces all lattice sums in momentum space to integrals.}. 

The amputated divergent diagrams for the lattice $d$ propagator are also log divergent. It is possible to extract the sum of these logarithmic divergences using the same tricks used for the fermions, evaluating the diagram in the naive continuum limit. The sum of all these diagrams, contracted with external $dB$ propagators, will then yield a log divergent term of the
form
\beq
C_{dd}=cf_{ACB}f_{DCB}\log{(\mu a)}~,
\eeq
where $c$ is a constant to be determined by explicitly evaluating the diagrams. However, it is not necessary to evaluate these diagrams, even in the continuum, to determine $\alpha_3$ -- the requirement that the continuum theory preserve full supersymmetry will automatically determine $\alpha_3$ in terms of the other $\alpha_i$ corresponding to the fermion propagator renormalization.

\subsection{Renormalized propagators}

\begin{figure}[pb]
\centerline{\psfig{file=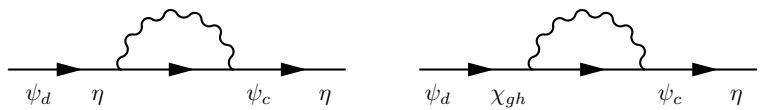,width=10cm}}
\vspace*{8pt}
\caption{\label{fig:full_p}Full $\eta \psi$ propagators.}
\end{figure}

Upon combining the divergent parts of the individual amputated diagrams computed above we can compute the leading logarithmic divergences appearing in the renormalized propagators. Several of the amputated fermion diagrams may appear as internal bubbles when correcting a given fermion propagator. For example, in the $\psi \eta$ diagram shown in figure \ref{fig:full_p}, naively three of the amputated diagrams contribute to the renormalization of this propagator. However, the underlying Lorentz structure of the propagators and integrals (at least in the case of the $\log$ divergences) restricts the contributions from some of the diagrams. As a result, only the $\eta \psi$ amputated diagram contributes to the renormalization of the $\eta \psi$ propagator.

The renormalization of the $\eta \psi$ propagator, denoting the full diagrams by $C$, is given by
\bea
C_{\psi_d \eta} &\sim& \frac{1}{4 \pi^2} f_{ABC}f_{BCD}\frac{2ip_d}{p^2} \log \mu a~.
\eea

For the case of other diagrams, it can be shown that only $I_{\psi \chi}$ contributes to $C_{\psi \chi}$ and $I_{\chi \chi}$ to $C_{\chi \chi}$.
\bea
C_{\psi_a \chi_{de}} &\sim& \frac{1}{4 \pi^2} f_{ABC}f_{BCD} \frac{ip_d \delta_{ae} -ip_e \delta_{ad}}{p^2}~\log \mu a~.
\eea
Noting that the internal propagator in $I_{\chi \chi}$ can be a $\psi \chi$ or $\chi \psi$, the full diagram $C_{\chi \chi}$ comes with a factor of 2.
\bea
C_{\chi_{ab} \chi_{de}} &\sim& \frac{1}{4 \pi^2} f_{ABC} f_{BCD}\log \mu a \sum_{k} \frac{ip_k}{2p^2} \epsilon_{abkde}~.
\eea

The coefficients of the propagators in the renormalized propagator amplitudes $C$ now determine the coefficients $\alpha_i$. Explicitly, they take the form
\beq
\alpha_i = 1 + b_i \log{\mu a} \quad i = 1, 2, 4~,
\eeq
where 
\beq
b_i = b = \frac{g^2N}{4\pi^2}~,
\eeq
and the property $f_{ABC}f_{BCD} = N \delta_{AD}$ is used for the case of $SU(N)$. This restriction is needed as the color structure of any counterterms must match that of the tree propagators. For the case of $U(N)$ it becomes $f_{ABC}f_{BCD} = N (\delta_{AD} - \delta_{A0} \delta_{D0})$. The $U(1)$ trace part can be ignored in the continuum as it simply decouples from the rest of the system. On the lattice, especially when performing simulations, a similar result can be achieved by giving the $U(1)$ mode a large mass of the order of the cut-off that will serve to decouple it from the $SU(N)$ modes at finite lattice spacing. This will result in breaking of supersymmetry in this sector but it may be removed by sending the $U(1)$ mass to zero {\it after} taking the continuum limit.

Naive expectations would lead us to conclude that the coefficients $b_i$ are all different. However, the above results indicate that, in fact, the log divergent parts of $b_i$ and, hence, $\alpha_i$ are actually all equal. The reasoning is, to untwist the continuum theory into a theory with four Majorana spinors requires that the continuum twisted fermions exhibit a common wavefunction renormalization. This just follows from the fact that the individual components of the spinors mix the different twisted fermions together. To achieve this requires that the corresponding renormalization constants of the kinetic terms $\alpha_i$ should all be equal -- agreeing with the findings above. Furthermore, since the leading log behavior of the lattice theory is the same as the continuum, one should expect that the log divergent part of the lattice couplings behave in the same way. Thus, a single wavefunction renormalization of the twisted lattice fermions is all that is needed to render the renormalized theory finite. 

The common anomalous dimension of the fermions in this twisted scheme is then given by 
\beq
\gamma=\frac{g^2N}{8\pi^2}.
\eeq

The leading log divergent contribution to the $\langle dd\rangle$ propagator can be computed from the naive continuum limit of the corresponding continuum expression for the sum of the $BB$ bubble diagrams given in diagram~\ref{AllBbubbles}. Combined with the fact that the tree level $\langle dB\rangle$ propagators required on the outside of these $BB$ amputated diagrams are the same as the continuum to $\mathcal{O}(a)$, the log divergence in the mass renormalization of the $d$ field must be the same on the lattice as in the continuum. Using this fact it can be argued that the log divergent part of $\alpha_3$ must actually be equal to that of the fermions, $\alpha_1$, for example. This follows from the fact that the bosonic action for general $\alpha_i$ can be rewritten as
\beq
\alpha_1 \Big(\cFb_{ab} \cF_{ab}\Big) + \frac{\alpha_2^2}{\alpha_3}\Big(\frac{1}{2}[\cDb_a,\cD_a]^2\Big)~.
\eeq
This renormalized bosonic action can be untwisted to yield the conventional gauge field plus scalar action in the continuum limit only for $\alpha_3 = \alpha_2 = \alpha_1$. This condition must be true since the continuum twisted theory exhibits full supersymmetry. Then the general arguments on renormalization described earlier indicate that the log divergence of $\alpha_3$ on the lattice must satisfy the same property. 

To conclude, the log divergent parts of the coefficients $\alpha_i, i = 1 \cdots 4$ must all be equal to one-loop order in the lattice theory. This implies that a common wavefunction renormalization of both twisted fermions and bosons is sufficient to render the renormalized theory finite at one-loop with all fields acquiring an anomalous dimension (in this scheme) given by $\gamma = \frac{g^2N}{8\pi^2}$. Physically, the equality of the couplings $\alpha_i, i = 1 \cdots 4$ means that {\it no} logarithmic fine tuning is required at weak coupling for the lattice theory to exhibit full supersymmetry as the lattice spacing is sent to zero.

\section{Applications to AdS/CFT}

The duality conjecture between gauge theories and string theories\cite{Itzhaki:1998dd} provides a very promising new direction for investigating the properties of gravitational theories. This conjecture may be used to describe certain black holes in terms of the world volume theories of the D-branes that compose them. The type II string theory reduces to a supergravity theory for low energies compared to the string scale $\alpha'^{-1/2}$. In this limit the gravitational theory contains black holes with finite temperature and $N$ units of D-brane charge. Then one can use the correspondence to describe these black holes in terms of the worldvolume theories of the D-branes, which are supersymmetric Yang--Mills theories with sixteen supercharges in various dimensions, taken in the large $N$ `t Hooft limit and at finite temperature. The string theory black holes are described by the strongly coupled sectors of these gauge theories. Solving the gauge theories would allow one to directly study the quantum properties of the dual black holes, including their thermodynamic properties. Since the gauge theories are strongly coupled, their direct computation needs nonperturbative techniques such as lattice simulations.

According to the AdS/CFT conjecture, type II superstring theory in $AdS_{d+1} \times M$ space, where $M$ is a compact manifold with positive curvature, should be equivalent to a superconformal field theory living on the $d$-dimensional boundary of $AdS_{d+1}$. Among the many generalizations of this conjecture we look closely at the two cases - those involving D0-branes and D1-branes - to apply the twisted field theory formulations to derive some interesting results. 

\subsection{D0-brane thermodynamics from lattice SYM}

The duality conjecture consists of a mapping between type II string theory containing $N$ Dp-branes and (p + 1)-dimensional supersymmetric gauge theories with gauge group $SU(N)$. Perhaps the simplest of these systems would be the one which describes the dynamics of D0-branes in type IIA string theory in terms of sixteen supercharge Yang--Mills quantum mechanics in the large $N$ `t Hooft limit. The low temperature string theory describes black holes with $N$ D0-brane charge, whose thermodynamics may hence be studied using the dual sixteen supercharge quantum mechanics with gauge group $SU(N)$. We need nonperturbative techniques such as lattice simulations to study the dual quantum mechanics since it is strongly coupled.

We consider a system of $N$ coincident D0-branes in the ``decoupling limit" with $N$ large and $Ng_s$ fixed, where $g_s$ is the string coupling. The decoupling limit is taken by considering excitations of the collection of D0-branes with fixed energy while sending the string length scale $\alpha'$ to zero. The degrees of freedom of the D0-brane system then split up into those localized near the branes (known as the `near horizon' excitations, which are of our interest) and those living far from the branes.

There are two ways to describe the degrees of freedom living near the D0-branes using perturbative approach depending on the value of the fixed quantity $Ng_s$. 

{\it Case 1. $Ng_s \ll 1$}

The D0-branes decouple from the ten-dimensional background gravitational theory. The degrees of freedom are described by the dynamics of the sixteen supercharge $SU(N)$ Yang--Mills quantum mechanics, which is the worldvolume theory of the D0-branes whose degrees of freedom are the open strings ending on the branes. At fixed energy excitations of the D0-branes and in the limit string length approaching zero, the higher string corrections become irrelevant. The Yang--Mills coupling is small in this regime and it is given by
\beq
g_{YM}^2 = g_s \alpha'^{-3/2}/(2\pi)^2~.
\eeq

{\it Case 2. $Ng_s \gg 1$}

The D0-branes couple strongly to gravity in this limit. The degrees of freedom are described by the target spacetime supergravity solution for $N$ D0-branes. Since the characteristic radius of curvature of the solution is much larger than the string length, the string worldsheet theory becomes weakly coupled. Thus supergravity is a good approximation in this regime. Geometrically this solution has two regions - the far horizon one, which is asymptotically flat 10-d spacetime and the near horizon one, which has a curvature characteristic to a black hole geometry. A potential energy barrier separates the finite energy excitations in these two regions. We focus on the `near-horizon' region excitations. Identifying $U$ and $U_0$ as the energy sales we wish to fix
\beq
U = \frac{r}{\alpha'}~~~U_0 = \frac{r_0}{\alpha'}
\eeq
with $U \geq U_0$. Then decoupling limit corresponds to taking $\alpha' \rightarrow 0$ keeping $U$, $U_0$ fixed. In this limit the entropy of the black hole becomes
\beq
\label{eq:entropy-D0}
S = \frac{1}{28 \sqrt{15}\pi^{7/2}} N^2 \Big(U_0/\lambda^{1/3}\Big)^{9/2}
\eeq
and the temperature is given by 
\beq
\label{eq:temp-D0}
T/\lambda^{1/3}= \frac{7}{16 \sqrt{15}\pi^{7/2}}\Big(U_0/\lambda^{1/3}\Big)^{5/2}
\eeq
where $\lambda = Ng_{YM}^2$.

It should be noted that both the above perturbative descriptions - the Yang--Mills for $Ng_s \ll 1$ and the stringy black hole for $Ng_s \gg 1$ - are not just limited to the regimes of $Ng_s$ where they are perturbatively good. In fact the Yang--Mills description ($Ng_s \ll 1$) is well defined for all $Ng_s$. The string black hole description ($Ng_s \gg 1$) is also valid for finite $Ng_s$ away from $Ng_s \gg 1$ but in that case one must take into account stringy $\alpha'$ corrections to the description of the black hole in supergravity.

The equations given in (\ref{eq:entropy-D0}) and (\ref{eq:temp-D0}) for entropy and temperature relate the Yang--Mills quantities to those of the string theory.

We define a parameter,
\beq
\beta = 1/t = \lambda^{1/3}/T~,
\eeq
which can be thought of as a dimensionless inverse temperature, characterizing the behavior of the theory. Then for large $\beta \gg 1$ (but still finite as compared with large $N$) the system of D0-branes should be well described by a supergravity black hole with negligible string corrections. Assuming the holographic correspondence is correct, we can predict (in the large $\beta$ limit) the precise form of the entropy and free energy of the Yang--Mills quantum mechanics,
\beq
S = 11.5 N^2 \beta^{-9/5},~~~f = -4.11N^2\beta^{-14/5}~,
\eeq
since it is known how to compute the Bekenstein--Hawking entropy of the supergravity black hole. Here $f$ is the dimensionless free energy, related to the usual free energy $F$ through $F = \lambda^{1/3}f$.

As $\beta$ decreases the curvature at the horizon radius becomes larger and the supergravity description receives string oscillator $\alpha'$ corrections. For small $\beta \ll 1$ the system can best be thought of as a highly excited hot ball of strings and branes. It has been argued that the hot ball of strings for $\beta \ll 1$ and the black hole at $\beta \gg 1$ are the same object, and the physics at the transition point $\beta \sim 1$ (the Horowitz--Polchinski `correspondence point') is therefore smooth\cite{Horowitz:1996nw}. The presence of a black hole in the dual string theory (geometrically, it is realized as the presence of a contractible Euclidean time circle in the theory) indicates the Yang--Mills theory is in a confined phase with a finite expectation value for the amplitude of the Polyakov loop $\langle | \frac{1}{N} \Tr e^{i\oint A d \tau} | \rangle$\cite{Witten:1998zw}. The appearance of a deconfined phase in the gauge theory would correspond to the absence of a black hole in the dual theory (geometrically, the presence of a non-contractible time circle) with a vanishing expectation value for the Polyakov loop. Thus we expect that the sixteen supercharge quantum mechanics at large $\beta \gg 1$ to be confined, as it is indeed dual to a black hole. At small $\beta \ll 1$ we expect that the sixteen supercharge theory is likely to be deconfined.

\begin{figure}[pb]
\centerline{\psfig{file=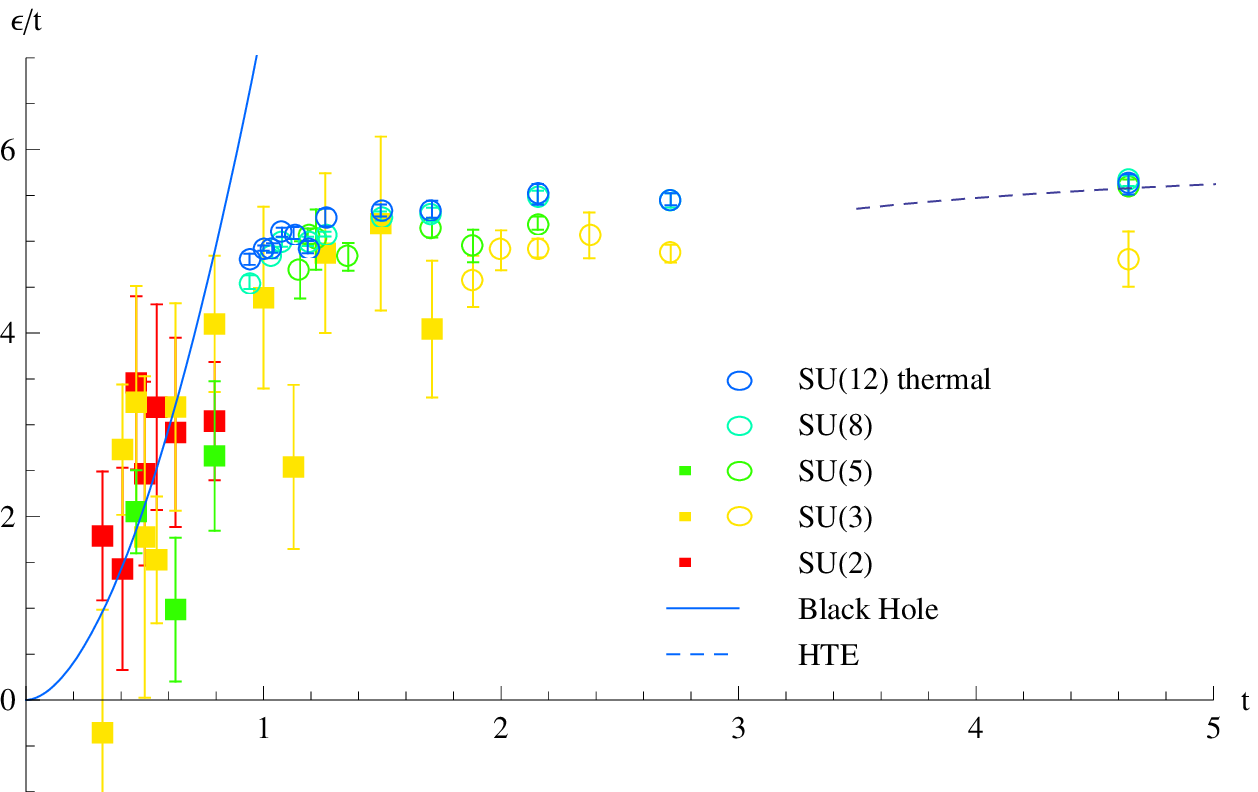,width=6cm}~~~\psfig{file=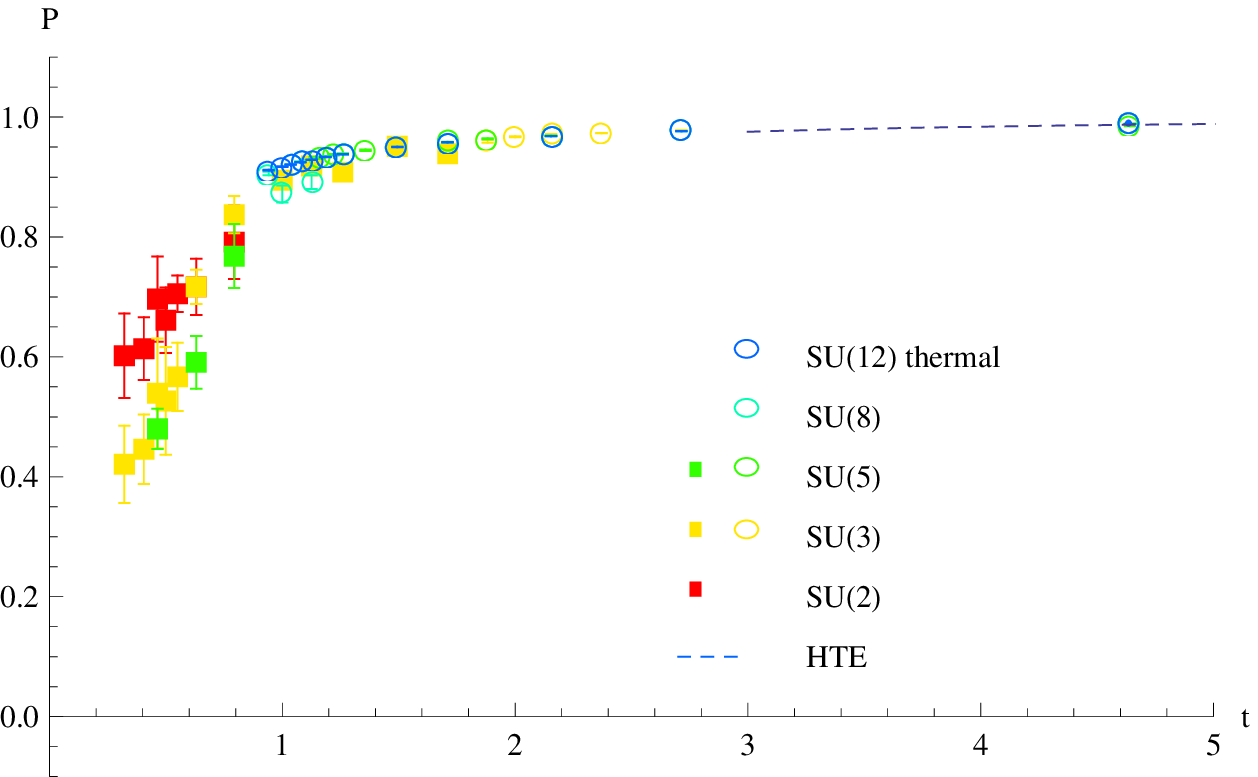,width=6cm}}
\vspace*{8pt}
\caption{\label{fig:energy-D0} {\it Left:} A plot of the dimensionless energy $\epsilon/t$ against dimensionless temperature $t$. The low temperature black hole prediction is shown using solid line. The high temperature expansion results are shown using asympotic dashed line. {\it Right:} A plot of the Polyakov loop observable $P$ against temperature $t$. For large $t$ the amplitude of the Polyakov loop has a finite expectation value and thus the sixteen supercharge quantum mechanics is in a deconfined phase. For small $t$ the theory is in a confined phase indicating the presence of a black hole in the dual gravitational theory. The two phases are connected through a smooth transition point, which agrees with the Horowitz--Polchinski correspondence point.}
\end{figure} 

Lattice simulations\footnote{See Ref. \refcite{Catterall:2011ce} for the details of the code for simulating SYM theoreis on the lattice.} of the sixteen supercharge $SU(N)$ Yang--Mills quantum mechanics have already been carried out in the `t Hooft limit\cite{Catterall:2008yz}. The simulations probed the low temperature regime for $N \leq 5$ and the intermediate and high temperature regimes for $N \leq 12$. The `t Hooft scaling of thermodynamic quantities has been observed and at low temperatures the numerical results are consistent with the dual black hole prediction. The intermediate temperature range is dual to the correspondence region, and the numerical results are consistent with smooth behavior there as well. 

The plots of dimensionless energy and expectation value of the Polyakov loop against temperature for various $N$ are shown in figure \ref{fig:energy-D0}. (Figures are from Ref. \refcite{Catterall:2008yz}.) The data points for energy approach a constant at very high temperatures corresponding to the result from classical equipartition assuming $N^2$ deconfined gluonic states. (In this limit the fermions acquire thermal masses and thus they are lifted out of the dynamics.) For low temperatures the energy approaches zero signaling the presence of a supersymmetric vacuum at vanishing temperature. The `t Hooft scaling sets in for small $N$, with $N = 3$ already giving results close to an extrapolated large $N$ result. The results obtained by the high temperature expansion\cite{Kawahara:2007ib} is denoted by the asymptotic dashed line and agree with the numerical data.

The simulations indicate that the data points appear to interpolate from high to low temperature smoothly. The intermediate temperature range $t \sim 1$ is dual to the Horowitz--Polchinski correspondence regime and the data indicate apparently smooth behavior at that regime. The low temperature behavior of the theory appears consistent with the prediction from supergravity (shown using solid curve near origin on the left plot).

There have been a series of numerical studies in recent years to explore the holographic principle between supersymmetric gauge theories and supergravity theories, focusing on the cases when the super Yang--Mills theory is one-dimensional and the dual gravitational theory describes the low energy dynamics of D0-branes \cite{Catterall:2008yz}$^,$\cite{Hanada:2007ti}$^,$\cite{Catterall:2007fp}$^,$\cite{Anagnostopoulos:2007fw}$^,$\cite{Hanada:2008gy}$^,$\cite{Hanada:2008ez}$^,$\cite{Catterall:2009xn}$^,$\cite{Hanada:2009ne} or the ${\cal N}=4$ theory compactified on $S^3\times {\mathbb R}$ \cite{Catterall:2010gf}$^,$\cite{Ishiki:2008te}$^,$\cite{Ishiki:2009sg}. In the next section we focus on the phase structure and thermodynamic properties of gravitational theories whose worldvolume theory is described by a collection of D1-branes\cite{Catterall:2010ya}$^,$\cite{Catterall:2010fx}.

\subsection{D1-brane thermodynamics from lattice SYM}

We consider the case where the low energy string theory consists of a large number of $N$ coincident D1-branes wrapped on a spatial circle, which, in the decoupling limit, is described by a two-dimensional maximally supersymmetric Yang--Mills theory on a circle\cite{Itzhaki:1998dd,Aharony:2004ig}. There exists a new dimensionless coupling in the gauge theory that can be varied in addition to the temperature when the spatial direction is compactified on a circle. Thus the two-dimensional Yang--Mills system possesses a richer structure at large $N$ than its one-dimensional counterpart. Arguments from a high temperature limit and also from strong coupling, using a dual supergravity description, indicate that the system should possess an interesting phase structure in the two-dimensional parameter space spanned by the temperature and this new coupling in the large $N$ limit\cite{Aharony:2004ig,Aharony:2005ew}. A large $N$ transition between confined and deconfined phases with respect to the spatial Polyakov line is expected, which interpolates between the high temperature region and the strongly coupled region. In particular, for the strongly coupled region, the dual D1-brane system can be described by certain black holes in supergravity, with a compact spatial circle. Then arguments from the dual gravity model indicate a first order Gregory--Laflamme\cite{Gregory:1993vy,Gregory:1994bj} phase transition between the black hole solutions localized on the circle and uniform black hole solutions which wrap the circle\cite{Aharony:2004ig,Aharony:2005ew,Susskind:1997dr,Li:1998jy,Martinec:1998ja,Kol:2002xz,Harmark:2004ws,Harmark:2005pq}. Translating back to the SYM, the dual gravity model predicts the parametric dependence of the transition temperature against dimensionless circle coupling -- a dependence that seemingly cannot be deduced by simple SYM considerations. Interestingly, since the relevant gravity solutions have not been constructed yet (analog solutions are known, but not in the correct dimension\cite{Kudoh:2003ki,Kudoh:2004hs,Headrick:2009pv}), the precise coefficient in this relation is not known, and determining it in SYM yields a prediction for the phase transition temperature that could be tested in the future when the gravity solutions (of a classical but non-trivial gravitational problem) are constructed.  

We are interested in studying large $N$ thermal two-dimensional maximally supersymmetric (16 supercharge) $SU(N)$ Yang--Mills theory, in the 't Hooft limit, with coupling $\lambda = N g_{YM}^2$, with the spatial direction compactified. Continuing the theory to Euclidean time, this implies the Yang--Mills theory is defined on a rectangular 2-torus, with time cycle size $\beta$, and space cycle size $R$. The fermion boundary conditions distinguish the two cycles, being anti-periodic on the time cycle so that $\beta$ has the interpretation of inverse temperature, and periodic boundary conditions on the space cycle. The action may then be written as:
\bea
S &=& \frac{N}{\lambda} \int_{T^2} d\tau dx \mathrm{Tr} \Big[~\tfrac{1}{4} F_{\mu\nu}^2 + \tfrac{1}{2} \sum_I [ D_\mu \phi^I , D^\mu \phi^I ]^2 - \tfrac{1}{4} \sum_{I , J} [ \phi^I , \phi^J ]^2 \nn \\
&&+ \mathrm{fermions}~\Big]~,
\eea
where $I,J = 1,\ldots,8$ and $\phi^I$ are the 8 adjoint scalars, and $\tau$ is the coordinate on the time circle, and $x$ the coordinate on the space circle. Since $\lambda$, $\beta$ and $R$ are dimensionful, it is convenient to work with the two dimensionless couplings:
\beq
r_\tau = \sqrt{\lambda} \, \beta~~~\textrm{and}~~~r_x = \sqrt{\lambda} \, R~,
\eeq
which give the dimensionless radii of the time and space circles, respectively, measured in units of the 't Hooft coupling. We will be interested in the expectation values of the trace of the Polyakov loops on the time and space circles,
\beq
P_\tau = \frac{1}{N} \Big{\langle} \Big| \Tr (P \exp (i \oint A_{\tau})) \Big| \Big{\rangle}~,~~P_x =  \frac{1}{N} \Big{\langle}  \Big| \Tr (P \exp (i \oint A_{x})) \Big| \Big{\rangle}~,
\eeq
as at large $N$, these give order parameters for confinement/deconfinement (or center symmetry breaking) phase transitions which we will discuss below.
\begin{figure}[pb]
\centerline{\psfig{file=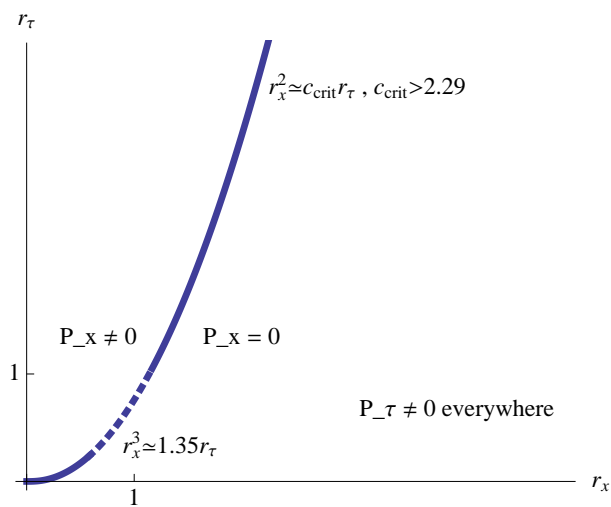,width=8cm}}
\vspace*{8pt}
\caption{\label{fig:phase}Cartoon of the expected large $N$, spatial Polyakov loop deconfinement transition line in coupling space. The simplest possibility is depicted, where the spatial deconfinement transition interpolates between the strong coupling Gregory-Laflamme transition parametric behavior $r_x^2 \sim r_\tau$, and the high temperature reduction deconfinement transition behavior $r_x^3 = 1.35 r_\tau$.}
\end{figure}
As discussed in\cite{Aharony:2004ig,Aharony:2005ew} there are several interesting limits for the theory. In the large torus limit, that is when $1 \ll r_x, r_\tau$, the string theory dual may be described by supergravity. For the weak coupling limit, $r_x , r_\tau \ll 1$, or asymmetric torus limits $r_\tau \ll r_x^3$ and $r_x \ll r_\tau^3$, we will find the dynamics are captured by a lower dimensional Yang--Mills theory. For $1 \ll r_\tau \ll r_x^2$, the string theory is in the Type IIB regime, where we expect $P_\tau \ne 0$ but $P_x = 0$. In the IIA regime, where $1 \ll r_\tau$ and $r_x^{4/3} \ll r_\tau$, we have $P_\tau \ne 0$, and $P_x \ne 0$ for $r_x^2 \leq c_{crit} \, r_\tau$ and $P_x = 0$ for $r_x^2 > c_{crit} \, r_\tau$, with $c_{crit}$ an order one constant with $c_{crit} > 2.29$. In the regime where both IIA and IIB apply, they give consistent results. Thus, in the large torus, supergravity regimes, the SYM is always deconfined in the time direction, and there is a first order deconfinement/confinement transition in the space direction at $r_x^2 = c_{crit} \, r_\tau$.

The simplest picture is then that the Gregory--Laflamme first order phase transition, $r_x^2 = c_{crit} \, r_\tau$ for $1 \ll r_\tau$, and the second order transition $r_\tau^3 \ll r_x$ and $r_x^3 = 1.35 \, r_\tau$ in the time reduced bosonic quantum mechanics are two ends of the same spatial Polyakov loop confinement/deconfinement phase transition line. At some point in-between, the order presumably changes, and here the new third order Gross--Witten phase transition emerges, although this is not measured by center symmetry breaking, but by more detailed information about the spatial Polyakov loop eigenvalue distribution. It is interesting\cite{Aharony:2004ig,Kol:2002xz,Harmark:2004ws} that the new phase at small $r_x$ also exists for $1 \ll r_x$ in the form of  non-uniform IIA black strings, but, unlike at weak coupling, these are never thermally dominant in the IIA supergravity region. In figure \ref{fig:phase}, the expected phase diagram for the spatial confinement/deconfinement transition is summarized.
\begin{figure}[pb]
\centerline{\psfig{file=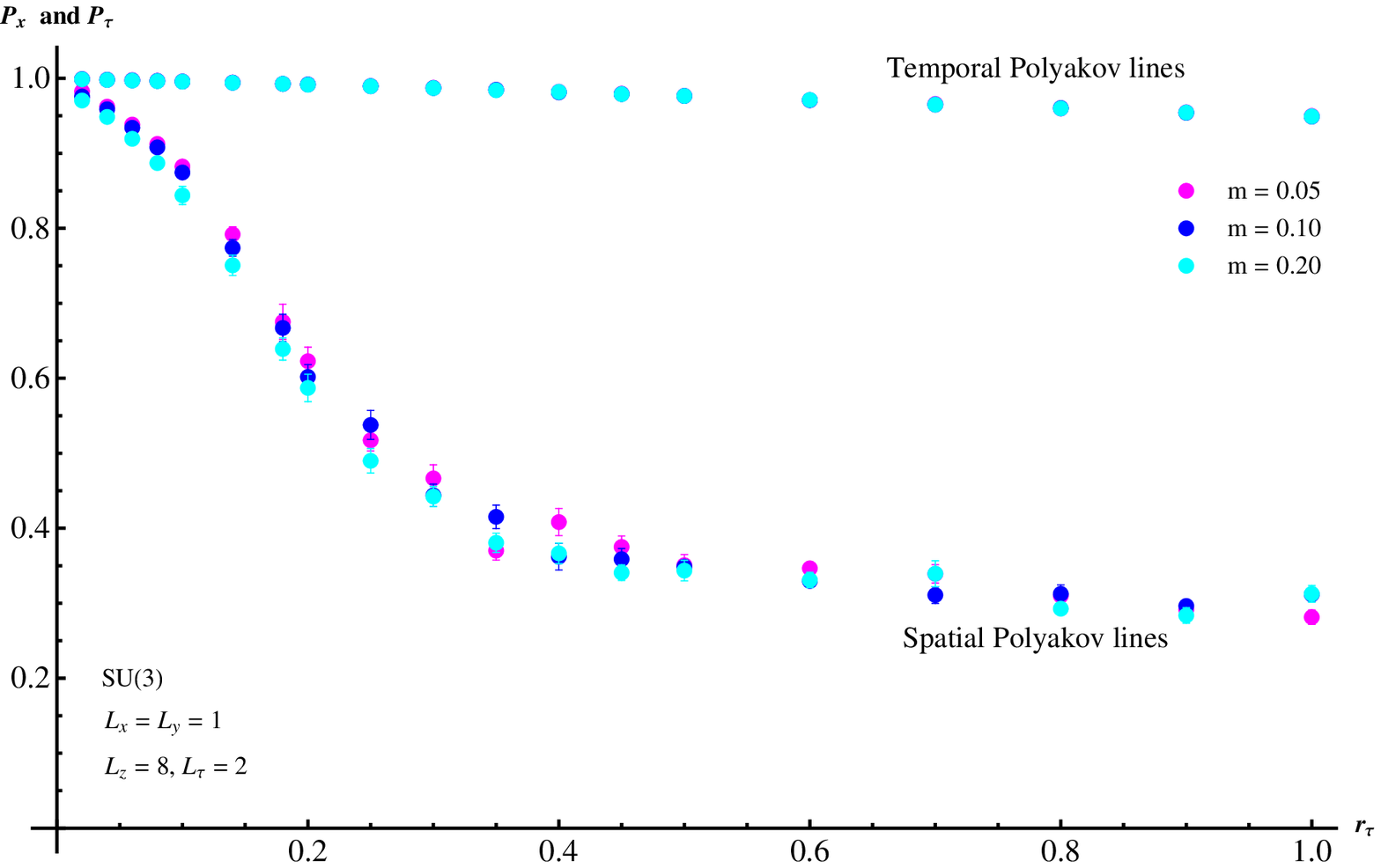,width=8cm}}
\centerline{\psfig{file=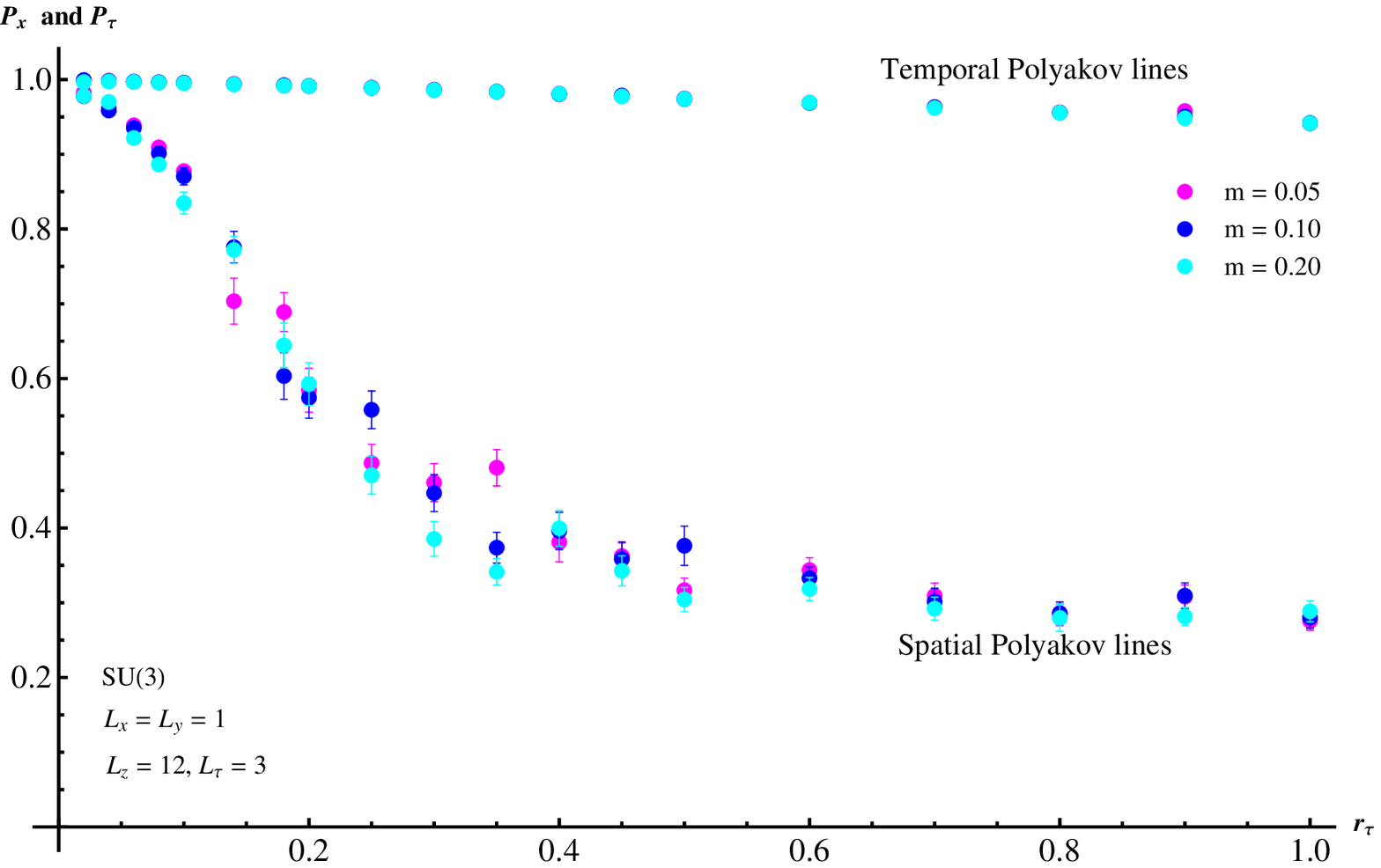,width=8cm}}
\vspace*{8pt}
\caption{\label{fig:transition}Spatial and temporal Polyakov lines ($P_x$ and $P_{\tau}$) against dimensionless time circle radius $r_\tau$ for maximally supersymmetric $SU(3)$ Yang--Mills on $2 \times 8$ and $3 \times 12$ lattices using different values of the infrared regulator m.}
\end{figure}

The numerical simulations of this theory focus on the Polyakov lines for both the thermal and spatial circle. These are defined on the lattice in the usual way
\beq
P_x = \frac{1}{N} \Big{\langle} \Big|\Tr \Pi_{a_x=0}^{L-1} U_{a_x}\Big| \Big{\rangle}~,~~P_\tau = \frac{1}{N} \Big{\langle} \Big|\Tr \Pi_{a_\tau=0}^{T-1} U_{a_\tau}\Big| \Big{\rangle}~,
\eeq
where the unitary piece of the complexified link $\cU_\mu$ is extracted to compute these expressions. The values of spatial and temporal Polyakov lines are evaluated as a function of $r_\tau$ for two different lattices with the same aspect ratio, a $2 \times 8$ lattice and a $3 \times 12$ lattice, for $N=3$ and with values of the infrared regulator $m = 0.05, 0.10$ and $0.20$. The use of two different lattices with the same aspect ratio would allow to test for and quantify finite lattice spacing effects. The simulations are performed for values of the dimensionless time circle radius in the range $0.02 \leq r_\tau \leq 1.0$. Figure~\ref{fig:transition} shows the numerical results.

\begin{figure}[pb]
\centerline{\psfig{file=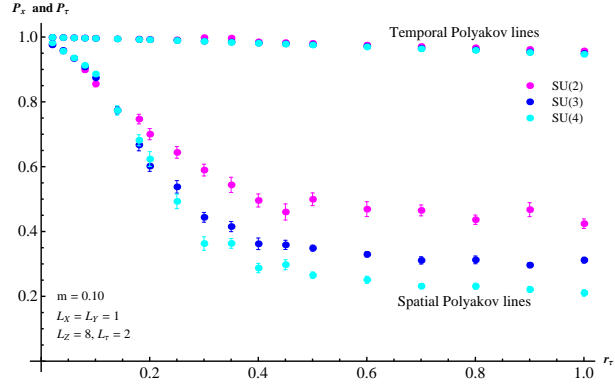,width=8cm}}
\vspace*{8pt}
\caption{\label{fig:color}\small{Plot of the absolute values of the spatial and temporal Polyakov lines ($P_x$ and $P_{\tau}$) against the dimensionless time circle radius $r_\tau$ for maximally supersymmetric $SU(N)$ Yang--Mills on a $2 \times 8$ lattice for $N = 2, 3, 4$, using the value of the infrared regulator m = 0.10.}}
\end{figure}

The temporal Polyakov remains close to unity over a wide range of $r_\tau$ indicating that the theory is (temporally) deconfined, and is consistent with expectations for the asymmetric torus limits, and the strong coupling regions, where there is a dual supergravity description in terms of black holes. However, the spatial Polyakov line has a different behavior taking values close to unity for small $r_\tau$ while falling rapidly to plateau at much smaller values for large $r_\tau$. It is tempting to see the rather rapid crossover around $r_\tau \sim 0.2$ as a signal for a would be thermal phase transition as the number of colors is increased. This conjecture is seen to be consistent with the data: in figure~\ref{fig:color} the Polyakov lines are shown for $N=2,3,4$ on $2\times 8$ lattices as a function of $r_\tau$. The plateau evident at large $r_\tau$ falls with increasing $N$ and the crossover sharpens. This is consistent with the system developing a sharp phase transition in the large $N$ limit. 

\begin{figure}[pb]
\centerline{\psfig{file=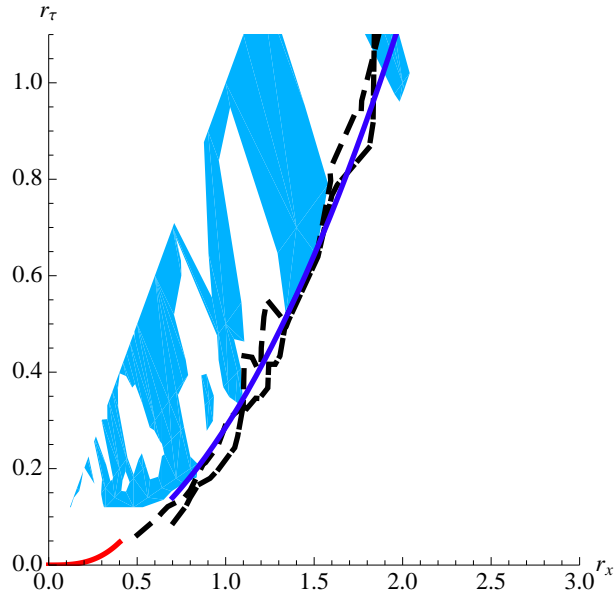,width=8cm}}
\vspace*{8pt}
\caption{\label{fig:contour2} \small{Plot showing a superposition of the $P_x = 0.5$ contours for $SU(3)$ and $SU(4)$ as dashed black lines. Plotted on the figure is the high temperature prediction for the transition ($r_x^3 = 1.35 r_\tau$, red curve). We note that the estimated large $N$ transition curve fits well both this high temperature prediction and also the strong coupling dual gravity predicted parametric behavior  $r_x^2 = c_{crit} r_\tau$. The data obtained through simulations suggests $c_{crit} \simeq 3.5$ (plotted as blue curve), which obeys the constraint from gravity $c_{crit} > 2.29$.}}
\end{figure}

Figure. \ref{fig:contour2} is obtained by analyzing the numerical data for $SU(3)$ and $SU(4)$. The superposition of the $P_x = 0.5$ contours for $SU(3)$ and $SU(4)$ is given as dashed black lines. The case where the $SU(4)$ loop $P_x$ is greater than the $SU(3)$ loop is denoted by shaded (blue) region, which is expected to estimate the large $N$ deconfined region for a first order transition (which gravity suggests at strong coupling). `Holes' in this blue region are due to statistical errors. The boundary of this region (ignoring `holes') seems to be matching well the $P_x = 0.5$ contours, and represents a best guess for where the large $N$ transition resides. This figure should be compared to figure \ref{fig:phase} giving a sketch of the expected phase structure. Plotted on the figure is the high temperature prediction for the transition ($r_x^3 = 1.35 r_\tau$, red curve). The estimated large $N$ transition curve fits well both this high temperature prediction and also the strong coupling dual gravity predicted parametric behavior  $r_x^2 = c_{crit} r_\tau$. The data obtained through simulations suggests $c_{crit} \simeq 3.5$ (plotted as blue curve), which obeys the constraint from gravity $c_{crit} > 2.29$.

The value of the ratio $\alpha \equiv c_{crit}/2.29$ gives the ratio of the GL thermal phase transition temperature to the Gregory--Laflamme dynamical instability temperature (the minimum temperature to which uniform strings can be supercooled), so $\alpha = T_{GL\; phase}/T_{GL\; instab}$. Although the Gregory-Laflamme instability temperature is known \cite{Aharony:2004ig} (corresponding to the behavior $r_x^2 = 2.29 r_\tau$ at strong coupling), the Gregory--Laflamme phase transition temperature is not known in the gravitational theory, as the localized solutions have not been constructed. The lattice estimation, $\alpha \simeq 1.5$, obtained through the analysis performed above provides a prediction for the thermal behavior of the gravity solutions. This is the first time a prediction about the properties of non-trivial classical gravity solutions has been made from the Yang--Mills side of a holographic correspondence. 

\section{Future Directions}

In this review we have focused on a set of of SYM theories with extended supersymmetries that exhibits compatibility with discretization on the lattice. These lattice theories are constructed based on the ideas drawn from topological twisting. The supersymmetries of certain Yang--Mills theories can be rewritten in a twisted form such that the resultant theory (twisting is just a change of variables in flat Euclidean spacetime) resembles like a topological field theory. The supersymmetries and fermions of the theory transform in integer spin representations of the twisted rotation group, and twisting always produces a nilpotent supersymmetry. The supersymmetry algebra associated with the nilpotent supersymmetry does not generate translations and thus we can implement this sector on the lattice hoping that the remaining broken supersymmetries on the lattice can be fine tuned, once we take the continuum limit of the lattice theory to approach the target theory. 

We have looked at the interesting case of four-dimensional $\cN=4$ SYM theory. At one-loop this theory needs only one wavefunction renormalization of the counterterms to achieve the continuum limit. The analysis was performed around a specific point in the coupling space, namely $\alpha_i = 1$. It would be interesting to do the one-loop analysis starting from an arbitrary point in coupling parameter space and study the flow of the couplings as the theory is renormalized. It would also be interesting to go beyond one-loop analysis on the lattice inspired by the topological field theory nature of the formalism. An interesting task would be to understand the perturbative lattice beta function, perhaps even to all orders. Also it would be interesting to see to what extent just some preserved supersymmetry charges guarantee scale invariance in the theory.\footnote{I thank Poul Damgaard for these suggestions.}

We could also extend the analysis in the context of AdS/CFT to three- and four-dimensional Yang--Mills systems, which are thought to be dual to D-2 and D-3 brane systems, using these exact supersymmetry lattice formulations.

\section{Acknowledgments}

I thank useful discussions with Simon Catterall, Eric Dzienkowski, Richard Galvez, Joel Giedt, Dhagash Mehta, Mithat \"Unsal, Robert Wells and Toby Wiseman. This work is supported in part by the LDRD program at the Los Alamos National Laboratory.

\end{document}